\documentclass[12pt]{article}
\pdfoutput=1
\newlength{\abstractwidth}
\usepackage{subcaption}
\usepackage{graphicx}
\usepackage{amsmath, nccmath}
\usepackage{amsfonts}
\usepackage{amssymb}
\usepackage{latexsym}
\usepackage{amsthm}

\usepackage{color}
\usepackage{graphicx}
\usepackage{tikz}
\usepackage{dsfont}

\usepackage{amssymb}  
\usepackage{physics}
\usepackage{mathtools}
\usepackage{amsmath}
\usepackage{tensor}
\usepackage{mathrsfs}
\usepackage{comment}

\usetikzlibrary{arrows,shapes,positioning}
\usetikzlibrary{decorations.markings}
\usepackage[rightcaption]{sidecap}
\tikzstyle arrowstyle=[scale=1]
\tikzstyle directed=[postaction={decorate,decoration={markings,
    mark=at position .65 with {\arrow[arrowstyle]{stealth}}}}]
\tikzstyle reverse directed=[postaction={decorate,decoration={markings,
    mark=at position .65 with {\arrowreversed[arrowstyle]{stealth};}}}]
\usetikzlibrary{positioning}

\usepackage{hyperref}
\definecolor{darkred}{rgb}{0.8,0.1,0.1}
\hypersetup{colorlinks=true, linkcolor=darkred, citecolor=blue, linktoc=page}

\usepackage{cite}

\renewcommand{\thefootnote}{\fnsymbol{footnote}}
\renewcommand{\thanks}[1]{\footnote{#1}}
\newcommand{\starttext}{
\setcounter{footnote}{0}
\renewcommand{\thefootnote}{\arabic{footnote}}}

\newcommand{\bea}{\begin{eqnarray}}
\newcommand{\eea}{\end{eqnarray}}
\newcommand{\be}{\begin{eqnarray}}
\newcommand{\ee}{\end{eqnarray}}
\newcommand{\bma}{\begin{matrix}}
\newcommand{\ema}{\cr\end{matrix}}

\newtheorem{thm}{Theorem}[section]

\newtheorem{prop}[thm]{Proposition}
\newtheorem{cor}[thm]{Corollary}

\newcommand{\SL}{ {\text {SL}}}

\newcommand{\SU}{ {\text {SU}}}
\newcommand{\Sp}{ {\text {Sp}}}

\newcommand{\U}{ {\text {U}}}
\newcommand{\AD}{ {\text {AD}}}

\newcommand{\N}{ {\mathcal N} }

\newcommand{\reals}{ {\mathbb R} }

\newcommand{\complex}{ {\mathbb C} }
\newcommand{\integers}{ {\mathbb Z} }

\def\cC{{\cal C}}
\def\cD{{\cal D}}

\def\cF{{\cal F}}

\def\cN{{\cal N}}
\def\cO{{\cal O}}

\def\cQ{{\cal Q}}

\def\cS{{\cal S}}
\def\cT{{\cal T}}

\def\bg{{\bf g}}

\def\bq{{\bf q}}

\def\mA{\mathfrak{A}}
\def\mB{\mathfrak{B}}

\def\mJ{\mathfrak{J}}

\def\ma{\mathfrak{a}}

\def\mr{\mathfrak{r}}

\def\ZZ{{\mathbb Z}}
\def\RR{{\mathbb R}}
\def\NN{{\mathbb N}}
\def\CC{{\mathbb C}}

\def\Re{{\rm Re \,}}
\def\Im{{\rm Im \,}}
\def\tr{{\rm tr}}

\def\det{{\rm det \,}}

\def\half{{1\over 2}}
\def\thalf{{\tfrac{1}{2}}}
\def\p{\partial}

\def\a{\alpha}
\def\b{\beta}
\def\g{\gamma}

\def\ep{\varepsilon}
\def\om{\omega}

\def\HE{{\sf E}}

\def\no{\nonumber}
\def\sm{\smallskip}

\begin{document}
\starttext
\setcounter{footnote}{0}

\begin{flushright}
2023  October 10  
\end{flushright}

\vskip 0.3in

\begin{center}

{\Large \bf Approaching Argyres-Douglas theories}
\footnote{This research was supported in part by the National Science Foundation under grant
PHY-22-09700.}

\vskip 0.1in

\vskip 0.3in

{\large  Sriram Bharadwaj, Eric D'Hoker} 

\vskip 0.1in

 { \sl Mani L. Bhaumik Institute for Theoretical Physics}\\
{\sl Department of Physics and Astronomy }\\
{\sl University of California, Los Angeles, CA 90095, USA} 

\vskip 0.05in

{\tt \small sbharadwaj@physics.ucla.edu, dhoker@physics.ucla.edu}

\end{center}

\vskip 0.5in
\begin{abstract}
The Seiberg-Witten solution to four-dimensional $\cN=2$ super-Yang-Mills theory with gauge group $\SU(N)$ and without  hypermultiplets is used to investigate the neighborhood  of the maximal Argyres-Douglas points of type $(\ma_1,\ma_{N-1})$. A convergent series expansion for the Seiberg-Witten periods near the Argyres-Douglas points is obtained by analytic continuation of the series expansion around the $\ZZ_{2N}$ symmetric point derived in {\tt arXiv:2208.11502}. Along with direct integration of the Picard-Fuchs equations for the periods, the expansion is used to determine the location of the walls of marginal stability for $\SU(3)$. The intrinsic periods and  K\"ahler potential of the $(\ma_1,\ma_{N-1})$ superconformal fixed point are computed by letting the strong coupling scale tend to infinity. We conjecture that the resulting intrinsic K\"ahler potential  is positive definite and convex, with a unique minimum at the Argyres-Douglas point, provided only intrinsic  Coulomb branch operators with unitary scaling dimensions  $\Delta>1$ acquire a vacuum expectation value, and provide both analytical and numerical evidence in support of this conjecture. In all the low rank examples considered here, it is found that turning on moduli dual to $\Delta \leq 1$ operators spoils the positivity and convexity of the  intrinsic K\"ahler potential.
\end{abstract}

\newpage

\baselineskip=15pt
\setcounter{equation}{0}
\setcounter{footnote}{0}

\newpage
\setcounter{tocdepth}{2}
\newpage

\pagenumbering{gobble}
\newpage
\tableofcontents
\newpage
\pagenumbering{arabic}

\newpage

\section{Introduction}
\label{sec:1}
\setcounter{equation}{0}

The Seiberg-Witten (SW) solution to four dimensional $\cN=2$ super-Yang-Mills theory provides the exact low energy effective action and BPS mass spectrum on the Coulomb branch. The vacuum expectation values (VEVs) of the gauge scalars and their magnetic duals are encoded in terms of a family of Riemann surfaces equipped with a  meromorphic differential, referred to as the SW curve and the SW differential, respectively.  The original construction was given for gauge group $\SU(2)$ in \cite{SW:1994rs, SW:1994aj}. In the present paper, we focus on theories with gauge group $\SU(N)$, primarily for $N\geq 3$ and without hypermultiplets, for which the SW curve and differential were constructed in \cite{Klemm:1994,Argyres:1994xh,Hanany:1995na,Argyres:1995wt}
 (see also \cite{Intriligator:1995au,Klemm:1995wp,DHoker:1996pva,Tachikawa:2013kta,Martone:2020hvy}).  

\sm

At generic points on the Coulomb branch, the gauge group $\SU(N)$ is spontaneously broken to its maximal Abelian subgroup $\U(1)^{N-1}$ and the low energy contents of the theory consist of $N-1$ massless Abelian $\cN=2$ gauge multiplets. The spectrum of massive BPS states includes the $N(N-1)$ gauge bosons and their magnetic counterparts. Remarkably, the SW solution predicts the existence of isolated points in the moduli space where the masses of one or several of these BPS state tends to zero. We now describe two distinct scenarios where a maximal number of massive BPS states become simultaneously massless.

\sm

There are $N$ \textit{multi-monopole points} in the moduli space, at each one of which $N-1$ \textit{mutually local} (i.e. with vanishing Dirac pairing) BPS states become simultaneously massless. At each multi-monopole point, there  exists an $\Sp(2N-2, \integers)$ electric-magnetic duality frame in which all the massless BPS states have purely  magnetic charges, whence the name. In view of their mutual locality, there exists an effective field theory description where the massless magnetic monopoles are described by hypermultiplets. Investigations into the behavior of the effective prepotential and periods in the neighborhood of a multi-monopole point may be found in \cite{Douglas:1995nw,DHoker:1997mlo,DHoker:2020qlp} as well as in \cite{DDN}, where the behavior of the K\"ahler potential and the walls of marginal stability are analyzed.

\sm

For gauge group $\SU(3)$, Argyres and Douglas discovered two so-called \textit{AD points} in the moduli space, where three \textit{mutually non-local} BPS states (i.e. with non-vanishing Dirac pairing) simultaneously become massless \cite{AD}. The corresponding AD theories are strongly  interacting $\cN=2$ superconformal field theories (SCFTs). For gauge group $\SU(N)$ with $N >3$ the maximal number of mutually non-local BPS states become massless at two \textit{maximal AD points}, thereby generalizing the case of $N=3$, while, for $N=2$, no AD points exist. Since the Dirac pairing is  invariant under $\Sp(2N-2, \integers)$, there is no electric-magnetic duality frame in which the massless BPS states are mutually local and simultaneously admit a standard local field theory description. 

\sm

While the absence of a standard local field theory formulation of the AD theories presents a considerable conceptual challenge, several indirect avenues of investigation have been explored, including the superconformal bootstrap and brane constructions in string theory. Here, we shall investigate the space of theories \textit{in the vicinity} of the maximal AD points, first by exploring the Coulomb branch of their embedding in $\SU(N)$ super-Yang-Mills and second by exploring their \textit{intrinsic Coulomb branch} obtained by sending the strong coupling scale $\Lambda$ of the $\SU(N)$ theory to infinity. The organization of the remainder of the paper and an overview of the results is presented in the subsections below. 

\subsubsection*{~ $\bullet$ Series expansion near the maximal Argyres-Douglas points}

In section \ref{sec2}, we compute the SW periods in a convergent series expansion around the maximal AD points. Our expansion provides a non-trivial analytic continuation of the strong-coupling expansion produced in \cite{DDN} around the unique $\integers_{2N}$-symmetric point. While the latter expansion contains the AD and the multi-monopole points on the boundary of its domain of convergence,  our expansion is centered at one or the other AD point and thereby provides a significant extension of the domain of convergence near the AD points.  On regions where they overlap, our expansion arguably has better convergence properties than the one given in \cite{DDN}. Finally, by taking the decoupling limit $\Lambda\rightarrow \infty$, where $\Lambda$ is the strong-coupling scale, we obtain the \textit{intrinsic} AD periods for $(\ma_1,\ma_{N-1})$ superconformal field theories in section \ref{sec4}.

\subsubsection*{~ $\bullet$ Charting candidate walls of marginal stability}

Two BPS states with central charges $Z_1$ and $Z_2$ and masses  $M_1= |Z_1|$ and $M_2=|Z_2|$ can form a  stable bound state provided its mass $M$  obeys $M< M_1+M_2$. The bound state is BPS when $M=|Z_1+Z_2|$, and becomes marginally stable when the binding energy vanishes, namely when $|Z_1+Z_2|=|Z_1|+|Z_2|$, which requires the ratio $Z_2/Z_1 $ bo be a real number. The reality of $Z_2/Z_1$ defines a real co-dimension one sub-variety of the Coulomb branch, referred to as a \textit{candidate wall of marginal stability}. Determining this sub-variety  was already undertaken in  \cite{SW:1994rs, SW:1994aj} for gauge group $\SU(2)$, and discussed in more detail in  \cite{Ferrari:1996sv,Bilal:1996sk,Bilal:1997st}. Candidate walls of marginal stability were investigated more recently in \cite{DDN} for gauge group $\SU(N)$ on restricted slices through the Coulomb branch. 
 
 \sm

In section \ref{sec3}, we shall map out candidate walls of marginal stability beyond the restricted slices of \cite{DDN}  for gauge group $\SU(3)$, and present partial results for  $N \geq 4$. The  series expansion around the AD points, discussed in the preceding subsection, will play a key role in gaining access to the walls of marginal stability beyond the special slices studied in \cite{DDN}. In addition, we shall adapt the numerical integration methods used in \cite{DDN} to  the computation of the SW periods and the central charges. These numerical computations will allow us to reach beyond the radius of convergence of either the $\integers_{2N}$ or the AD expansion, and to complete the charting of candidate walls of marginal stability.

\subsubsection*{~ $\bullet$ Exploring the intrinsic K\"ahler potential of the $(\ma_1,\ma_{N-1})$ AD theories}

Interest in the behavior of the K\"ahler potential, within the context of the SW solution, has recently been rekindled by the role it  may play in the soft breaking of $\cN=2$ super Yang-Mills theory and the renormalization group flow of this theory to adjoint QCD \cite{Cordova:2018acb, DHoker:2020qlp, DDN, DDGN}.  Specifically, the flow of the mass operator $M^2 \tr(\phi^\dagger \phi)$ for the gauge scalar $\phi$ purely within the $\cN=2$ super Yang-Mills theory is to the K\"ahler potential of the SW solution. Motivated in part by future work on soft supersymmetry breaking in or near AD theories, we initiate here a study of the intrinsic K\"ahler potential of the maximal AD theories. Key questions concern its positivity, convexity, and global minima properties.  

\sm

In section \ref{sec4}, we shall study the intrinsic periods of the AD theories;  calculate the intrinsic  K\"ahler potential; investigate the location of its minima; and understand its positivity and convexity properties.  We will  find that,  in the absence of deformations, namely moduli corresponding to operators with dimension $\Delta \leq 1$, but allowing the VEVs of genuine Coulomb branch operators with dimension  $\Delta > 1$ to be non-zero, the intrinsic K\"ahler potential exhibits positivity and convexity. This distinction between the dimensions coincides precisely with the unitarity bound on scaling dimensions of operators in any $\cN=2$ SCFT: $\Delta \geq 1$, where free bosonic fields saturate this bound. We shall gather compelling evidence that turning on Coulomb branch operators with unitary scaling dimensions is compatible with  maintaining positivity and convexity of the K\"ahler potential, and its unique global minimum being at the AD point.

\subsection*{Acknowledgements}

We gratefully acknowledge useful conversations with Thomas Dumitrescu and Emily Nardoni. SB is happy to thank  Lukas Lindwasser for conceptual discussions and Amey Gaikwad for suggestions on numerical calculations.  This research was supported in part by the National Science Foundation under grant PHY-22-09700.

\newpage

\section{Series expansion near a maximal AD point}
\label{sec2}
\setcounter{equation}{0}

We begin this section with a brief summary of the salient features of the Seiberg-Witten (SW) solution for four-dimensional $\cN=2$ super Yang-Mills theory  gauge group $\SU(N)$ without hyper-multiplets \cite{Klemm:1994,Argyres:1994xh,Klemm:1995wp}. We also review the expansion of the SW solution around the $\ZZ_{2N}$ symmetric point obtained in \cite{DDN}. A similar set-up is then used to derive a convergent expansion of the SW periods near one of the $\ZZ_N$ symmetric maximal AD points for $N \geq 3$, and to show that this expansion coincides with the analytic continuation of the $\ZZ_{2N}$ expansion of \cite{DDN}. A key tool in matching the expansions is the Gauss-Kummer quadratic transformation on hypergeometric functions. A detailed analysis of the domain of convergence  is undertaken and its results are presented graphically.

\subsection{Summary of the Seiberg-Witten solution}

The SW solution determines the vacuum expectation values of the gauge scalars~$a_I(u)$ and their magnetic duals~$a_{D,I}(u)$ as locally holomorphic functions of the gauge invariant Coulomb branch moduli~$u_n$ for $I=1,\cdots, N-1$ and $n = 0, 1, \ldots, N-2$. The SW solution is constructed from a family of Riemann surfaces $\cC(u)$ that depends holomorphically on the moduli $u_n$ and is referred to as the Seiberg-Witten curve.  For gauge group $\SU(N)$ and no hyper-multiplets, the SW curve is given by,
\bea
\label{2.SWC}
y^2 = A(x)^2 -\Lambda ^{2N} \, ,
\hskip 0.8in 
A(x) = x^N - \sum _{n=0}^{N-2} u_n x^n\, ,
\eea
where $\Lambda$ is the strong-coupling scale of the non-Abelian $\SU(N)$ super Yang-Mills theory.\footnote{~\label{fn:Lrel} Our conventions for the curve differ from those in \cite{DHoker:1997mlo,DHoker:2020qlp} by an $N$-dependent redefinition of the strong-coupling scale $4 \Lambda_\text{there}^{2N} = \Lambda_\text{here}^{2N}$. Henceforth we shall set $\Lambda =1$, unless otherwise stated.}  Each Riemann surface in the family is hyper-elliptic and has genus $N-1$. Choosing a basis of homology cycles $\mA_I$ and $\mB_I$ with canonical intersection pairing $\mJ$,
\bea
\label{2.CHB}
 \mJ(\mA_I, \mB_J) = - \mJ(\mB_I, \mA_J) & = & \delta_{IJ}
 \no \\ 
\mJ(\mA_I, \mA_J) \, = \, \mJ(\mB_I,  \mB_J) & = & 0~,
\eea
the vacuum expectation values of the gauge scalars~$a_I(u)$ and their magnetic duals~$a_{D,I}(u)$ are obtained as the periods of a meromorphic Abelian differential $\lambda$ as follows, 
\bea
\label{2.SWP}
2 \pi i \, a_I = \oint _{\mA_I} \lambda \, ,
\hskip 0.91in 
2 \pi i \, a_{D,I} = \oint _{\mB_I} \lambda \, 
\hskip 0.91in 
\lambda = { x A'(x) dx \over y}\, .
\eea
The matrix $\tau$ of $\U(1)^{N-1}$ gauge couplings and mixings is given by, 
\bea
\tau_{IJ} = { \p a_{D,I} \over \p a_J} = { \p a_{D,J} \over \p a_I}
\eea
The matrix $\tau$ is symmetric and its imaginary part is positive definite. The symmetry of $\tau$  implies the existence of a pre-potential $\cF$, determined by  $a_{D,I} = \p \cF / \p a_I$, which will not be needed in the sequel.  The imaginary part of $\tau$ is the matrix of inverse gauge couplings squared and must be positive on physical grounds. This property is automatic in the SW solution. Indeed, the partial derivatives $\p \lambda / \p u_n$ are holomorphic Abelian differentials on $\cC(u)$, up to exact differentials of single-valued functions, so that the partial derivatives $\p a_I / \p u_n$ and $\p a_{D,I} / \p u_n$ are periods of holomorphic differentials and $\tau$ is the period matrix of the Riemann surface $\cC(u)$. The Riemann bilinear relations automatically imply that $\tau$ has positive imaginary part. Finally, modular transformations on the cycles $\mA, \mB$ leave the canonical intersection pairing $\mJ$ invariant and  form the duality group  $\Sp(2N-2,\ZZ)$.

\subsection{Review of the expansion around the $\integers_{2N}$ point}

For gauge group $\SU(2)$, the SW curve has genus one, namely it is a torus,  so that the periods may be solved in terms of elliptic functions and modular forms \cite{SW:1994rs}. For gauge group $\SU(3)$, the periods are given by hyper-elliptic integrals which may be reduced to linear combinations of the Appell $F_4$ functions \cite{Klemm:1994}. For gauge group $\SU(N)$ with $N \geq 4$, however, the periods are given by hyper-elliptic  integrals that are no longer tabulated special functions. Nonetheless, a relatively simple convergent Taylor series expansion of the periods was obtained in \cite{DDN}  around the $\integers_{2N}$ symmetric point $u_n=0$ for all $n= 0,1,\dots, N-2$ for arbitrary $N$. This expansion, which we shall briefly review below,  will serve as a guide to obtaining a similar expansion around the maximal AD points.

\sm

At the $\ZZ_{2N}$ point we have $A(x) = x^N$ so that the SW curve $y^2=x^{2N}- 1$ manifestly exhibits the $\ZZ_{2N}$ symmetry $ x \to \ep x$ where $\ep =e^{2\pi i / 2N}$. The curve contains the two $\ZZ_{2N}$ symmetric  points $(x,y)=(0, \pm i)$. The $2N$ branch points are given by the $2N$-th roots of unity $(x,y)=(\ep^k,0)$ for $k=0 , \cdots, 2N-1$, and are shown in {the left panel of} figure~{\ref{fig:1} for the case of $N=3$. They are mapped into one another by $\ZZ_{2N}$.  The expansion around the $\ZZ_{2N}$ point is obtained by Taylor expanding the SW periods in powers of the moduli $u_n$. In practice, the expansion may be organized   by setting,
\bea
y^2 = x^{2N}  -1 +U(x)^2 - 2 x^N U(x) \hskip 1in U(x) = \sum _{n=0}^{N-2} u_n x^n
\eea
and Taylor expanding $\lambda $ in powers of $U(x)$, 
\bea
\lambda = \sum _{k=0}^\infty { \Gamma (k+\half) \over \Gamma (\half) k!} 
{ (2x^NU(x) - U(x)^2)^k \over (x^{2N}-1)^{k+\half}} \Big ( N x^N - x U'(x) \Big ) dx
\eea
The integrals of $\lambda$ along the homology cycles $\mA_I$ and $\mB_I$, needed in the calculation of the SW periods  in (\ref{2.SWP}), may be computed by integrating term by term in powers of $U(x)$. The homology cycles for all terms may then be chosen along the  line segments of the branch cuts of the $\ZZ_{2N}$ symmetric curve $y^2=x^{2N}-1$, as illustrated in figure \ref{fig:1} for the case $N=3$. 

\sm

As shown in \cite{DDN}, all such integrals may be obtained by evaluating the function $Q(\xi)$  which is defined as the Abelian integral of the SW differential, given by (\ref{2.SWP}),  
\begin{align}
  { \pi i} \,  Q(\xi) = \int_0 ^\xi \lambda
  \hskip 1in 
  \xi^{2N}=1
\end{align}
between either $\ZZ_{2N}$ symmetric point $(x,y)=(0,\pm i)$, denoted hereby $0$, and an arbitrary branch point {$(x,y)=(\xi,0)$} denoted here by $\xi$. The paths of integration are indicated in green  in {the left panel of} figure \ref{fig:1}. In terms of $Q(\xi)$ the SW periods are,
\begin{align}
\label{2.aaD}
 a_I &= \sum_{J=1}^I \Big \{Q(\ep^{2J-1})-Q(\ep^{2J-2})  \Big \} & a_{D,I} &=Q(\ep^{2I}) - Q(\ep^{2I-1}) 
\end{align}
Swapping the roles of the $\ZZ_{2N}$ {symmetric} points $(0,\pm i)$ reverses the signs of $Q$ and all the periods which, in turn, is equivalent to a modular transformation  by $-I \in \Sp(2N-2,\ZZ)$.   The Taylor series expansion of $Q(\xi)$ in powers of the moduli $u_n$ is given by,\footnote{{The notation used here is related to the notation used in \cite{DDN} by letting $L+1 \to L$, $M_0 \to M$, and $Y_M(\xi^N,L) \to Y(\xi^N,\a;u_0)$, as defined in (\ref{2.LMa}) and (\ref{2.Yhyper}), and will be convenient when matching with the expansion around the AD points in the sequel.}} 
\begin{align}
\label{Qexp}
 Q(\xi) = \sum_{\ell_i=0}^\infty \frac{2^{\frac{M-{L}}{N}}}{2\pi^2 N} \, \xi^{N M+{L}+N} \, 
 \Gamma \left (\tfrac{{L}}{N} \right ) Y(\xi^N, \a;u_0) \, \frac{u_1^{\ell_1}\dots u_{N-2}^{\ell_{N-2}}}{\ell_1!\dots \ell_{N-2}!}
\end{align}
where {we shall use the following combinations throughout}, 
\bea
\label{2.LMa}
L = {1+} \sum_{j=1}^{N-2}j\ell_j 
\hskip 0.8in 
M = \sum_{j=1}^{N-2}\ell_j
\hskip 0.8in 
\a= \frac{NM  -{L}}{2N}
\eea
The function $Y(\xi^N,\a;u_0)$ is given by the following linear combination of Gauss hypergeometric functions $F={}_2F_1$, 
\begin{align}
\label{2.Yhyper}
Y (\xi^N, \a;u_0)  =  & \,
2  u_0 \, \xi^N  \cos^2  ( \pi  \a  )  \Gamma \left (\a +\thalf  \right )^2 
 F \left (\a +\thalf, \a +\thalf ; \tfrac{3}{2};  u_0^2 \right )
\no \\ & + 
\sin^2 ( \pi  \a  )  \Gamma(\a )^2 \, F \left (\a, \a; \thalf; u_0^2 \right )
\end{align}
Alternatively, the hypergeometric functions may themselves be expanded in  Taylor series in $u_0$ \cite{DDN}, but the above formulation will be more pertinent to the expansion around the maximal AD points, to which we now turn.

\subsection{Expansion around a maximal  AD point}

For gauge group $\SU(N)$ with $N \geq 3$, the maximal AD points are characterized by $u_n=0$ for all $n>0$ and $u_0 = \pm 1$, recalling that we set the strong coupling scale $\Lambda =1$. Without loss of generality we may concentrate on the AD point with $u_0=1$  so that $A(x) = x^N -1 $. The {corresponding} SW curve $y^2=x^N(x^N-2)$  manifestly exhibits $\ZZ_N$ symmetry $x \to \ep^2 x$, recalling that $\ep =e^{2\pi i / 2N}$ while the SW differential $\lambda = N x^N dx/y$ transforms as $\lambda \to \ep^2 \lambda$. 

\subsubsection{Expansion of the SW differential}

The neighborhood of the AD point $u_0=1$, inside of which we shall obtain a convergent series expansion, may be parametrized by first taking $u_0$ away from the value 1 and then turning on the moduli $u_n$ for $n>0$. To do so we introduce the shifted variable $v=1-u_0$ keeping $u_n=0$ for $n>0$. In terms of $v$ the SW curve is given by,
\bea
y^2 = (x^N+v) (x^N+v-2)
\eea
Its branch points exhibit $\ZZ_N$ symmetry but, for $v \not= 1$, do not exhibit $\ZZ_{2N}$ symmetry. Instead,  they are given as follows for $k=0,1,\cdots, N-1$,
\begin{align}
   x_k^+ = (2-v)^\frac{1}{N}\ep^{2k} 
   \hskip 1in 
    x_k^- = v^\frac{1}{N} \ep^{2k+1} 
\end{align}
For sufficiently small $|v|\ll 1$ the distance from  branch points to the origin is of order $\Lambda =1$ for $x^+_k$, but  of order $|v|^{\frac{1}{N}} \ll 1$ for   $x_k^-$, as illustarted  in {the right panel of} figure \ref{fig:1} for the case $N=3$. The \textit{small branch points} $x_k^-$ correspond to the intrinsic data of the AD theory, while the \textit{large branch points} $x_k^+$ correspond to its embedding into $\SU(N)$. The small branch points dominate the dynamics of the AD theory as  $\Lambda\rightarrow \infty$  since the heavy states have masses of order $\Lambda$ and  decouple. For $v \not=0,2$, the SW curve also contains two points that are invariant under $\ZZ_N$ given by {$(x,y)=(0,\pm \sqrt{v(v-2)})$}  and that are referred to simply as $0$ in figure \ref{fig:1}.

\sm

Turning on the remaining moduli $u_n$ for $n >0$ will modify the disposition of the branch points from the one for the $\ZZ_N$ symmetric curve. As long as the $u_n$ for $n >0$ remain sufficiently small, the branch points will remain well-separated and a convergent Taylor series expansion should be expected. 

\sm

In this subsection we shall evaluate the periods spanned by the small branch points, {which are represented by the cycles $\mA_1^{(s)}$ and $\mB_1^{(s)}$ in the right panel of figure \ref{fig:1} for $\SU(3)$, and will be defined for arbitrary $\SU(N)$ in (\ref{Rperiod}).  The periods involving the large branch points, which are represented by the cycles $\mA_1^{(\ell)}$ and $\mB_1^{(\ell)}$ in the right panel of figure \ref{fig:1} for $\SU(3)$, will be defined for arbitrary $\SU(N)$ in (\ref{2.long}) and will be evaluated in subsection 2.3.5.}

\begin{figure}[htb]
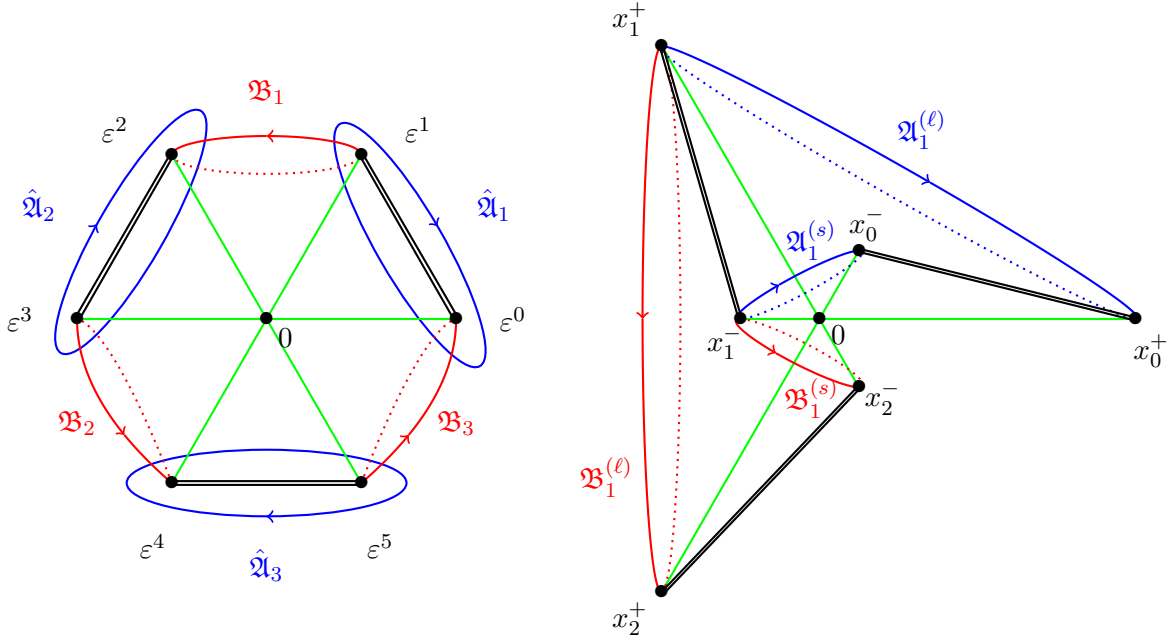

\begin{center}
\tikzpicture[scale=2.1]
\scope[xshift=0cm,yshift=0cm, scale=1.2]
\draw [thick] (-0.5,-0.856) -- (0.5,-0.856);
\draw [thick] (-0.5,-0.876) -- (0.5,-0.876);
\draw [thick] (-0.495,0.856) -- (-0.995,-0.015);
\draw [thick] (-0.505,0.876) -- (-1.005,0.015);
\draw [thick] (0.495,0.856) -- (0.995,-0.015);
\draw [thick] (0.505,0.876) -- (1.005,0.015);
\draw[thick, color=blue, rotate=0] (0,-0.866)  ellipse (21pt and 5pt);
\draw[thick, color=blue, rotate=60] (0.04,0.85)  ellipse (21pt and 5pt);
\draw[thick, color=blue, rotate=-60] (0.04,0.85)  ellipse (21pt and 5pt);
\draw  [red, thick, domain=0:90,->] plot ({0.5*cos(\x)},{0.866+0.1*sin(\x)});
\draw  [red, thick, domain=90:180] plot ({0.5*cos(\x)},{0.866+0.1*sin(\x)});
\draw  [red, thick, dotted, domain=180:360] plot ({0.5*cos(\x)},{0.866+0.1*sin(\x)});
\draw  [red, thick, domain=0:90,->] plot ({-0.75-0.25*cos(\x)},{-0.866*sin(\x/2)});
\draw  [red, thick, domain=90:180] plot ({-0.75-0.25*cos(\x)},{-0.866*sin(\x/2)});
\draw  [red, thick, dotted, domain=180:360] plot ({-0.75-0.25*cos(\x)},{-0.866*sin(\x/2)^3});
\draw  [red, thick, domain=0:90] plot ({0.75+0.25*cos(\x)},{-0.866*sin(\x/2)});
\draw  [red, thick, domain=90:180,<-] plot ({0.75+0.25*cos(\x)},{-0.866*sin(\x/2)});
\draw  [red, thick, dotted, domain=180:360] plot ({0.75+0.25*cos(\x)},{-0.866*sin(\x/2)^3});
\draw [thick, blue,->] (0.895,0.5)--(0.902,0.49);
\draw [thick, blue,->] (0.01,-1.043)--(0,-1.043);
\draw [thick, blue,->] (-0.902,0.49) -- (-0.895,0.5);
\draw [thick, green] (0,0) -- (1,0);
\draw [thick, green] (0,0) -- (0.5,0.866);
\draw [thick, green] (0,0) -- (-0.5,0.866);
\draw [thick, green] (0,0) -- (-1,0);
\draw [thick, green] (0,0) -- (-0.5,-0.866);
\draw [thick, green] (0,0) -- (0.5,-0.866);
\draw (0,0) node{$\bullet$};
\draw (1,0) node{$\bullet$};
\draw (-1,0) node{$\bullet$};
\draw (0.5,0.866) node{$\bullet$};
\draw (-0.5,0.866) node{$\bullet$};
\draw (0.5,-0.866) node{$\bullet$};
\draw (-0.5,-0.866) node{$\bullet$};
\draw (0.1,-0.1) node{\small $0$};
\draw (1.3,0) node{\small $\ep^0$};
\draw (0.8,1) node{\small $\ep^1$};
\draw (-0.8,1) node{\small $\ep^2$};
\draw (-1.3,0) node{\small $\ep^3$};
\draw (-0.6,-1.2) node{\small $\ep^4$};
\draw (0.6,-1.2) node{\small $\ep^5$};
\draw [blue] (1.2,0.6) node{\small $\hat \mA_1$};
\draw [blue] (-1.2,0.6) node{\small $\hat \mA_2$};
\draw [blue] (0,-1.3) node{\small $\hat \mA_3$};
\draw [red] (0,1.2) node{\small $\mB_1$};
\draw [red] (-1,-0.55) node{\small $\mB_2$};
\draw [red] (1,-0.55) node{\small $\mB_3$};
\endscope
\scope[xshift=3.5cm,yshift=0cm]
\draw [thick, green] (0,0) -- (2,0);
\draw [thick, green] (0,0) -- (0.25,0.433);
\draw [thick, green] (0,0) -- (-1,1.732);
\draw [thick, green] (0,0) -- (-0.5,0);
\draw [thick, green] (0,0) -- (-1,-1.732);
\draw [thick, green] (0,0) -- (0.25,-0.433);
\draw  [red, thick,  domain=0:90,->] plot ({-1-0.12*sin(\x)},{1.73*cos(\x)});
\draw  [red, thick,  domain=90:180] plot ({-1-0.12*sin(\x)},{1.73*cos(\x)});
\draw  [red, thick, dotted, domain=180:360] plot ({-1-0.12*sin(\x)},{1.73*cos(\x)});
\draw  [blue, thick,  domain=180:90,->] plot ({0.5+0.2*sin(\x)+3/2*cos(\x)},{-0.865*cos(\x)+0.865});
\draw  [blue, thick,  domain=90:0] plot ({0.5+0.2*sin(\x)+3/2*cos(\x)},{-0.865*cos(\x)+0.865});
\draw  [blue, thick,  dotted, domain=180:360] plot ({0.5+0.2*sin(\x)+3/2*cos(\x)},{-0.865*cos(\x)+0.865});
\draw  [blue, thick,  domain=180:90,->] plot ({-0.125-0.13*sin(\x)+0.375*cos(\x)},{0.2165*cos(\x)+0.2165});
\draw  [blue, thick,  domain=90:0] plot ({-0.125-0.13*sin(\x)+0.375*cos(\x)},{0.2165*cos(\x)+0.2165});
\draw  [blue, thick,  dotted, domain=0:180] plot ({-0.125+0.13*sin(\x)+0.375*cos(\x)},{0.2165*cos(\x)+0.2165});
\draw  [red, thick,  domain=90:180,<-] plot ({-0.125-0.15*sin(\x)+0.375*cos(\x)},{-0.2165*cos(\x)-0.2165});
\draw  [red, thick,  domain=90:0] plot ({-0.125-0.15*sin(\x)+0.375*cos(\x)},{-0.2165*cos(\x)-0.2165});
\draw  [red, thick,  dotted, domain=180:360] plot ({-0.125-0.15*sin(\x)+0.375*cos(\x)},{-0.2165*cos(\x)-0.2165});
\draw [thick] (-1.005,-1.735) -- (0.24,-0.43);
\draw [thick] (-0.990,-1.745) -- (0.26,-0.436);
\draw [thick] (-1.013,1.732) -- (-0.512,0);
\draw [thick] (-0.992,1.732) -- (-0.492,0);
\draw [thick] (0.25,0.445) -- (2,0.01);
\draw [thick] (0.25,0.425) -- (2,-0.01);
\draw (0,0) node{$\bullet$};
\draw (2,0) node{$\bullet$};
\draw (-0.5,0) node{$\bullet$};
\draw (0.25,0.433) node{$\bullet$};
\draw (-1,1.732) node{$\bullet$};
\draw (0.25,-0.433) node{$\bullet$};
\draw (-1,-1.732) node{$\bullet$};
\draw (0.12,-0.1) node{\small $0$};
\draw (2.1,-0.2) node{\small $x_0^+$};
\draw (0.3,0.6) node{\small $x_0^-$};
\draw (-1.2,1.9) node{\small $x_1^+$};
\draw (-0.6,-0.15) node{\small $x_1^-$};
\draw (-1.2,-1.9) node{\small $x_2^+$};
\draw (0.4,-0.5) node{\small $x_2^-$};

\draw [blue] (0.65,1.2) node{\small $\mA_1^{(\ell)}$};
\draw [red] (-1.35,-1) node{\small $\mB_1^{(\ell)} $};
\draw [blue] (-0.05,0.5) node{\small $\mA_1^{(s)}$};
\draw [red] (-0.05,-0.5) node{\small $\mB_1^{(s)} $};

\endscope
\endtikzpicture
\caption{The $\ZZ_6$ symmetric curve $y^2=x^6-1$ is shown in the left panel, while the $\ZZ_3$ symmetric curve 
$y^2=(x^{3}+v)(x^3+v-2)$ is shown in the right panel. In each case, the  branch cuts are shown in black double lines; the integration paths for $Q(\ep^n)$ and $R(\zeta)$ are shown in green; the cycles of the canonical homology bases $\mA$ and $\mB$ are shown in blue and red, respectively.  \label{fig:1}}
\end{center}
\end{figure}

To proceed {with the evaluation of the small periods,} we set $A(x) = \hat A(x) -1$ in terms of which the SW curve and differential become, 
\bea
\label{2.SW.A}
y^2 = \hat A (x) ( \hat A (x)-2) 
\hskip 1in 
\lambda = { x \hat A '(x) dx \over y}
\eea
The AD point $u_0=1$ corresponds to $\hat A(x)=x^N$ and the small branch points correspond to $|x|\ll 1$. Thus, to evaluate the periods spanned by the small branch points, we expand the denominator in powers of $\hat A(x)$, as follows, 
\bea
\lambda = { 1 \over \sqrt{-2}} \sum _{k=0}^ \infty { \Gamma (k+\half) \over 2^k \, \Gamma (\half) \, k!} \, { x \, \hat A '(x)  dx \over \hat A (x)^{\half -k}}
\eea
By setting $u_0=1-v$ and rescaling the variable $x$ and the moduli $u_n$ as follows, 
\bea
\label{2.scale}
x= v^\frac{1}{N}  z \hskip 1in u_n = v^{1- \frac{n} {N}} v_n \quad \hbox{ for } n > 0
\eea
the function $\hat A(x)$ decomposes into a factor of $v$ times a factor that only depends on the remaining rescaled moduli  $v_n$ for $n>0$, but is independent of $v$,
\bea 
\label{2.SW.B}
\hat A (x) =  v \Big ( z^N - V(z) + 1 \Big )
\hskip 1in 
V(z) = \sum _{n=1}^{N-2} v_n z^n
\eea
Clearly, the expansion (\ref{2.SW.A}) in powers of $\hat A(x)$ is equivalent to an expansion in powers of $v$. To proceed, we expand $\lambda$ in powers of the remaining moduli $v_n$ with $n>0$ as follows, 
\bea
\label{2.lambda}
\lambda = { v^{\half + \frac{1}{N}} \over \sqrt{-2}} \sum _{k=0}^ \infty { \Gamma (k+\half) \over \Gamma (\half) \, k!} \, 
\left ( {v \over 2} \right ) ^{k}
 \sum_{M=0} ^\infty { \Gamma (\half -k+M) \over \Gamma (\half -k) \, M!}
{ V(z)^{M}   (N z^N - z V'(z) )  dz \over \big ( z^N  + 1 \big ) ^{\half -k +M} }
\eea
Note that all reference to the larger branch points has been translated into analytic dependence in $z$, and the above expression for $\lambda$ can be used only to calculate the periods spanned by the small branch points.

\subsubsection{The short SW periods in terms of $R(\zeta)$}

The short SW periods may  be evaluated in terms of the function $R(\zeta)$ defined for $\zeta^N=-1$\footnote{We shall use the symbol $\zeta$ for the $N$-th roots of unity satisfying $\zeta^N=-1$ here in order to clearly distinguish them {from} the arbitrary $2N$-th roots of unity denoted by $\xi$ in the preceding subsection.}  as the integral  from either one of the $\ZZ_N$ symmetric points, denoted here by $z=0$,  to the small branch point $z=\zeta$ by,
\bea
\label{Rdef}
 i \pi R(\zeta) = \int_0  ^\zeta  \lambda \hskip 1in \zeta ^N=-1
\eea
The integral  is taken along a path from $z=0$ to $z=\zeta$ that does not intersect any of the branch cuts produced by the square root, as shown in green in {the right panel of } figure~\ref{fig:1}.  As in the case of the expansion around the $\ZZ_{2N}$ symmetric point, swapping the sign of the $\ZZ_N$ symmetric point amounts to reversing the sign of all periods and is equivalent to the modular transformation $-I \in \Sp(2N-2,\ZZ)$. The integrals $R(\zeta)$ will soon be related by analytic continuation to the integrals $Q(\xi)$ and hence {to the} periods $a_I$ and $a_{D,I}$ considered in \cite{DDN}. {Choosing a basis for the short homology one-cycles, }
\begin{align}
    \hat{\mA}_j^{(s)} &= [\ep^{4j-3}, \ep^{4j-1}] &\mA_i^{(s)} &= \bigcup_{j=1}^i\hat{\mA}_j^{(s)} 
    \no \\ 
    \mB^{(s)}_i & = [\ep^{4i-1},\ep^{4i+1}] & i &=1, \cdots \left [ \tfrac{N-1}{2} \right ]
\end{align}
the short periods $a_i^{(s)}$ and $a_{D,i}^{(s)}$ may be expressed in terms of the periods $a_I$ and $a_{D,I}$ {and  in terms of the function $ Q(\xi)$ as follows, 
\begin{align} 
\label{Rperiod}
\hat{a}_i^{(s)} &=  \hat{a}_{2i} + a_{D, 2i-1} = R(\ep^{4i-1}) - R(\ep^{4i-3}) & a_i^{(s)} &= \sum_{j=1}^i \hat{a}^{(s)}_j
\no \\
 a_{D,i} ^{(s)} &=  \hat{a}_{2i+1}+a_{D, 2i} = R(\ep^{4i+1})-R(\ep^{4i-1}) & i &=1, \cdots \left [ \tfrac{N-1}{2} \right ]
\end{align}
The short homology cycles $\mA_1^{(s)}$ and $\mB_1^{(s)}$ are indicated in figure \ref{fig:1} for $\SU(3)$. }

\subsubsection{Expansion of $R(\zeta)$ and the short periods}

We are now ready to formulate and prove one of the fundamental results of this paper, namely the expansion of the short periods around the maximal AD points for arbitrary gauge group $\SU(N)$. As shown in the preceding subsection, the periods are given by (\ref{Rperiod}) in terms of the function $R(\zeta)$ defined in (\ref{Rdef}). The results below give the expansion of the function $R(\zeta)$ around the maximal AD points.

\begin{thm}
\label{thm:2.1}
The function $R(\zeta)$ for the small branch points $\zeta$ with $\zeta^N = -1$ admits the following series expansion around $v_n=0$ for all $n=1, \dots, N-2$ and $v_0\equiv v = 1-u_0\neq 0$:
\begin{align}
\label{2.thm1a}
R(\zeta) &= \frac{v^{\half+\frac{1}{N}}}{\sqrt{2\pi} N} 
\sum _{{\ell_n=0 \atop n=0,\cdots, N-2}} ^\infty 
\frac{(-)^{M+1} \zeta^L\Gamma(\frac{L}{N})\Gamma(\ell_0+\half)^2}{\Gamma(\half)^2\Gamma(\frac{3}{2}- 2 \a +\ell_0)}\, 
\frac{v_0^{\ell_0}\dots v_{N-2}^{\ell_{N-2}}}{2^{\ell_0} \, \ell_0!\dots\ell_{N-2}!}    
        \end{align}
where the combinations $L, M$ and $\a$ were defined in (\ref{2.LMa}).
\end{thm}

The proof proceeds from the SW differential $\lambda$ in (\ref{2.lambda}) and is relegated to Appendix A. 

\begin{cor}
\label{Rhyp}
    The summation over $\ell_0$ in the Taylor series expansion for $R(\zeta)$ for the small branch points with $\zeta^N=-1$ in Theorem \ref{thm:2.1} may be carried out in terms of an infinite series of Gauss hypergeometric functions $_2 F_1 = F$ and the result is given by,
\begin{align}
\label{2.Rxi1}
        R(\zeta) &= 
        \frac{v^{\half+\frac{1}{N}}}{\sqrt{2\pi} N} 
        \sum _{{\ell_n=0 \atop n=1,\cdots, N-2}} ^\infty W_{L, M}(\zeta, v) \, 
        \frac{{v}_1^{\ell_1}\dots {v}_{N-2}^{\ell_{N-2}}}{\ell_1!\dots\ell_{N-2}!}
\end{align}
where $L,M$ were defined in (\ref{2.LMa}) and the coefficient functions $W_{L, M}(\zeta, v)$ are given by,
    \begin{align}
    \label{2.Rxi2}
        W_{L, M}(\zeta, v) &= \frac{(-)^{M+1} \, \zeta^L \, \Gamma(\tfrac{L}{N})}{\Gamma(\tfrac{3}{2} -2\a)}F(\thalf, \thalf;\tfrac{3}{2}-2\a;\tfrac{v}{2}).
    \end{align}
In the special case where ${v}_n=0$ for all $n\neq 0$, the function $R(\zeta)$  reduces to, 
    \begin{align}
    \label{2.Rxi3}
        R(\zeta) = -\frac{\zeta \, v^{\half+\frac{1}{N}} \, \Gamma(\frac{1}{N})}{\sqrt{2\pi} N \, \Gamma(\frac{3}{2}+\frac{1}{N})} F(\thalf,\thalf;\tfrac{3}{2}+\tfrac{1}{N};\tfrac{v}{2}).
    \end{align}
\end{cor}
The corollary readily follows from Theorem \ref{thm:2.1}, and its proof is left to the reader.

\subsubsection{Evaluating $R(\zeta)$ by analytic continuation of $Q(\xi)$ for $\xi^N=-1$}

Before addressing the calculation of the long periods, we show that $R(\zeta)$ may be obtained from $Q(\xi)$ by analytic continuation in the variable $v=1-u_0$ for the small branch points, namely $\xi=\zeta$ for which $\zeta ^N=-1$. Using these results, we shall then use the same analytic continuation to obtain the long periods in the next subsection. We begin by proving the following corollary of Theorem \ref{thm:2.1}.

\begin{cor}
\label{thm:2.2}
    The function $R(\zeta)$ is the analytic continuation in the modulus $v=1-u_0$ of $Q(\xi)$ for the small branch points specified by $\xi=\zeta$ and $\xi^N=-1$.
\end{cor}

To prove the theorem, we start from the Taylor series expansion (\ref{Qexp}) for $Q(\xi)$ and re-express the coefficient functions $Y$ of (\ref{2.Yhyper}) using the reflection formula for the $\Gamma$-function,
\bea
\label{2.Gamma}
\Gamma (z) \Gamma (1-z) \sin (\pi z) = \pi
\eea
 as well as the change of variables $v=1-u_0$,
\bea
Y(\xi^N, \a;u_0) = - \xi^N { \pi^2 \, f_3(\a,v) \over \Gamma (\thalf -\a)^2}  
+ { \pi^2 \, f_4(\a,v) \over \Gamma (1- \a)^2} 
\eea
{in terms of the functions  $f_3$ and $f_4$}  given by,
\bea
f_3(\a,v) & = & - 2 \, (1-v) F \left ( \a+\thalf, \a+\thalf; \tfrac{3}{2}; (1-v)^2 \right )
\no \\ 
f_4(\a,v) & = & F \left ( \a, \a; \tfrac{1}{2}; (1-v)^2 \right )
\eea
By construction, both functions admit a convergent Taylor series expansion around the point $v=1$, which is the $\ZZ_{2N}$ symmetric point. Our goal is to perform an analytic continuation to functions that admit convergent Taylor series expansions around the point $v=0$, which is one of the AD points. To proceed, it is readily verified that both 
functions $f_3, f_4$ are solutions to the same hypergeometric differential equation,
\begin{align} 
\label{DiffEqn}
v(2-v) \frac{d^2 f}{dv^2} + ( 4\a+1) (1-v) \frac{df}{dv} - 4 \a^2 f=0
\end{align}
The solutions to this equations may alternatively be expressed in terms of hypergeometric functions with argument $v/2$, whose normalizations are conveniently chosen as follows, 
\bea
\label{2.f1f2}
        f_1(\a,v) & = & {2\,  (2v)^{\half -2 \a} \over 1-4\a} F \left ( \thalf, \thalf; \tfrac{3}{2} -2 \a ;  \tfrac{v}{2} \right )
            \no \\
        f_2 (\a,v) & = & F(2\a, 2 \a; 2\a +\thalf; \tfrac{v}{2})
\eea
The two bases of solutions {to (\ref{DiffEqn})} are related by a matrix $\cS \in \SL(2,\RR)$, 
\bea
\label{f1}
\left ( \bma f_1(\a,v) \cr f_2(\a,v)  \ema \right ) = \cS 
\left ( \bma f_3(\a,v) \cr f_4(\a,v)  \ema \right ) 
\hskip 1in \cS = \left ( \bma \cS_{13} & \cS_{14} \\ \cS_{23} & \cS_{24} \ema \right )
\eea
The resulting expression for $f_2$ is the Gauss-Kummer quadratic transformation of hypergeometric functions. The matrix elements of $\cS$ are given as follows,\cite{BatemanI}
\begin{align}
\cS_{13} (\a) & =    { \Gamma (\half) \Gamma(\half -2\a)  \over  \Gamma(\half-\a)^2}
& 
\cS_{14} (\a) & = { \Gamma (\half)  \Gamma(\half -2\a)  \over  \Gamma (1-\a)^2} 
\no \\
\cS_{23} (\a) & =  {  \Gamma (\half)  \Gamma(\half +2\a)  \over \Gamma (\a)^2 } 
&
\cS_{24} (\a) & = {  \Gamma (\half)  \Gamma(\half +2\a)  \over \Gamma (\half + \a)^2 } 
\end{align}
One verifies that indeed $\det (\cS)=1$ by using the reflection relation (\ref{2.Gamma}).  Inspection of the coefficients $\cS_{13}$ and $\cS_{14}$ reveals that, for the special case where $\xi^N=-1$, the combination $Y(-1, \a; u_0)$ is proportional to the function $f_1(\a,v)$, namely,
\bea
Y(-1, \a;u_0) = { \pi^2 \, f_1(\a, v) \over \Gamma (\thalf) \Gamma (\half -2 \a)} 
\eea
Substituting this expression into (\ref{Qexp}) readily produces the expressions of Theorem \ref{thm:2.1} in (\ref{2.Rxi1}) and (\ref{2.Rxi2}), which concludes the proof of Theorem \ref{thm:2.2}. Henceforth, we shall set $R(\zeta) = Q(\zeta)$ for $\zeta ^N=-1$ and express the short SW periods in terms of $Q(\zeta)$.

\subsubsection{Expansion of $Q(\xi)$ for the long periods by analytic continuation}

Obtaining the expansion of the long SW periods around the maximal AD points directly from the integral  representation of the periods is considerably more involved than the evaluation for the short periods given in Theorem \ref{thm:2.1}. Instead of proceeding directly here, we shall take advantage of the analytic continuation of the expansion for $Q(\xi)$ for $\xi^N=1$ around the $\ZZ_{2N}$ symmetric point to obtain the long periods. The long homology cycles are chosen to be,
\begin{align}
    \hat{\mA}_i^{(\ell)} &= \hat{\mA}_{2i-1} -\mB_{2i-1} & \mA_i^{(\ell)} &= \bigcup_{j=1}^i \hat{\mA}_j^{(\ell)} 
    \no \\
    \mB_i^{(\ell)} &= -\hat{\mA}_{2i} + \mB_{2i} & i & =1, \cdots \left [ \tfrac{N}{2} \right ]
\end{align}
The long periods will be denoted by $a_i^{(\ell)}$ and $a_{D,i}^{(\ell)}$ and may be expressed as follows,
\begin{align}
\label{2.long}
\hat{a}_i^{(\ell)} &= \hat{a}_{2i-1}-a_{D,2i-1}  &a_i^{(\ell)} &=\sum_{j=1}^i \hat{a}_j^{(\ell)}
\no \\
a_{D,i} ^{(\ell)} &= -\hat{a}_{2i} + a_{D, 2i} & i & =1, \cdots \left [ \tfrac{N}{2} \right ]
\end{align}
{We alert the reader to the fact that the ranges of the index $i$ labelling the short and long cycles (and periods) coincide for odd values of $N$ but differ when $N$ is even, in which case there is one more pair of long periods than short periods. Using these definitions and (\ref{2.CHB}) one verifies that the short and long cycles satisfy the following canonical intersection pairings, 
\begin{align}
\label{2.canls}
    \mJ(\mA^{(\ell)}_i, \mB_j^{(\ell)}) & = \delta _{i,j} & i,j & =1, \cdots \left [ \tfrac{N}{2} \right ]
    \no \\
    \mJ(\mA^{(s)}_i, \mB_j^{(s)}) & = \delta_{i,j} & i,j & =1, \cdots \left [ \tfrac{N-1}{2} \right ]
\end{align}
while all other pairings of the cycles $\mA^{(\ell)}_i, \mB_j^{(\ell)}, \mA^{(s)}_i$, and $ \mB_j^{(s)} $ vanish. The long and short homology cycles $\mA_1^{(\ell)}, \mB_1^{(\ell)}, \mA_1^{(\ell)}$ and $\mB_1^{(\ell)}$ are indicated in figure \ref{fig:1} for $\SU(3)$.}

\sm

The {Taylor series expansion of $Q(\xi)$ for $\xi^N=1$ at the maximal AD point in powers of the moduli $v_n$ for $n>0$} is given by the following theorem.
\begin{thm}
\label{+1}
The Taylor series expansion near the AD point of the function $Q(\xi)$ for $\xi^N = 1$ is given by the following expression, 
\begin{align}
Q(\xi) = \sum_{{\ell_n=0 \atop n=1, \cdots, N-2}}^\infty \frac{2^{M - \frac{L}{N}}}{2\pi^2 N} 
\xi^{L} \Gamma(\tfrac{L}{N}) Y(1, \a;u_0) v^\frac{NM-L+1}{N}
\frac{{v}_1^{\ell_1}\dots {v}_{N-2}^{\ell_{N-2}}}{\ell_1!\dots \ell_{N-2}!},
\end{align}
where {$u_0=1-v$,} the combinations $L,\a$ were defined in (\ref{2.LMa}), and $Y(1,\a;u_0)$ is given by,
\begin{align}
Y(1, \a;u_0) = -\Gamma(\thalf)\Gamma(\thalf+2\alpha)f_1(\alpha,v) 
+ { 2\pi^2 \Gamma(\thalf)\Gamma(\thalf-2\alpha) \over {\Gamma (1-\a)^2 \Gamma (\half -\a)^2}} f_2(\alpha,v)
\end{align}
The functions $f_1(\a,v)$ and $f_2(\a,v)$ were defined in   (\ref{2.f1f2}).   In the special case where $v_n=0$ for $n>0$, only the term with $\ell_n=0$ for $n>0$ contributes so that $L=1$, and we have, 
\bea
{Q(\xi)} =  \frac{2^{ - \frac{1}{N}}}{2\pi^2 N} \,  \xi \, \Gamma(\tfrac{1}{N}) Y(1, -\tfrac{1}{2N} ;u_0)
\eea
\end{thm}
The proof of the theorem  follows  from using the relation (\ref{f1}) to re-express $Q(\xi)$.

\subsection{The case of gauge group $\SU(3)$}

The $\cN=2$ super-Yang-Mills theory with gauge group $\SU(3)$ offers one of the simplest settings in which the AD theories arise. For this reason, and because one {can} at the same time obtain simplified and more explicit formulas for the periods than in the case of arbitrary~$N$, we shall study the behavior of the $\SU(3)$ theory in detail here. The formulas we shall obtain may also be compared with various known results available in the literature \cite{Klemm:1994}. In terms of the moduli $u=u_1$ and $v=1-u_0$ the SW curve and differential are given by,
\bea
\label{2.N=3curve}
y^2 = (x^3-ux+v)(x^3-ux+v-2) \hskip 1in \lambda = { (3x^3-ux)dx \over y}
\eea
Since the two factors in $y^2$ have no common roots, the zeros of the discriminant of this curve obey either $4u^3=27v^2$ or $4u^3=27(v-2)^2$.

\subsubsection{Series expansion of the short and long periods} 

Using Corollary \ref{Rhyp} for $N=3$ and for the case $\xi^3=-1$ we obtain the following expansion for the small branch points, 
\bea    
\label{2.Q-}
Q(\xi) =  - { v^\frac{5}{6} \over 3 \sqrt{2} \pi } \sum_{m=0}^\infty 
        \xi ^{m+1} {\Gamma(\tfrac{m+1}{3}) \Gamma (\half) \over \Gamma (\tfrac{11-4m}{6}) \, m!} \left ( -{ u \over v^\frac{2}{3}} \right )^m 
         \, F(\thalf, \thalf; \tfrac{11-4m}{6} ; \tfrac{v}{2})
\eea
while using Theorem \ref{+1} for $N=3$ and for the case $\xi^3=1$ we obtain the expansion for the large branch points,
\bea 
\label{2.Q+}      
Q(\xi) & = & \sum_{m=0}^\infty {2^{ (2m - 1)/3} \over 6 } 
        \xi ^{m+1} \Gamma(\tfrac{m+1}{3}) {u^m \over m!}
        \bigg [  {\Gamma (\frac{4m-5}{6}) \over \Gamma(\half)^3} (2v)^{\frac{5-4m}{6} } 
        F(\thalf, \thalf; \tfrac{11-4m}{6} ; \tfrac{v}{2})
\no \\ & & \hskip 1.6in
        + { 2 \Gamma (\half) \Gamma (\frac{5-4m}{6}) \over \Gamma (\frac{2-m}{3})^2 \Gamma (\frac{7-2m}{6})^2} F(\tfrac{2m-1}{3}, \tfrac{2m-1}{3}; \tfrac{4m+1}{6}  ; \tfrac{v}{2})  \bigg ]
\eea
We note that $Q(\xi)$ is analytic in $u$, but non-analytic in $v$ as it contains powers of $v^\frac{1}{6}$ for all values of $\xi$. For $u=0$, its dependence on $v$ is through a factor of $v^\frac{5}{6}$ times integer powers of $v$. This scaling behavior for small $v/\Lambda^N$ is consistent with the predictions of the scaling dimension $\Delta =\tfrac{6}{5}$  for the intrinsic Coulomb branch of rank 1 AD theories \cite{Gaiotto:2009we,Martone:2020hvy,Argyres:2015ffa,Argyres:2015gha,Argyres:2016xua,Argyres:2016xmc,Argyres:2018urp,Kaidi:2022sng}.  
We shall return to this point in later sections.

\subsubsection{Analyticity of the long periods}

On physical grounds, the long periods, namely those associated with the embedding of the AD theory into the $\SU(3)$ super Yang-Mills theory, are expected to be analytic in all moduli $u,v$ for $|u|, |v| \ll 1$. The fact that this is the case is borne out by the following proposition. 

\begin{prop}
    There exists an $\Sp(4, \integers)$ electric-magnetic duality frame such that one pair of periods is analytic in the moduli $(u, v)$, while the other pair of periods carry the non-analyticities associated with the AD point.
\end{prop}

\tikzset{
    partial ellipse/.style args={#1:#2:#3}{
        insert path={+ (#1:#3) arc (#1:#2:#3)}
    }
}

To prove the proposition, consider the following $\Sp(4,\integers)$ duality transformation, {which implements the relations  (\ref{Rperiod}) and (\ref{2.long}) for the special case of $N=3$,}
\begin{align}
    \begin{pmatrix}
        a^{(\ell)}\\
        a^{(s)}\\
        a_{D}^{(\ell)}\\
        a_D^{(s)}
    \end{pmatrix}
    &= \begin{pmatrix}
        1&0&-1&0\\
        -1&1&1&0\\
        1&-1&0&1\\
        0&-1&0&1
    \end{pmatrix}
    \begin{pmatrix}
    a_1\\
    a_2\\
    a_{D,1}\\
    a_{D,2}
    \end{pmatrix}
\end{align}
where we have suppressed the sole index $i=1$. One may verify that the corresponding cycles satisfy the canonical intersection relations of (\ref{2.canls}).  It will be convenient to use the decomposition of $Q(\xi)$  into characters of $\ZZ_6$, familiar from \cite{DDN}, 
\begin{align}
    \label{2.char}
        Q(\xi) &= \sum_{n = 0}^5\xi^n Q_n &Q_3&=0
    \end{align}
where we recall that $Q_0$ actually drops out of all SW periods (see Appendix \ref{ExplicitSU(3)}). Expressing the long periods in terms of the functions $Q_n$, we obtain, 
\begin{align}
    a^{(\ell)} &= (1+\rho) (Q_1- 3Q_4) -\rho (Q_5 - 3 Q_2)
    \no \\
    a_D^{(\ell)} &= (Q_1-3Q_4) + (Q_5 - 3 Q_2)
\end{align}
{where $\rho =\ep^2= e^{2 \pi i /3}$.} The combinations $3Q_4-Q_1$ and $3Q_2-Q_5$ may be obtained using equations (\ref{bp}) and (\ref{bm}) {of appendix B, and are given by, }
\begin{align}
3Q_4-Q_1 &= { 2^{1/3} \over 2 \pi^\frac{3}{2} }  \sum_{\mu=0}^\infty 
{ \Gamma (2 \mu -{ \frac{1}{3}})^2 \Gamma(\mu+\tfrac{1}{3})  \over  \Gamma (2\mu +\frac{1}{6}) \, (3\mu)!} 
F( 2 \mu - \tfrac{1}{3} , 2 \mu - \tfrac{1}{3} ; 2 \mu + \tfrac{1}{6}; \tfrac{v}{2})  \,  { u^{3\mu} \over 2^{2\mu} } 
\no \\
3Q_2-Q_5 &= 
{ 2^{-1/3}  \over 2 \pi^\frac{3}{2} } \sum_{\mu=0}^\infty 
 { \Gamma (2 \mu +\frac{1}{3})^2  \Gamma(\mu+\tfrac{2}{3})  \over   \Gamma (2\mu +\frac{5}{6}) (3\mu+1)!} 
 F(2 \mu+\tfrac{1}{3} , 2 \mu +\tfrac{1}{3}; 2 \mu +\tfrac{5}{6} ; \tfrac{v}{2})  \, { u^{3\mu+1} \over 2^{2\mu} }  
\end{align}
Hence, {it is is clear by inspection that} all non-analytic behavior of the long periods completely cancels, as is expected on physical grounds.

\subsubsection{$\SU(3)$ periods via elliptic functions and modular forms}

The expansion of the SW periods near the AD points for $\SU(3)$ gauge group  may be obtained in terms of elliptic functions and modular form, in parallel to the results of \cite{DDN} for the expansion around the $\ZZ_6$ symmetric point. We shall adopt the notations and conventions of appendix C in \cite{DDN}. The derivation of these results is analogous to the one used in section 15 of \cite{DK}, and will not be presented here. 

\sm

We begin by parametrizing the genus 2 curve for $N=3$ given in (\ref{2.N=3curve}) as follows, 
\bea
\label{2.param}
(2\om)^2 x & = &  \wp (z|\tau) 
\no \\
4 (2 \om)^4 u & = & g_2(\tau) = {{\HE_4(\tau) \over 12}}
\no \\ 
- 4(2 \om)^6 v & = &  g_3(\tau) = - { {  \HE_6(\tau) \over 216}}
\eea
where the Weierstrass function $\wp(z|\tau)$ has periods $2 \om$ and $2 \om \tau$, and satisfies the relation $\wp'(z|\tau)^2 = 4 \wp(z|\tau)^3 - g_2(\tau) \wp(z|\tau) - g_3(\tau)$, while $\HE_4(\tau)$ and $\HE_6(\tau)$ are the modular forms of weight 4 and 6 respectively, normalized to the value 1 at the cusp $ i \infty$. 
In terms of the parametrization (\ref{2.param}), the SW curve (\ref{2.N=3curve}) becomes, 
\bea
y^2 & = & (4 \om )^{-12} (4 \wp^3 -g_3 \wp -g_3) (4 \wp^3 - g_2 \wp - g_3 - 4 (2 \om)^6)
\eea
{Suitably deformining the short cycles {$ \mA= \mA^{(s)}_1  = [0,2\pi i]$ and $ \mB= \mB^{(s)}_1 =[0, 2 \pi i\tau]$} 
in order to avoid the double-pole of $\wp(z)$ at $z=0$, the elliptic integrals defined by, }
\begin{align}
    A_n &= \frac{1}{2\pi i}\oint_\mA dz\;\wp(z)^n & B_n &= \frac{1}{2\pi i}\oint_\mB dz\;\wp(z)^n
\end{align}
satisfy the following recursion that holds for $J_k\in\{A_k, B_k\}$:
\begin{align}
\label{recursion}
    (8n-4)J_n = (2n-3)g_2J_{n-2} + (2n-4)g_3 J_{n-3}
\end{align}
with the initial conditions  $A_0 = 1$ and $B_0 = \tau$. The solution is given by,
\begin{align}
\label{linearity}
A_n = K_n + \frac{\HE_2}{12} L_{n-1} \hskip 1in  B_n = \tau A_n + \frac{1}{2\pi i}L_{n-1}
\end{align}
where $K_n$ and $L_n$ are modular forms of weight $2n$, determined by the recursion relation (\ref{recursion}) and the initial conditions. {We have $K_1=L_1=0$ and, }
\begin{align}
    K_0&= 1&K_2&= \tfrac{\HE_4}{144}&K_3&=-\tfrac{\HE_6}{2160} &K_4&= \tfrac{5\, \HE_4^2}{48384} &K_5&=-\tfrac{\HE_4 \, \HE_6}{77760}
    \no \\
    L_0&= 1 &L_2&= \tfrac{\HE_4}{80} &L_3&= -\tfrac{\HE_6}{1512} &L_4&= \tfrac{7\, \HE_4^2}{48384} &L_5&=-\tfrac{29\, \HE_4\, \HE_6}{1330560}
\end{align}
The short periods are then given by expanding the SW differential in powers of $1/\omega$ as given by the following theorem, which offers a non-trivial extension of the calculation of short periods carried out to leading order in large $\Lambda$ in \cite{AD}.  
\begin{thm}
\label{2.thm:4}
    The small periods {$a(\tau)= a^{(s)}_1(\tau)$ and $a_D(\tau)= a^{(s)}_{D,1}(\tau)$} admit the following Taylor series-expansion in terms of the basis $\{\HE_2, \HE_4, \HE_6\}$ of the ring of quasi-modular forms, along with the variable $\om$, 
    \bea
    a & = &  {-i\over\sqrt{2 \pi }} \sum_{k,\ell,m=0}^\infty 
    {  \Gamma (k+\ell+m+\half) \over  2^{k+3\ell+3m}  k! \, \ell ! \, m!} 
    { (-)^{\ell+m} \, g_2 ^\ell \, g_3^m \over  ( 2 \om)^{6(k+\ell+m) +5} }
     \left ( 3A_{3k+\ell+3} - {g_2 \over 4}  A_{3k+\ell+1}  \right ) 
     \no \\
    a_D & = & {-i\over\sqrt{2 \pi }} \sum_{k,\ell,m=0}^\infty 
    {  \Gamma (k+\ell+m+\half) \over  2^{k+3\ell+3m}  k! \, \ell ! \, m!} 
    { (-)^{\ell+m} \, g_2 ^\ell \, g_3^m \over  ( 2 \om)^{6(k+\ell+m) +5} }
     \left ( 3B_{3k+\ell+3} - {g_2 \over 4} B_{3k+\ell+1}  \right )  
 \qquad
 \eea
 As a result, the following combination can be expressed in terms of modular forms $\{\HE_4, \HE_6\}$
 \begin{align}\label{aDa}
     a_D - \tau a = 
 { -1 \over  (2\pi)^\frac{3}{2} } \sum_{k,\ell,m=0}^\infty 
{  \Gamma (k+\ell+m+\half) \over   2^{k+3\ell+3m}  k! \, \ell ! \, m!} 
{ (-)^{\ell+m} \, g_2 ^\ell \, g_3^m \over  ( 2 \om)^{6(k+\ell+m) +5} }
 \left ( 3L_{3k+\ell+2} - {g_2 \over 4} L_{3k+\ell}  \right )
 \end{align}
 which is a locally holomorphic modular form of weight $-1$ provided $\omega$ is assigned holomorphic weight $1$ and $L_n$ has weight $2n$.
\end{thm}

\subsubsection*{Remarks}

\begin{enumerate}
\itemsep=0in
\item {The factor $2\om$ in the denominator of each formula in Theorem \ref{2.thm:4} may be eliminated in favor of either the variables $(u, g_2(\tau))$ or the variables $(v, g_3(\tau))$ using (\ref{2.param}) depending on whether the expansion is sought near the points $\tau=i$ or $\tau = \rho=e^{2\pi i/3}$ respectively.}
\item {The combination $a_D-\tau a$ vanishes at the AD point $\tau = \rho$, since we have $g_2(\rho)=0$, and the recursion relation implies $L_{3k+2}=0$ for all $k\geq 0$.} Thus, at the AD point, we have $a_D = \tau a$, as must be the case for any rank-1 $\N = 2$ SCFT.
\item {The $\ZZ_3$ symmetry of the AD point  fixes the modulus $\tau=\rho$ and $g_2(\rho)=0$. In the neighborhood of the AD point, the combination $|g_2(\tau)^3/g_3(\tau)^2|=4|u|^3/|v|^2$ is small,    a condition that coincides with the original assumption for the validity of the expansion. }
\item  There exists a different potential superconformal fixed point at $\tau=i$ that preserves $\integers_2$ symmetry, and where $v=g_3(i)=0$ and $g_2(i) = (2\om)^4 4 u$. The sum over $m$ in (\ref{aDa}) then collapses to the $m=0$ contribution,  {the recursion relation (\ref{recursion}) for $L_n$ may be solved, and the remaining dependence on $\om$ may be eliminated in favor of $u$, }
 \begin{align}
     a_D-\tau a = - \frac{u^\frac{5}{4}}{9 \sqrt{\pi}}\sum_{n=0}^\infty  
\frac{ 2^{-10n} \, \Gamma(n+\frac{1}{4})^2 \Gamma(n+\frac{3}{4})^2 }{\Gamma(\frac{7}{12})\Gamma(\frac{11}{12})\Gamma(n+\frac{13}{12})\Gamma(n+\frac{17}{12})\Gamma(n+\frac{1}{2})n!} \left ( { u \over 3} \right )^{3n}
 \end{align}
{The above series has radius of convergence $\abs{u}<3$. The scaling dimension of the operator corresponding to the modulus $u$ is $\Delta(u) = \frac{4}{5} < 1$ is below the unitarity bound. This implies that there is no consistent way to take $\Lambda\rightarrow\infty$ such that the resulting theory is unitary and superconformal.}
\end{enumerate}

\subsection{Remarks on the convergence of the $\integers_N$ expansion}

In this subsection {we shall discuss the convergence properties of the expansion around the maximal AD point, briefly for the $\SU(N)$ case, and in more detail for $\SU(3)$.}

\subsubsection{Heuristic analysis of the $\SU(N)$ case}

As illustrated in the right panel of figure \ref{fig:1} for the case of $\SU(3)$, the branch points split into a set of  small branch points  of order $\cO(v^\frac{1}{N})$  and large branch points of order $\cO(\Lambda)$. The starting point for our expansion is a point in the moduli space of the Coulomb branch where $v \neq 0$ and $v_n=0$ for all $n>0$. The non-vanishing of $v$ guarantees that the small branch points $x^-_k$ remain well-separated. Turning on the moduli $v_n$ for $n \neq 0$ we observe that the parameter entering from the Taylor expansion in corollary \ref{Rhyp} is $v_n$. The expansion will remain convergent as long as no two branch points are brought to coincide with one another, which requires the parameter $u_n$ to remain sufficiently small with respect to $v^{(N-n)/N}$. Thus, the conditions for convergence, derived on heuristic grounds, are as follow,  
\bea
|v| \ll |\Lambda|^N 
\hskip 1in 
{|v_n| \ll 1}
\eea
For the case of $\SU(3)$ we can make these bounds more precise.

\subsubsection{Detailed analysis of the $\SU(3)$ case}

In appendix \ref{sec:3}, we provide a detailed derivation of the convergence conditions for the $\integers_3$ series. {Here, we remark on its consequences and compare it with the $\integers_6$ series of \cite{DDN}. 
\begin{enumerate}
\item The $\ZZ_6$ expansion converges provided the moduli satisfy the following inequalities \cite{DDN}, 
\begin{align}
    \tfrac{2}{\sqrt{27}} \abs{u}^\frac{3}{2} +\abs{1-v} <1
\end{align}
In figure \ref{z6}, the green translucent region shows the domain of convergence of the $\integers_6$ expansion. The  AD points are located on the boundary of this domain and the three multi-monopole points are mapped to the red dot at the peak of the conical region. 
\item The $\integers_3$ expansion converges provided the moduli satisfy the inequalities,
\begin{align}
    \abs{\tfrac{4u^3}{27}}<\abs{v}^2 < 1
\end{align}
In figure \ref{z3}, the red translucent cylindrical region  minus the solid cone shows the domain of convergence of the $\integers_3$ expansion. The AD point is at the peak of the cone. The multi-monopole and $\integers_6$ points are on the boundary of the domain of convergence. 
\item Figure \ref{combined} shows the total region that we can access with the combined expansions.
\end{enumerate}}
\begin{figure}
    \centering
    \begin{subfigure}[t]{0.45\textwidth}
        \centering
        \includegraphics[scale=0.5]{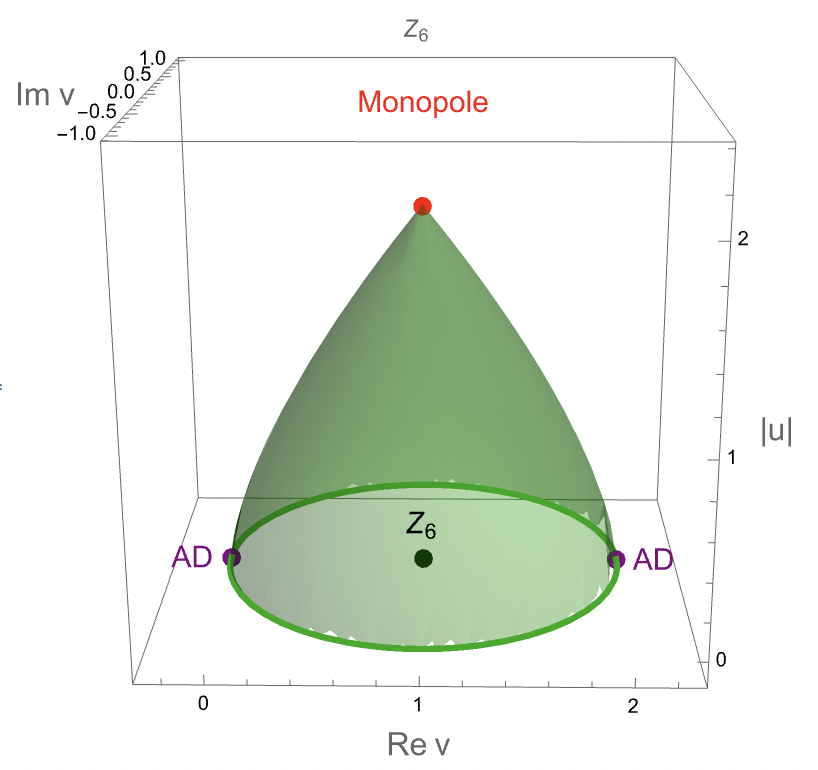}
    \caption{Convergence region of $\integers_6$ expansion
    \label{z6}}
    \end{subfigure}
    \hfill
    \begin{subfigure}[t]{0.45\textwidth}
        \centering
        \includegraphics[scale=0.5]{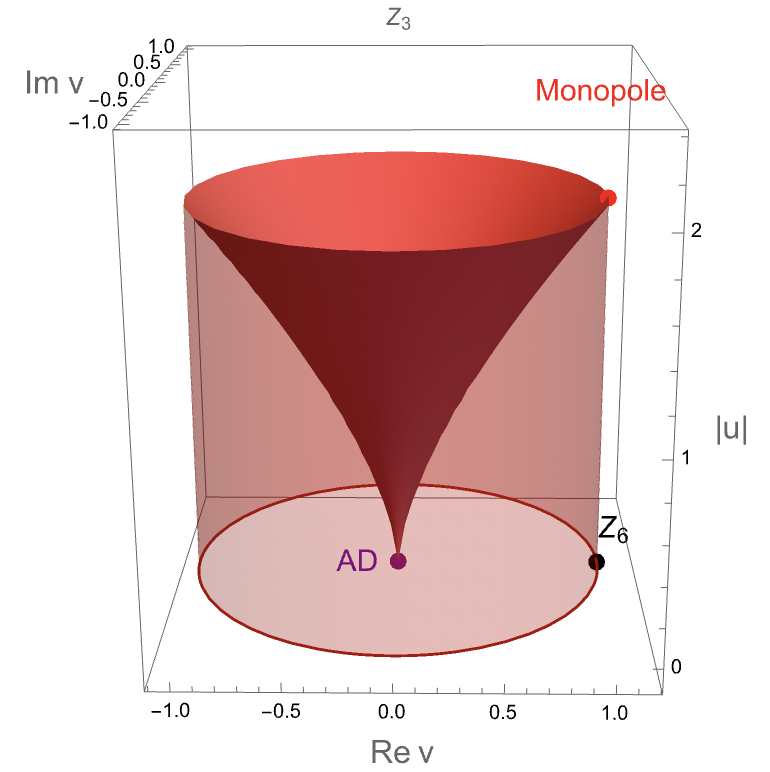}
        \caption{Convergence region of $\integers_3$ expansion
        \label{z3}}
    \end{subfigure}
    \caption{All the translucent colored regions denote convergence in the coordinates $(\Re v, \Im v, \abs{u})$, but the opaque colored regions are excluded by convergence.}
\end{figure}

\vskip -0.3in

\begin{figure}
    \centering
        \includegraphics[scale=0.8]{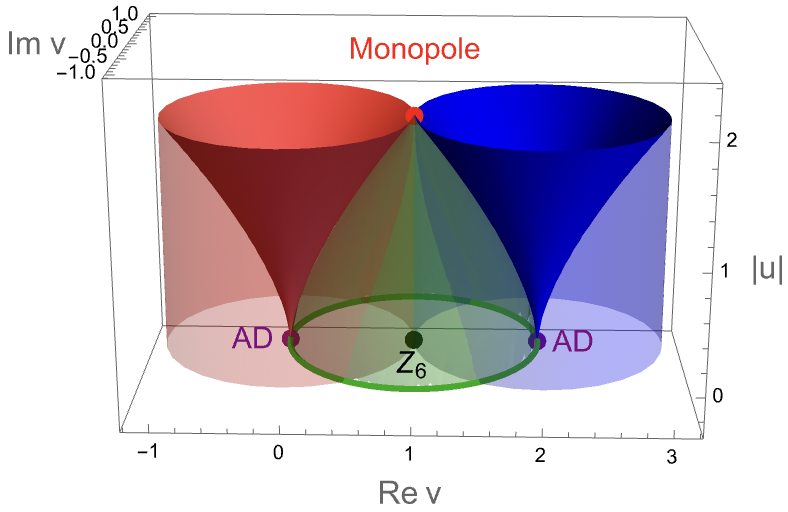}
        \caption{ This plot shows the total region of moduli-space that we can access using the two $\integers_3$ and $\integers_6$ expansions: the solid blue and red cones are excluded from the regions of convergence of the $\integers_3$ expansions, and the translucent regions denote convergence.} \label{combined}
\end{figure}

\newpage

\section{Candidate walls of marginal stability revisited }
\label{sec3}
\setcounter{equation}{0}

In this section, we shall reanalyze the candidate walls of marginal stability proposed in \cite{DDN} for $\SU(3)$, this time  from the perspective of the expansion of the periods around the AD points. We shall check agreement of the results obtained by the two expansions on the slice with $u=0$ and map out more walls of marginal stability beyond the $v$-plane for $\SU(3)$, and  analyze marginal stability in the $\SU(4)$ case on the slice $u_1=u_2=0$.

\subsection{Setup}

In this subsection, we shall briefly summarize the setup of \cite{DDN} used to analyze the marginal stability of BPS states. At a generic point on the Coulomb branch, the $\SU(N)$ gauge-group is spontaneously broken to its maximal Abelian subgroup  $\U(1)^{N-1}$. With respect to this unbroken gauge group, the states in the theory carry both electric charges $\bq = (q_1, \cdots, q_{N-1})$ and magnetic charges $\bg=(g_1, \cdots, g_{N-1})$, which we shall assemble into a single multiplet, 
\begin{align}
    \mu = (\bq, \bg) = (q_1, \dots, q_{N-1};g_1, \dots, g_{N-1})\in\integers^{N-1}\times \integers^{N-1}
\end{align}
The central charge $Z[\mu]$ of the $\cN =2$ supersymmetry  algebra, evaluated  in a state with charge vector $\mu$, is a linear function of $\mu$ given by \cite{SW:1994rs},
\begin{align}
    Z[\mu] = \sum_{I=1}^{N-1} (q_Ia_I+g_I a_{D,I})
\end{align}
In a unitary  theory the mass $M$ of any  state with charge vector $\mu$ satisfies the BPS bound $\abs{Z[\mu]} \leq M$. 
The state is a BPS state provided its mass $M$ saturates the BPS bound,
\begin{align}
    M = \abs{Z[\mu]}
\end{align}
Two BPS states with charges $\mu = (\bq,\bg) $ and $\mu'= (\bq',\bg')$  obey the Dirac quantization condition $ D \in \ZZ$ where $D$ is given by the symplectic pairing of the charges $\mu$ and $\mu'$, 
\bea
D= \bq \cdot \bg ' - \bg \cdot \bq' = \sum _{I=1}^{N-1} \Big ( q_I g_I ' - q_I' g_I \big )
\eea
When $D=0$, one may perform an $\Sp(2N-2,\ZZ)$ duality transformation to new charges $\tilde \mu$ and $\tilde \mu'$ that have vanishing magnetic components, and are therefore \textit{mutually local}. By contrast, when $D \not= 0$, the corresponding BPS states are \textit{mutually non-local}. This non-locality is, of course, familiar in the semi-classical limit where electric charges are light and magnetic monopoles are heavy soliton states such as the `t Hooft-Polyakov monopole. The novelty of the AD theories is the presence of massless mutually non-local states and fields.

\sm

Two BPS states with charges $\mu, \mu'$ and masses $M= \abs{Z[\mu]} ,M'=\abs{Z[\mu']}$, respectively,   can form a bound state of charge $\mu+ \mu'$  provided the mass $M_{b}$ of the bound state satisfies $M_{b} < M+M'$. The mass $M_{b}$ satisfies the BPS bound $\abs{Z[\mu+\mu']} \leq M_{b}$. In general, the inequality will be a strict one and the resulting bound state will not be a BPS state. For special charge arrangements and for special values of the vacuum expectation values $a_I$ and $a_{D,I}$, however, two BPS states can form a BPS bound state, namely when,
\begin{align}
\label{3.zeta}
    Z[\mu'] = r \, Z[\mu] \hbox{ for some } r \in \RR
\end{align}
For a given pair $\mu, \mu'$, the solutions to this equation carve out a real co-dimension one slice of the Coulomb branch that we refer to as a \textit{candidate wall of marginal stability}. Having equality of the mass $M$ of the bound state with its BPS bound $\abs{Z[\mu+\mu']}$ on the wall does open the option of forming a stable non-BPS bound state on either side of the wall. Whether this option is actually adopted by the theory is a dynamical question that goes beyond the purely kinematical considerations used here. For this reason the terminology \textit{candidate} wall of marginal stability will be used throughout.

\subsection{Marginal stability of BPS states in $\SU(3)$}

{Candidate walls of marginal stability were analyzed in \cite{DDN} for $\SU(3)$ on the slice $u=0$ {for arbitrary $v$} using the $\ZZ_6$ expansion of the periods. In this subsection, we re-examine these candidate walls of marginal stability using the $\ZZ_3$ expansion around one or the other of the AD points, first for $u=0$ and then for arbitrary $u,v$. }

\subsubsection{The $u=0$ slice}

In terms of our expansion around the AD point $u_0=1$, given in (\ref{2.Q+}), (\ref{2.Q-}) and (\ref{2.char}), the expressions for the SW periods of (\ref{2.aaD}) for the case $N=3$ are as follows,
\begin{align}
    a_1 &= (\rho - 1) (Q_1-Q_4)  - (3Q_4-Q_1)  & a_2 &= (1+\rho) a_1
    \no \\ 
    a_{D,1} &= (\rho - 1)(Q_1-Q_4) + \rho (3Q_4-Q_1) & a_{D,2} &= \rho a_{D,1}
\end{align}
where  $\rho= e^\frac{2\pi i}{3}$. Consider BPS states with charge vectors $\mu=(q_1, q_2; g_1, g_2)$ and $\mu'=(q_1', q_2'; g_1', g_2')$ and corresponding central charges,
\begin{align}
Z[\mu'] &= \a Q_1 + \b Q_4
\no \\
Z[\mu] &= \g Q_1 + \delta Q_4
\end{align}
where we have defined the following integers of the ring $\ZZ[\rho]$, 
\bea
\a & = & \rho \, (q_1'-g_2') - (q_2' + g_1')  
\no \\
\b & = & -(2+\rho)(q_1'+g_2') -(1+2 \rho)(q_2'-g_1')
\no \\
\g & = & \rho \, (q_1-g_2) - (q_2 + g_1)  
\no \\ 
\delta & = & -(2+\rho)(q_1+g_2) -(1+2 \rho)(q_2-g_1)
\eea
We may parametrize the solutions to the equation (\ref{3.zeta}) for the candidate wall of marginal stability in terms of the real variable $r$ as follows,
\begin{align}
    r &= {Z[\mu'] \over Z[\mu] } = { \a z + \b \over \gamma z + \delta} &z(v) &= \frac{Q_1(v)}{Q_4(v)}
\end{align}
Inverting the relation between $z$ and the real parameter $r$, for given charge assignments, will make $z$ trace arcs of circles  in the complex $z$-plane   that depend on the particular charge assignments of $\mu$ and $\mu'$. The strong-coupling spectrum of $\SU(N)$ SW theory has been worked out in \cite{Alim:2011kw}, and the $\SU(3)$ case is explained in Appendix E of \cite{DDN}. We summarize the results in the following table.

\begin{center}
\begin{tabular}{ |p{4cm}||p{4cm}|p{2.5cm}|p{2.5cm}| }
 \hline
 \multicolumn{4}{|c|}{Central charges and masses of BPS dyons near the $\SU(3)$ AD points} \\
 \hline
 Dyon charge & Central charge $Z[\mu_{kI}]$ & $M(u_0 = +1)$ & $M(u_0 = -1)$ \\
 \hline
     $\mu_{01} = (-1, 0;-1, 0)$   & $-(\rho-1) (Q_1+Q_4)$  & 1.55632 & 0\\
     $\mu_{12} = (-1,1;0,1)$ & $-(2\rho+1) (Q_1+Q_4)$ & 1.55632 & 0\\
     $\mu_{21} = (0,1;1,1)$ & $-(\rho+2) (Q_1+Q_4)$ & 1.55632 & 0 \\ \hline
     $\mu_{02} = (0,1;0,-1)$ & $ (\rho-1) (Q_1-Q_4)$ & 0 & 1.55632\\
     $\mu_{11} = (1,0;-1,-1)$  & $(2\rho + 1) (Q_1-Q_4)$ & 0 & 1.55632\\
     $\mu_{22} = (1,-1;-1,0)$ & $(\rho + 2) (Q_1-Q_4) $ & 0 & 1.55632\\
 \hline
\end{tabular}
\end{center}

As indicated in the table, each AD point has three mutually non-local dyons that become simultaneously massless. The equation (\ref{3.zeta}) has been solved for all possible pairs on the $u=0$ plane in \cite{DDN}. We summarize these results below, and then build on them.

\sm

There are 15 distinct pairwise ratios of the six BPS states. Two sets of three of these ratios are between massless mutually non-local dyons. These ratios are independent of $Q_1$ and $Q_4$ and necessarily complex, such as for example $Z[\mu_{11}]/Z[\mu_{02}] = -\rho$. There can be no walls of marginal stability between such pairs. The remaining nine ratios are between one massive and one massless dyon, they do depend on $Q_1$ and $Q_4$ through the ratio $z=Q_1/Q_4$, and can lead to candidate walls of marginal stability. To analyze the ratios systematically, we compare the central charges in the first triplet of mutually non-local dyons $(Z[\mu_{01}], Z[\mu_{12}], Z[\mu_{21}])$ with cyclic permutations of the second triplet of mutually non-local dyons $(Z[\mu_{02}], Z[\mu_{11}], Z[\mu_{22}])$, as follows,
\bea
(Z[\mu_{01}], Z[\mu_{12}], Z[\mu_{21}]) & = & r_1 \, (Z[\mu_{02}], Z[\mu_{11}], Z[\mu_{22}]) 
\no \\
(Z[\mu_{01}], Z[\mu_{12}], Z[\mu_{21}]) & = & r_2 \, (Z[\mu_{22}], Z[\mu_{02}], Z[\mu_{11}])
 \\
(Z[\mu_{01}], Z[\mu_{12}], Z[\mu_{21}]) & = & r_3 \, (Z[\mu_{11}], Z[\mu_{22}], Z[\mu_{02}])\no
\eea
where $z=z(v)$ and,
\bea
r _1 = - { z+1 \over z-1} 
\hskip 0.8in 
r _2 = - \rho { z+1 \over z-1} 
\hskip 0.8in 
r _3 = \rho^2 { z+1 \over z-1} 
\eea
\begin{enumerate}
\itemsep=0in
        \item The reality of $r_1$ parametrizes a straight line segment in the complex $z$-plane that lies on the real-axis. We will not consider such walls because they are non-compact.
        \item  The reality of $r_2$ parametrizes a continuous subset of the circle $|z  + \frac{i}{\sqrt{3}}|^2 = \frac{4}{3}$ in the complex $z$-plane for a continuous range of values of $r_2 \in \reals$.
        \item The reality of $r_2$ parametrizes a continuous subset of the circle $|z  - \frac{i}{\sqrt{3}}|^2 = \frac{4}{3}$ in the complex $z$-plane for a continuous range of values of $r_3 \in \reals$.
    \end{enumerate}
   
    In this sense, each candidate wall of marginal stability on the $v$-plane has a three-fold degeneracy, i.e. each wall can be obtained from three distinct pairs of dyons. 
    Finally, we can numerically map $z(v)$ to the $u_0$-plane by using the expression for $w(v)$ in terms of $v$ or $u_0$. This is displayed in {the right panel of figure \ref{u0}}, which reproduces the result of \cite{DDN}
    \begin{figure}
    \centering
    \begin{subfigure}[t]{0.5\textwidth}
        \centering
        \includegraphics[scale=0.35]{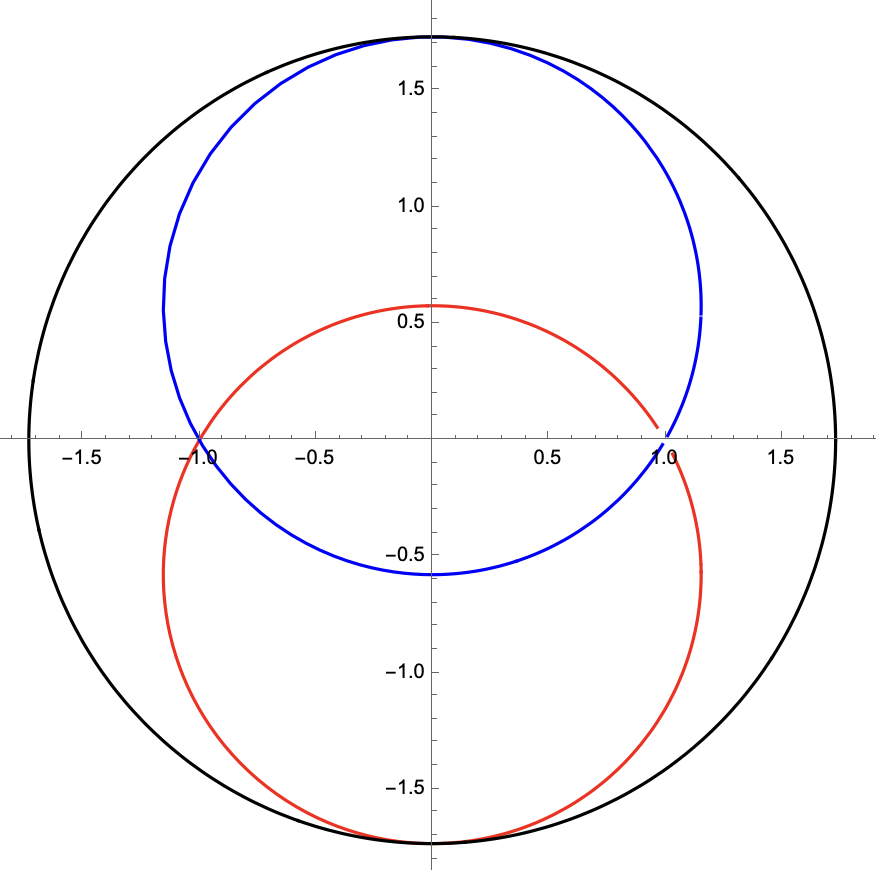}
    \end{subfigure}
    \hfill
    \begin{subfigure}[t]{0.45\textwidth}
        \centering
        \includegraphics[scale=0.5]{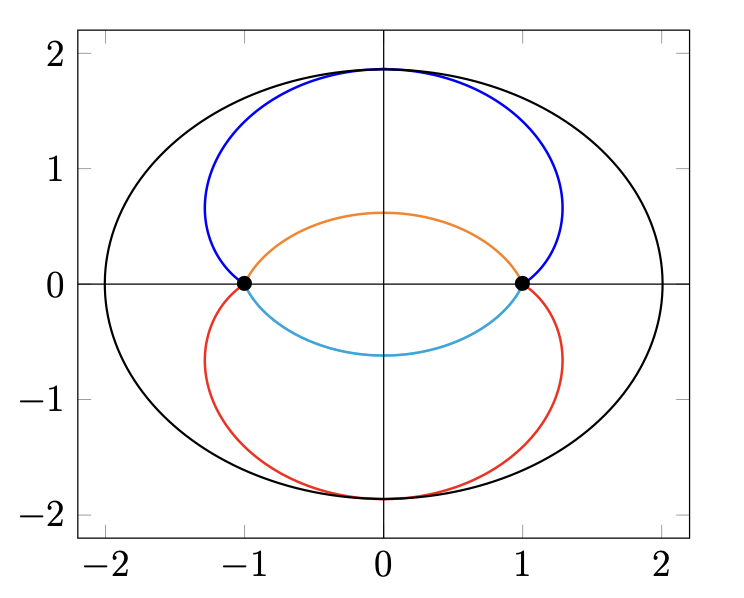}
    \end{subfigure}
    \caption{The candidate walls of marginal stability for $\SU(3)$ on the slice $u=0$ are represented by colored arcs  in the $z$-plane in the left panel and in the $u_0$ plane in the right panel. The contour of vanishing K\"ahler potential is drawn in black \cite{DDN}.\label{u0}}
    \end{figure}

\subsubsection{Existence of walls of marginal stability away from the $v$-plane}

In this sub-subsection, we give an argument for the existence of walls of marginal stability in the $\SU(3)$ Coulomb branch with $u\neq 0$ over the curves of marginal stability restricted to the $v$-plane. We expect to find a marginal stability subspace of real dimension three inside the Coulomb branch, since we have 2 complex degrees of freedom $u$ and $v$ parametrizing the Coulomb branch, and one real constraint $Z_2/Z_1\in \reals$. Since such a surface lives in the four-dimensional moduli space, it is rather hard to visualize and we shall focus instead on the marginal stability subspace of a three-dimensional slice of the Coulomb branch that we can visualize more easily. The following proposition shows the existence of walls in the $u$-plane for fixed values of $v$ that lie on a wall. 
\begin{prop}
\label{propWall}
    For any point $v_0$ on a curve of marginal stability in the $v$-plane, there exists a curve of marginal stability in the $u$-plane that goes through the point $(u, v) = (0, v_0)$.
\end{prop}
We hold $v=v_0$ fixed and consider two central charges $Z_1(u)$ and $Z_2(u)$ evaluated at the point $v=v_0$, which may be regarded as a (locally) holomorphic function of a single complex variable $u$. Then we may define the relative phase of these two central charges as,
\begin{align}
     e^{2 i \phi(u, \bar{u})} = \frac{Z_1(u)\Bar{Z}_2(\Bar{u})}{Z_2(u)\Bar{Z}_1(\Bar{u})}
\end{align} 
In what follows, we will use the fact that the phase of a holomorphic function $f$ is harmonic for all points where the function is non-zero; this assumption is necessary since one takes the logarithm of $f$ in the proof of this fact. However, (potential) vanishing of the central charge $Z_k(u)$ does not pose an issue since we take its complex modulus in the definition of the relative phase. Then it follows that $\phi(u, \bar{u})$ is \textit{locally} a harmonic function on any open set that contains $u=0$. But a harmonic function on a connected domain $\cD\subset\complex$ can never attain its extreme values in the interior. Any point $u_0\in\complex$ that lies on a wall of marginal stability, i.e. $Z_2/Z_1\in\reals$, satisfies $\phi(u_0, \bar{u}_0)=0$. In particular, $\phi(0,0)=0$. This implies that zero is neither a minimum nor a maximum of $\phi$, and that there exists a closed subset $\cC\subset\partial\cD$ such that $\phi(\cC)< 0$ and $\phi(\cC^c)\geq0$. Hence, there exists a curve of marginal stability that goes through the interior and is continuously connected to $u=0$.

\sm

\textbf{Remark.} The above proposition applies locally in a neighborhood of $u=0$, where the central charges are analytic. In particular, $Z(u)$ is not required to be analytic \textit{globally}, and hence the function $\phi$ would \textit{not} be globally harmonic due to the potential non-analyticities in $Z(u)$. Therefore, this local existence result for walls of marginal stability does not pose an obstruction to the compactness of the walls of marginal stability.

\subsubsection{Numerically finding walls of marginal stability beyond the $v$-plane}

In this sub-subsection, we will find candidate walls of marginal stability using two distinct methods: perturbation theory and numerical integration. 

\sm

The first method involves first-order perturbation theory in $u$ for a fixed value of $v$ on the orange arc in figure \ref{u0} and a mesh of values of $r$. We note that the figure does not appreciably change even if we go to high orders in perturbation theory, as long as we are inside the radius of convergence.

\begin{figure}
    \centering
    \begin{subfigure}[t]{0.4\textwidth}
        \centering
        \includegraphics[scale=0.3]{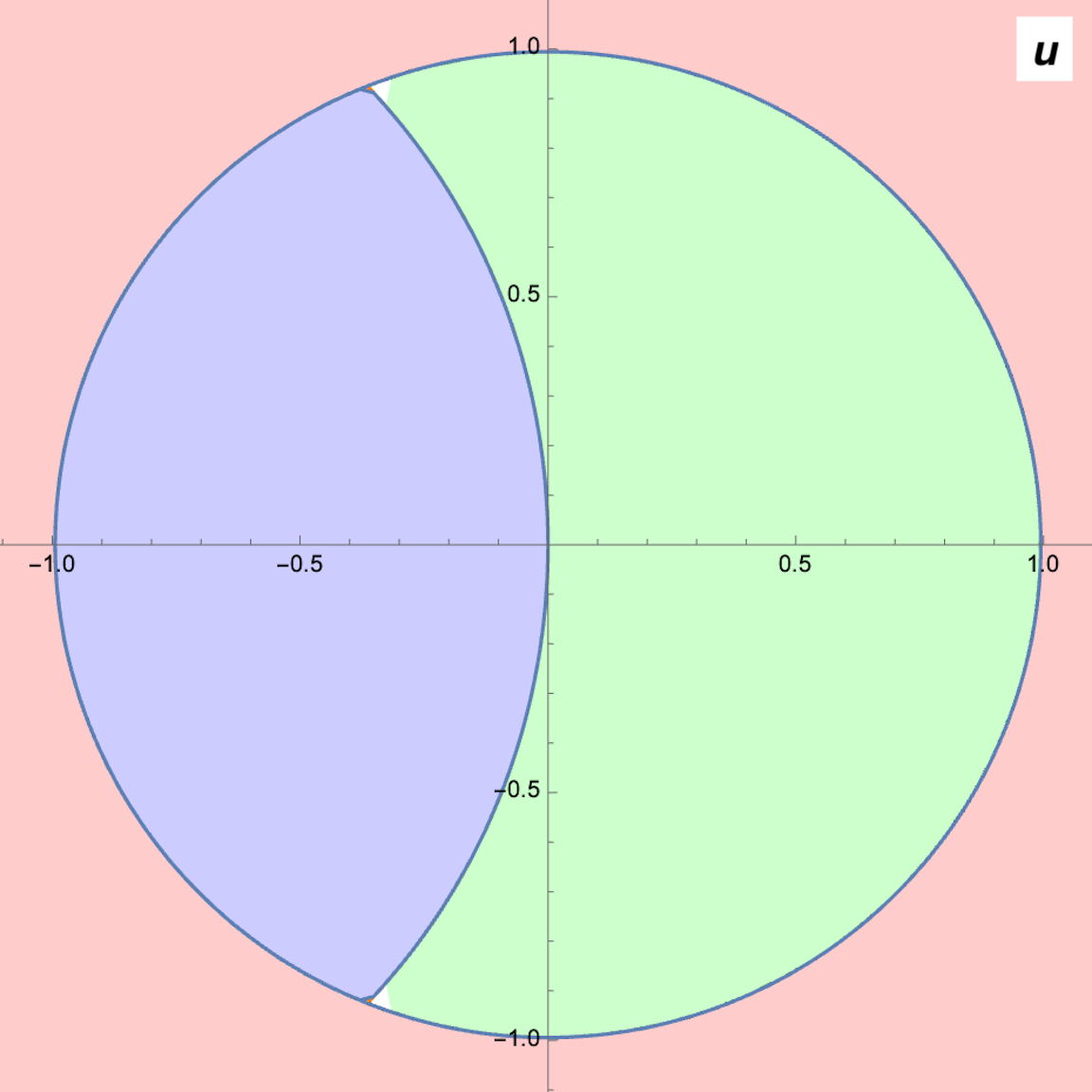}
    \caption{Walls for the a single pair $(\mu_{01},\mu_{22})$}
    \end{subfigure}
    \hfill
    \begin{subfigure}[t]{0.5\textwidth}
        \centering
        \includegraphics[scale=0.3]{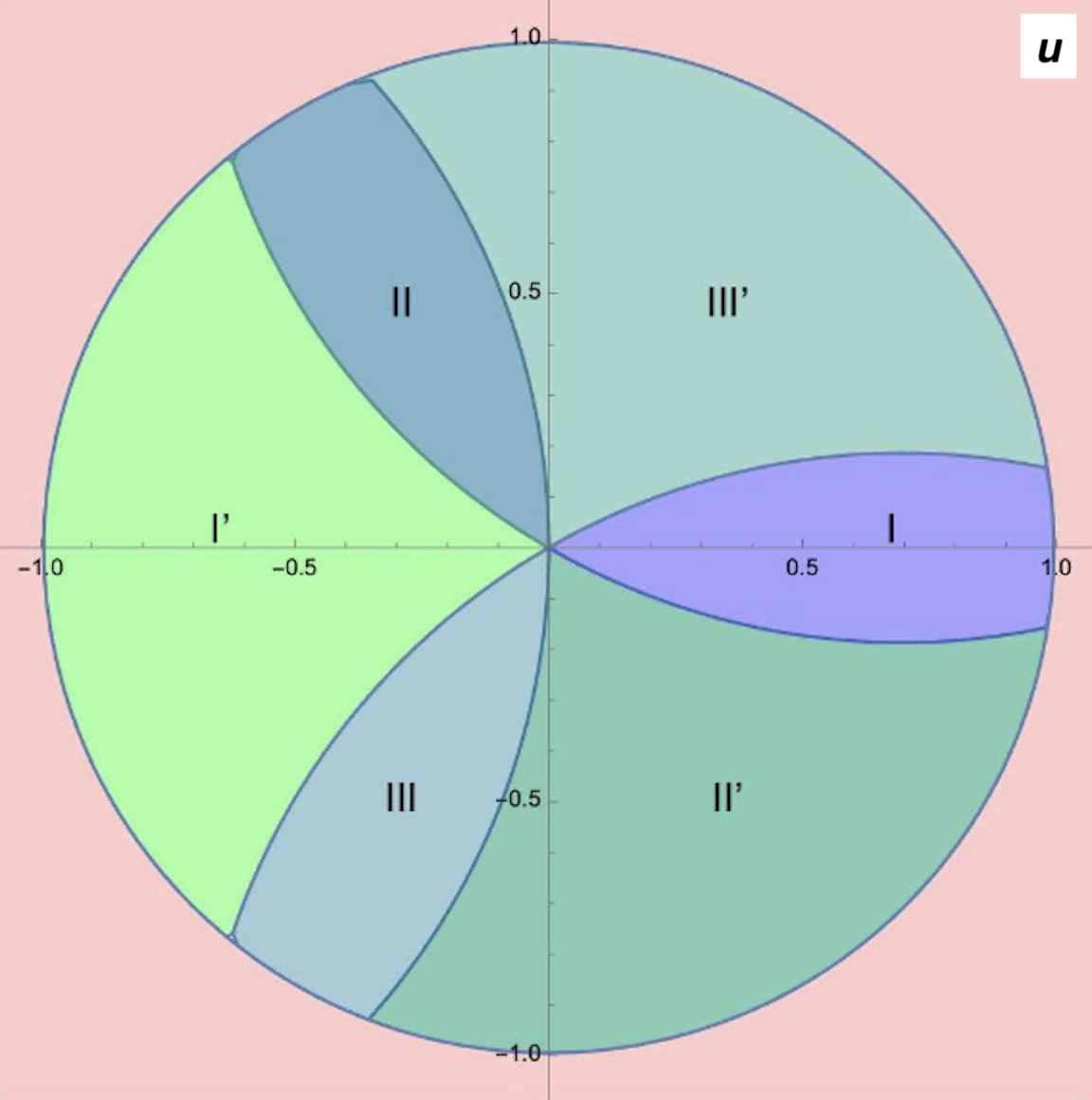}
        \caption{Walls for all three pairs }
    \end{subfigure}
    \hfill
    \begin{subfigure}[t]{0.8\textwidth}
        \centering
        \includegraphics[scale=0.5]{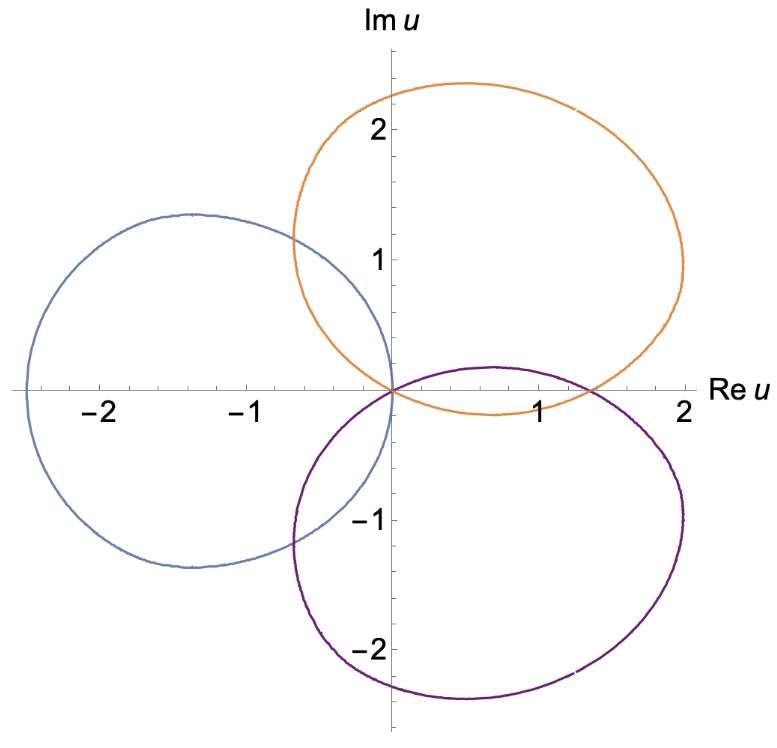}
        \caption{Full picture of the three walls beyond the radius of convergence\label{cross}}
    \end{subfigure}
    \caption{The figure above shows the curves of marginal stability on the $u$-plane for $u_0 \sim 0.6184 i$ (the midpoint of the orange arc in figure \ref{u0}). The left panel shows the walls for a single pair of dyons. The red region is inaccessible by perturbation theory since our $\integers_6$-expansion does not converge there, the blue region is where the BPS states $\mu_{01}$ and $\mu_{22}$ are stable with $\phi < 0$, and the green region is where bound states may be formed, i.e. $\phi>0$. The right panel shows a plot of the curves of marginal stability on the $u$-plane for all three pairs of dyons simultaneously. Again, the red region is inaccessible by perturbation theory. The regions of mutual stability ($\phi < 0$) and instability ($\phi > 0$) for 2 pairs are: II and II$'$ for $(\mu_{01},\mu_{22})$ and $(\mu_{12}, \mu_{02})$, I and I$'$ for $(\mu_{12}, \mu_{02})$ and $(\mu_{21}, \mu_{11})$, III and III$'$ for $(\mu_{21}, \mu_{11})$ and $(\mu_{01},\mu_{22})$. The panel \ref{cross} displays the full walls beyond the radius of convergence obtained from numerical integration. Blue walls correspond to $(\mu_{01},\mu_{22})$, orange walls to $(\mu_{12},\mu_{02})$, and purple walls to $(\mu_{21},\mu_{11})$. \label{u-plane SU(3)}}
\end{figure}

We will focus on the segment of the curve produced by the pairs $(\mu_{01},\mu_{22})$, $(\mu_{12}, \mu_{02})$, and $(\mu_{21}, \mu_{11})$ that has positive imaginary part, i.e. the orange curve in figure \ref{u0}. For any $u_0$ with $\abs{u_0}<1$ on this arc, we have an absolutely convergent expansion in $u$, and we can get arbitrarily close to the AD points at $u_0=\pm 1$. Recall that our expansions have the following radii of convergence in the $u$-plane.
\begin{align}
    \integers_3\text{ points}: ~ 2^\frac{1}{3} \abs{u}&<3\abs{\pm 1\mp u_0}^\frac{2}{3}<3 \text{ for }u_0=\pm 1\text{ respectively}
    \no \\
    \integers_6 \text{ point}: ~ 2^\frac{2}{3} \abs{u} & <3 \left( 1-\abs{u_0}\right)^\frac{2}{3} 
    \quad \text{ and } \quad \abs{u_0}<1.
\end{align}
On regions of overlapping convergence, recall that $\text{Radius}_u(\integers_3)\geq\text{Radius}_u(\integers_6)$. Hence, it is more fruitful to apply the $\integers_3$ expansions near the AD points because this inequality is significant. On the other hand, we apply the $\integers_6$ expansion on the imaginary $v$-axis because that is the boundary of convergence for the $\integers_3$ expansion. Precisely at the AD points, neither expansion converges but we can get arbitrarily close.

\sm

However, this method is limited by the radius of convergence of the expansion. To circumvent this limitation, we recall that the periods of pure $\SU(3)$ SW theory satisfy Picard-Fuchs equations in the variables \cite{Klemm:1995wp}
\begin{align}
    (x_1,x_2) = \left(\frac{4u^3}{27}, (1-v)^2\right)
\end{align}
This system of second-order PDEs can be transformed into a system of first-order ODEs which are numerically integrable (see Appendix D of \cite{DDN}). We use this method to compute the periods, central charges, and walls of marginal stability, by  scanning for points on the $u$-plane, for fixed values of $v$ on the orange arc in figure \ref{u0} where $Z[\mu_2] / Z[\mu_1]$ is real-valued. 
\subsubsection*{Remarks}
\begin{figure}
    \centering
    \begin{subfigure}[t]{0.49\textwidth}
        \centering
        \includegraphics[scale=0.26]{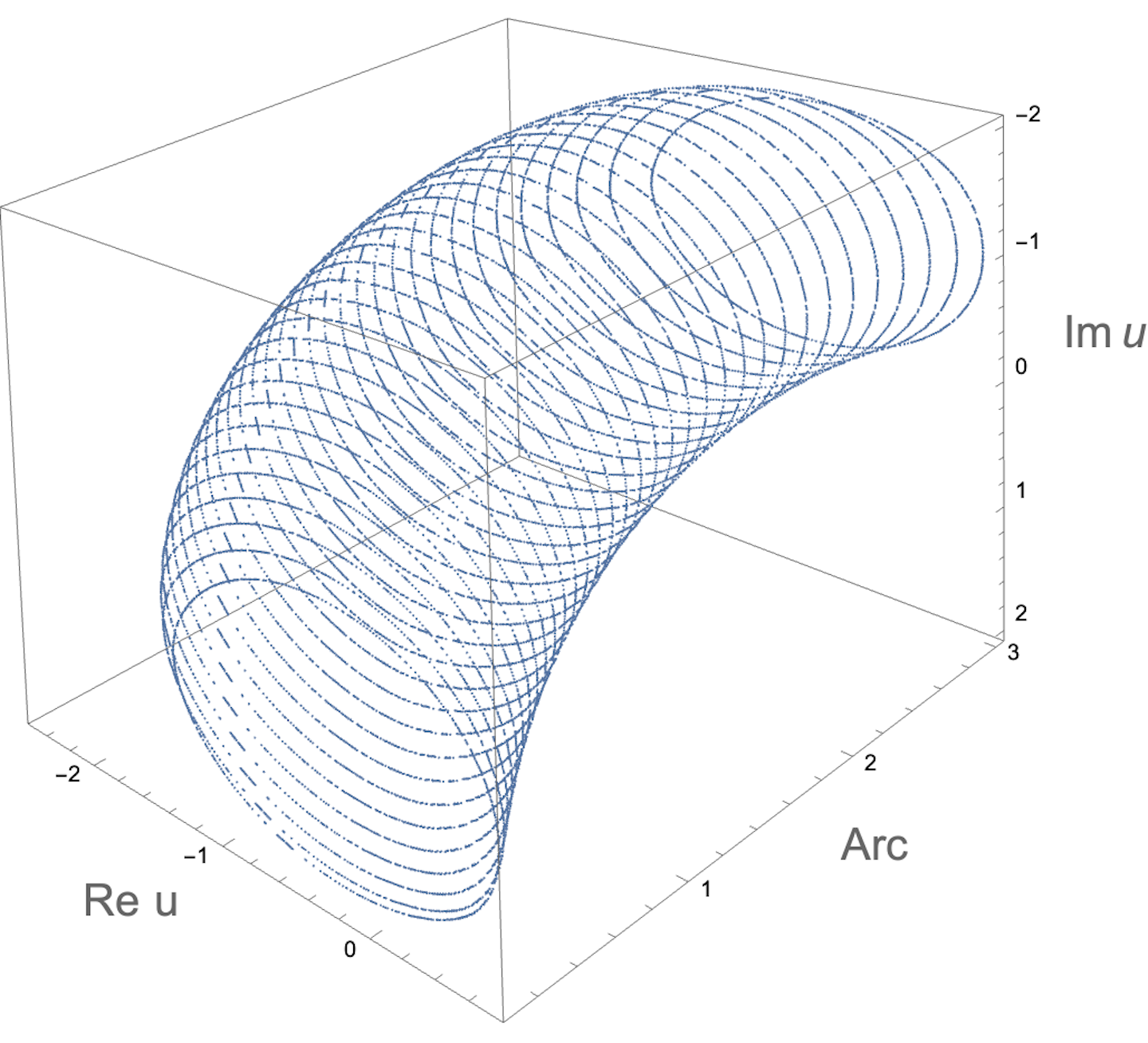}
    \caption{1 pair $(\mu_{01},\mu_{22})$\label{one pair}}
    \end{subfigure}
    \begin{subfigure}[t]{0.49\textwidth}
        \centering
        \includegraphics[scale=0.4]{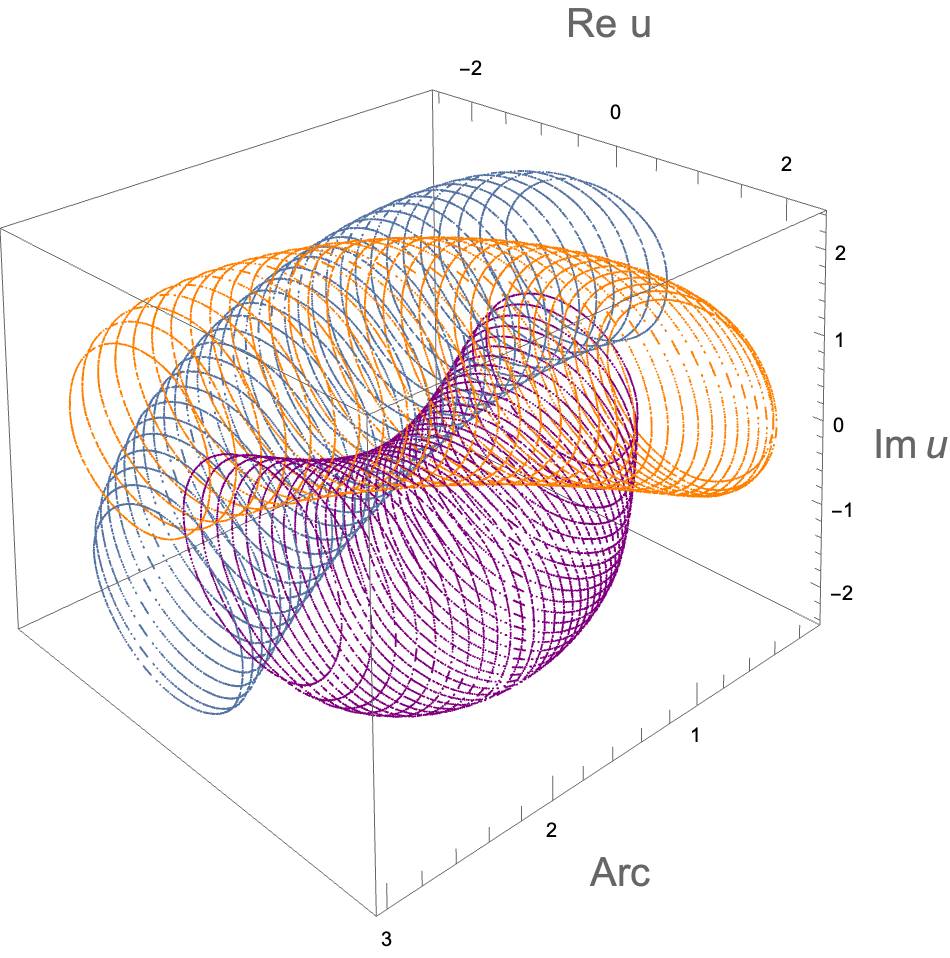}
    \caption{All $3$ pairs -- degeneracy lifted\label{three pairs}}
    \end{subfigure}
    \hfill
    \caption{A three-dimensional view of the candidate walls of marginal stability. \label{full walls}}
\end{figure}
\begin{enumerate}
\itemsep=0in
    \item The three pairs correspond to the  three walls in figure \ref{u-plane SU(3)}, both related by $\integers_3$. At the fixed point $u=0$ of this $\integers_3$ symmetry the degeneracy among the three pairs is restored.
    \item The arcs in figure \ref{u-plane SU(3)} with $\phi=0$ are approximately circular near the origin because, inside the radius of convergence, equation (\ref{3.zeta}) was truncated to first order to give,
    \begin{align*}
        (d u-b) &\approx r (-cu+a) \Rightarrow u(r) = \frac{ar+b}{cr+d}+\dots    
        \end{align*}
for some $ a,b,c,d \in \CC$ with $ad-bc\neq 0$. An equation of this form traces out an arc of a circle inside the radius of convergence. However, the arcs are slightly deformed from circles outside the radius of convergence, as is clear from figure \ref{cross}.
    \item The plots in figure \ref{full walls} were created by picking 33 evenly spaced points on the orange arc of marginal stability in figure \ref{u0} and then solving numerically for $\phi=0$.
    \item The region of stability is a tubular neighborhood, and there are three such compact regions corresponding to the three walls that are related by $\integers_3$.
\end{enumerate}

\subsection{Marginal stability of BPS states in $\SU(4)$}

On the slice that contains the AD points (i.e. $u_k = 0$ for $k\neq 0$), we have 
\begin{align}
    a_1 &= \varepsilon Q_- - Q_+ &a_{D,1} &= \varepsilon^2 Q_+ - \varepsilon Q_-\nonumber\\
    a_2 &= (\varepsilon^2 +1) a_1 &a_{D,2} &= \varepsilon^2 a_{D,1}\\
    a_3 &= \varepsilon^2 a_1 &a_{D,3} &= -a_{D,1} \nonumber
\end{align}
The central charge can be shown to take the following form ($Q_\pm = Q_1 \pm Q_5$):
\begin{align}
    Z[\mu] &= (m_0 + \varepsilon^2 m_2) Q_+ + \varepsilon (n_0 + \varepsilon^2 n_2) Q_-,    
\end{align}
where 
\begin{align*}
    m_0 &= -q_1-q_2-g_2 & n_0 &= q_1+q_2 - g_1 + g_3\\
    m_2 &= -q_2-q_3 + g_1 -g_3 & n_2 &= q_2 + q_3 - g_2
\end{align*}
As for $\SU(3)$, the following coordinate will be convenient
\begin{align}
    z(v)= \frac{Q_1}{Q_5}
\end{align}

The strong-coupling spectrum of BPS dyons in pure $\N = 2\;\SU(N)$ gauge theory was worked out in \cite{Alim:2011kw}. Following their algorithm for the $N=4$ case, we see that there are 12 stable BPS dyons at strong coupling which split into 2 sets of 6, each of which  becomes massless at one or the other of the two AD points corresponding to $(u_0,u_1,u_2) = (\pm 1,0,0)$. Two dyons within each set of 6 are mutually local, while two dyons belonging to different sets of 6 are mutually non-local.  The electromagnetic charge vectors $\mu_{kJ} = (q_1, q_2, q_3; g_1, g_2, g_3)$ for  the IR gauge-group $\U(1)_1\times \U(1)_2 \times \U(1)_3$ are displayed in the table that follows. 

\begin{center}
\begin{tabular}{ |p{5cm}||p{2.5cm}|p{2.5cm}|p{2.5cm}| }
 \hline
 \multicolumn{4}{|c|}{Masses and central charges of BPS dyons near the $\SU(4)$ maximal AD points} \\
 \hline
 Dyon & $Z[\mu_{kI}]$ & $M(u_0 = +1)$ & $M(u_0 = -1)$ \\
 \hline
     $\mu_{01} = (-1,0,0;-1,0,0)$   & $(1-\varepsilon^2)Q_+$    & 1.2828 & 0\\
     $\mu_{03} = (0,1,-1; 0,0,-1)$ & $- (1-\varepsilon^2)Q_+$ & 1.2828 & 0 \\
     $\mu_{12} = (-1,1,0;0,1,0) $ & $-(1+\varepsilon^2) Q_+$ & 1.2828 & 0\\
      $\mu_{32} = (0,0,1; 1,1,1)$ & $-(1+\varepsilon^2) Q_+$ & 1.2828 & 0 \\
     $\mu_{21} = (0,1,0; 1,1,0)$ & $-2 Q_+$ & 1.8142 & 0\\
     $\mu_{23} = (-1,0,1; 0,1,1)$ & $-2\varepsilon^2 Q_+$ & 1.8142 & 0 \\
 
     \hline
     $\mu_{02} = (0,1,-1; 0,-1,0)$ & $\varepsilon(1+\varepsilon^2) Q_-$ & 0 & 1.2828\\
     $\mu_{22} = (1,0,0; -1,-1,-1)$ & $\varepsilon(1+\varepsilon^2) Q_-$ & 0 & 1.2828\\
     $\mu_{31} = (0,0,1; 0,0,-1)$ & $-\varepsilon (1-\varepsilon^2) Q_-$ & 0 & 1.2828\\
     $\mu_{33} = (1,-1,0; -1,0,0)$ & $\varepsilon(1-\varepsilon^2)Q_-$& 0 & 1.2828\\
      $\mu_{11} = (1,0,-1; -1,-1,0)$  & $2\varepsilon Q_-$ & 0 & 1.8142\\
     $\mu_{13} = (0,1,0; 0,-1,-1)$ & $2\varepsilon^3 Q_-$ & 0 & 1.8142\\
 \hline
\end{tabular}
\end{center}

We observe a phenomenon that did not occur in $\SU(3)$: at each AD point, the massive  BPS states belong to two different multiplets with unequal masses $m_0 = 1.2828$ and $m_1=1.8142$, listed explicitly below.  
\begin{itemize}
\itemsep=0in
    \item \underline{IA}. Massless at $u_0 = 1$ but have mass $m_0$ at $u_0 = -1$: $\{\mu_{02}, \mu_{22}, \mu_{31}, \mu_{33}\}$.
    \item \underline{IB}. Massless at $u_0 = 1$ but have mass $m_1$ at $u_0 = -1$: $\{\mu_{11}, \mu_{13}\}$.
    \item \underline{IIA}. Massless at $u_0 = -1$ but have mass $m_0$ at $u_0 = 1$: $\{\mu_{01}, \mu_{03}, \mu_{12}, \mu_{32}\}$.
    \item \underline{IIB}. Massless at $u_0 = -1$ but have mass $m_1$ at $u_0 = 1$: $\{\mu_{21}, \mu_{23}\}$.
\end{itemize}
Na\"ively, there are $\binom{12}{2}$ distinct pairs. But ratios of central charges within a single category always give rise to trivial walls of complex-co-dimension 1 in the $z$-plane. So, it suffices to consider walls between distinct types: $\{$IA-IIA, IA-IIB, IB-IIA, IB-IIB$\}$. Hence, there are only $36$ pairs between mutually local BPS states that could give rise to genuine walls. 
\begin{enumerate}
\itemsep=0in
    \item \underline{IA-IIA.} We have the folllowing candidate walls for this case:
    \begin{align}
        z_1(r) = \frac{1+\alpha r}{1-\alpha r}, \; \text{where } \alpha\in \{\pm \varepsilon, \pm i\varepsilon\}. 
    \end{align}
    
    \item \underline{IB-IIA.} We have the folllowing candidate walls for this case:
    \begin{align}
        z_2(r) = \frac{1+\beta r}{1-\beta r}, \; \text{where } \beta\in \{\pm \tfrac{1}{\sqrt{2}}, \pm \tfrac{i}{\sqrt{2}}\}. 
    \end{align}
    \item \underline{IA-IIB.} We have the folllowing candidate walls for this case:
    \begin{align}
        z_3(r) = \frac{1+\gamma r}{1-\gamma r}, \; \text{where } \gamma\in \{\pm \sqrt{2}, \pm i\sqrt{2}\}. 
    \end{align}
    \item \underline{IB-IIB.} We have the folllowing candidate walls for this case:
    \begin{align}
        z_4(r) = \frac{1+\delta r}{1-\delta r}, \; \text{where } \delta\in \{- \varepsilon^{\pm 1}\}. 
    \end{align}
\end{enumerate}
\begin{figure}\label{SU(4) BPS}
    \centering
    \begin{subfigure}[t]{0.4\textwidth}
        \centering
        \includegraphics[scale=0.45]{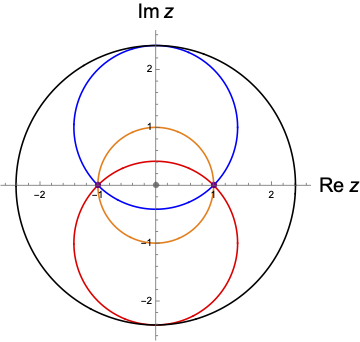}
    \end{subfigure}
    \hfill
    \begin{subfigure}[t]{0.5\textwidth}
        \centering
        \includegraphics[scale=0.45]{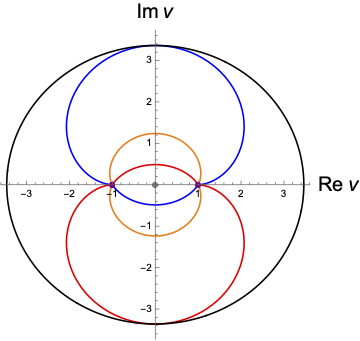}
    \end{subfigure}
    \caption{The candidate walls of marginal stability for the $\SU(4)$ case on the slice $u_1=u_2=0$ represented in the  $z$-plane in the left panel and in the $v$-plane in the right panel. The purple dots denote the AD points, and the gray dot is the $\integers_{8}$-point. The black curve is the contour of vanishing K\"ahler potential.}
\end{figure}

This set of curves is still highly degenerate: there are only three distinct walls of marginal stability.
\begin{enumerate}
\itemsep=0in
    \item A circle of radius $\sqrt{2}$ centred at $z = i$ with degeneracy $10$. 
    \item A circle of radius $\sqrt{2}$ centred at $z = -i$ with degeneracy $10$.
    \item A circle of radius $1$ centred at $z = 0$ with degeneracy $12$.
\end{enumerate}
Note that the above degeneracy adds up to $32$ instead of $36$ since we do not consider the straight lines with degeneracy $4$ corresponding to $\beta = \pm\frac{1}{\sqrt{2}}$ and $\gamma=\pm\sqrt{2}$.
The last contour to plot is of vanishing K\"ahler potential:
\begin{align}
    K(v) = 0 \Rightarrow \abs{z} = \cot(\frac{\pi}{8}) = 1+\sqrt{2}.
\end{align}

\subsection{Comments on the $\SU(N)$ case for $u_{k}=0$ for $k>0$}

We have computed candidate walls of marginal stability up to $\SU(7)$ on the $v$-plane with $u_{k}=0$ for $k>0$. We do not give the pictures here explicitly since they share many features. Instead, we close this section with some general remarks on the $\SU(N)$ case, always assuming $u_{k}=0$ for $k>0$ in what follows. A generic wall of marginal stability on this slice satisfies 
\begin{align}
    z(r) = \frac{Q_1}{Q_{N+1}} = \frac{\alpha r + 1}{\alpha r - 1}, \text{ for some }\alpha\in\complex\text{ and any }r \in \reals.
\end{align}
The K\"ahler potential on such a wall of marginal stability satisfies \cite{DDN}
\begin{align}\label{Kwall}
    K(v)|_{\text{wall}} = \frac{N}{2\pi} \abs{Q_{N+1}}^2 \tan(\tfrac{\pi}{2N}) \left[\abs{\frac{\alpha r+1}{\alpha r-1}}^2 - \cot^2\left(\tfrac{\pi}{2N}\right)\right].
\end{align}
Since $r$ can be arbitrarily rescaled, without loss of generality, we may take $\alpha = e^{i\phi} \in S^1$, for $\phi\in [0,\pi]$ (we identify $\alpha\sim -\alpha$ since they are equivalent under $\zeta\mapsto -\zeta$). For the present discussion, we suppose that $\alpha\notin\{0,\pi\}$ so that we exclude straight lines on the real axis.

Consider the special case $\alpha=i$. Then we have $\abs{z}=1$. This is the origin-centred circle in the $z$-plane that exist for $N=2, 4, 6$. For the $N=2$ case, this is precisely the $K=0$ contour found by Seiberg and Witten \cite{SW:1994rs}. However, such an origin-centred circle seems to be absent for odd $N$ --  we checked this up to $\SU(7)$ but lack a proof.
    
If a wall of marginal stability that is confined to the region $K\leq 0$ contains a point where $K=0$, then we claim that it must be tangential to the $K=0$ contour at one of $z=\pm i\cot(\frac{\pi}{2N})$.  This point, together with both the AD points, uniquely specifies a circle in the $z$-plane. This can be seen explicitly by setting (\ref{Kwall}) to zero, which amounts to
    \begin{align}
        (1-\cot^2(\tfrac{\pi}{2N}))r^2 + 2\csc^2(\tfrac{\pi}{2N})\cos\phi r + (1-\cot^2(\tfrac{\pi}{2N})) = 0.
    \end{align}
    This discriminant of this quadratic polynomial is 
    \begin{align}
        D = 2\csc^4(\tfrac{\pi}{2N})(\cos(2\phi) - \cos(\tfrac{2\pi}{N})).
    \end{align}
    This equation admits real solutions provided $\phi \in [0,\frac{\pi}{N}]\cup [\pi-\frac{\pi}{N},\pi]$. If $\phi \notin \{\frac{\pi}{N}, \pi-\frac{\pi}{N}\}$, the wall intersects the $K = 0$ contour twice. Such a wall goes beyond the $K\leq 0$ region since the circle also goes through the AD points; i.e. this is a circle through four specified points. If $\phi\in\{\frac{\pi}{N}, \pi-\frac{\pi}{N}\}$, then $D=0$ and we have a unique intersection corresponding to $\alpha=e^{\pm i\frac{\pi}{N}}=\ep^{\pm 1}$. Solving the quadratic for these values of $\alpha$ fixes $r = \pm 1$, i.e. $z = \pm i \cot(\tfrac{\pi}{2N})$. Hence, there exist only two distinct walls of marginal stability that are confined to the region $K\leq 0$, and are tangential to the contour $K=0$. Furthermore, such walls exist for all $N\geq 3$. However, the above statement does not preclude the existence of walls that are confined to the strong-coupling region but always have $K<0$. This pair of walls is explicitly realized in the cases $N=3,4,5,6,7$. 
    
    \sm
    
The contour of vanishing K\"ahler potential is also a universal feature, and always has radius $\cot(\frac{\pi}{2N})$ in the $z$-plane, which scales like $\sim N$ as $N\rightarrow\infty$. This $N$-scaling of the radius of the $K=0$ contour implies that the universal curves of marginal stability that are tangential to the $K=0$ contour do not exist at large-$N$ since they tend to straight lines. On the other hand, for any even $N$, we always expect to find the contour with $\abs{z}=1$ since its radius is independent of $N$.  The fate of the other contours (in the strict region $K<0$) for even or odd $N$ is not completely clear. We expect that the even and odd cases converge to the same picture as $N\rightarrow\infty$. On these grounds, we suspect that the contours that lie in the (strict) $K<0$ region for any $N$ converge to the contour with $\abs{z}=1$ as $N\rightarrow\infty$, but we do not have a proof. We close with an open question: is $\abs{z}=1$ the unique contour of marginal stability as $N\rightarrow \infty$ on the $z$-plane with $u_{k}=0$ for $k\neq 0$?

\newpage

\section{The intrinsic Coulomb branch of AD theories}
\label{sec4}
\setcounter{equation}{0}

In this section we shall study certain properties of the \textit{intrinsic Coulomb branch} of the AD theories by taking the $\Lambda \to \infty$ limit of the asymptotically free embedding $\SU(N)$ super Yang-Mills theory near one of the maximal AD points. The resulting intrinsic AD theory is $\cN=2$ superconformal, its operators transform under representations of the superconformal Lie algebra  $\SU(2,2|2)$, and have definite scaling dimensions. In particular, we shall study the behavior of the K\"ahler potential in this limit, and show that it is positive definite (vanishing only at the AD point) and convex provided only genuine intrinsic Coulomb branch operators $\cO_n$  are turned on away from the AD point, whose operator dimension  satisfies the unitarity bound $\Delta (\cO_n) >1$ in a unitary superconformal field theory.

\subsection{The $(\ma_1, \ma_{N-1})$ intrinsic Coulomb branch}

The intrinsic AD theories exist independently of their embedding into the Coulomb branch of an asymptotically-free parent theory, and a given AD theory may be reached from different parent theories. For example, the AD theories obtained in the $\Lambda \to \infty$ limit of the parent theories $\SU(3)_{N_f=0}$ and $\SU(2)_{N_f=1}$ are the same and referred to as the $(\ma_1, \ma_2)$ theory. The nomenclature originates with yet another parent theory, namely the $(\ma_1, \ma_2)$ theory may be constructed by compactifying the six-dimensional $\cN=(2,0)$ theory with gauge-algebra~$\ma_1$. The BPS quiver for such a theory is given by the product of the $\ma_1$ and $\ma_{N-1}$ Dynkin diagrams, whence the nomenclature $(\ma_1, \ma_{N-1})$ \cite{Gaiotto:2009hg,Alim:2023doi,Cecotti:2010fi,Cecotti:2011rv}.
 
\sm

In this section, we shall consider the intrinsic theory obtained near the maximal AD points of the $\cN=2$ super Yang-Mills theory with gauge group $\SU(N)$ without hyper-multiplets. These theories can also be constructed  The SW curve and one-form are given by, \cite{Eguchi:1996ds}\footnote{The relation may be derived by temporarily restoring the  $\Lambda$-dependence in (\ref{2.SW.A}) to $A(x) = \hat A(x) - \Lambda ^N$ and $y^2=\hat A(x) (\hat A(x) - 2 \Lambda ^N)$ and taking the limit $-2 \hat y^2 = \lim _{\Lambda \to \infty} y^2/\Lambda ^N$, while   keeping $x, u_n$ constant.}
\begin{align}
    \hat{y}^2 &= x^N - \sum_{n = 1}^{N-2} u_n x^n + v &\lambda &= \sqrt{-2} \, \hat{y}\;dx
\end{align}
The resulting SW curve and differential are scale covariant in the following sense. Since the SW periods are given by integrals of $\lambda$, we must assign to $\lambda$ the scaling dimension one, which means that under a scale transformation by a factor of $s \in \CC^*$ we have $\lambda \to \lambda ' = s \lambda$. This scaling relation may be derived from the following scale transformations on $x, u_n, v\sim u_0$,
\bea
x \to x' = s^{ \tfrac{2}{N+2}} x
\hskip 0.6in 
\hat y \to \hat y' = s^{\frac{N}{N+2}} \hat y
\hskip 0.6in
u_n \to u_n' = s^{2 \frac{N-n} {N+2}} u_n
\eea 
for $n=0, 1, \cdots , N-2$. The scaling dimension of the operator $\cO_n$ whose expectation value is $u_n$ has the same scaling dimension as $u_n$ and therefore is given by,
\bea
\label{4.dim}
\Delta (\cO_n) = 2 { N-n \over N+2}
\eea
In a unitary superconformal field theory the dimension of every physical operator must be larger than one in view of the unitarity bound imposed by the representation theory of the supersymmetry algebra, which requires $\Delta (\cO_n) >1$ and thus,
\bea
\label{4.rank}
n+1 \leq \left[\frac{N-1}{2}\right] = \mr
\eea
The parameter $\mr$ is referred to as the \textit{rank} of the $(\ma_1, \ma_{N-1})$ AD theory, and is defined to be the dimension over $\CC$ of the intrinsic Coulomb branch. By contrast, the parameters $u_n$ for $n \geq \mr$ do not correspond to intrinsic moduli of the AD theory.

\subsection{The intrinsic K\"ahler potential}

The K\"ahler potential of $\SU(N)$ SW theory is defined by,
 \begin{align}
    K_{\SU(N)} = \frac{i}{4\pi} \sum_{I=1}^{N-1} \Big ( a_I \, \bar{a}_{D,I} - \bar{a}_{I} \, a_{D,I} \Big ) 
    \end{align}
We may conveniently re-express $K_{\SU(N)}$ in  terms of the long and short periods with the help of an $\Sp(2N-2,\ZZ)$ change of duality frame, which results in the following expression,  
\begin{align}
    K_{\SU(N)} = 
\frac{i}{4\pi} \sum_{i=1}^{\mr}\left(a_i^{(s)}\bar{a}_{D,i}^{(s)} - \bar{a}_{i}^{(s)}a_{D,i}^{(s)}\right) 
+ \frac{i}{4\pi} \sum_{i=1}^{\left [\tfrac{N}{2} \right ]}\left(a_i^{(\ell)}\bar{a}_{D,i}^{(\ell)} - \bar{a}_{i}^{(\ell)}a_{D,i}^{(\ell)}\right)
\end{align}
where the rank $\mr$ was defined in (\ref{4.rank}).
We consider the decoupling limit of the $\SU(N)$ super Yang-Mills theory near the AD point as $\Lambda \rightarrow \infty$ to obtain the intrinsic periods and the intrinsic K\"ahler potential. Doing so causes the term in the K\"ahler potential for the long periods to vanish, leaving the intrinsic K\"ahler potential $K_{\AD}$ of the AD theory expressed solely  in terms of  the short periods, 
\begin{align}
K_\text{AD}  &= \frac{i}{4\pi}\sum_{i = 1}^{\mr} (a_i^{(s)}  \Bar{a}_{D,i}^{(s)} - \Bar{a}_i ^{(s)} a_{D,i}^{(s)} )
\label{KAD}
\end{align}
 in the limit where $v/\Lambda^N \to 0$. 
 In the remainder of this section we shall analyze the intrinsic periods and the K\"ahler potential using a combination of analytical and numerical methods.

\subsection{Analytical results for the periods}

In this subsection, we obtain analytical results for the periods and the K\"ahler potential, in terms of an expansion in the parameters $v$ and  $v_n$ already encountered in (\ref{2.scale}),
\bea
v_n = v^{\frac{n}{N} -1} \, u_n \hskip 1in n=1,\cdots N-2
\eea
Note that these parameters include the moduli of the intrinsic Coulomb branch, namely $v$ and $v_n$ for $n=1, 2, \cdots, \mr-1$, but also the non-intrinsic parameters $v_n$ for $n \geq \mr$. We consider both for the sake of completeness, but also to contrast the difference of the behavior of the K\"ahler potential under both deformations. We first turn to evaluating the short periods.

\begin{cor}\label{simpR}
The short periods of the $(\ma_1, \ma_{N-1})$ theory may be expressed in terms of the  function $R(\zeta)$ that admits the following infinite series expansion in $v_1,\dots, v_{N-2}$,
    \begin{align}
R(\zeta) & = - { v^{\half + \frac{1}{N}} \over \sqrt{2 \pi}\, N} \sum _{{\ell_n=0 \atop n=1,\cdots, N-2}} ^\infty 
{(-)^M}\zeta^{L}  { {v}_1^{\ell_1} \cdots {v}_{N-2}^{\ell_{N-2} } \over \ell_1! \cdots \ell_{N-2} !} 
{\Gamma (\tfrac{L}{N})  \over  \Gamma ( \frac{3}{2}+\tfrac{L}{N} -M)  }
        \end{align}
        where $L$ and $M$ were defined in (\ref{2.LMa}). 
\end{cor}
This expression for $R(\zeta)$ follows from Corollary \ref{Rhyp}, upon restoring the $\Lambda$-dependence and taking the limit $\Lambda\rightarrow\infty$. The result represents a major simplification of the expressions obtained in (\ref{2.Rxi1}) and (\ref{2.Rxi2}) for the embedded theory.  The short periods may be expressed in terms of the decomposition of the function $R$ in terms of characters of $\ZZ_N$ as follows, 
\begin{align} 
\label{4.char}
  R(\zeta)&= \sum_{n=0}^{N-1}\zeta^n R_n 
  \hskip 1in
\zeta ^N=-1 
\end{align}
Throughout, it will be convenient to use the  abbreviations, 
\bea
s_n = \sin \left (\tfrac{2\pi n}{2N} \right ) \hskip 1in c_n=\cos \left (\tfrac{2\pi n}{2N} \right )
\eea

\begin{cor}\label{ADper}
    The short periods of $(\ma_1,\ma_{N-1})$ theories have the following expressions in terms of the characters $R_n$ for any $N\geq 3$ and $j=1,\dots,\mr$,
\bea
a_{D,j}^{(s)} = 2j\sum_{n=1}^{N-1}\ep^{4nj}s_nR_n 
\eea   
and, 
\begin{align}
        a_j^{(s)} &=\begin{dcases}
            \sum_{n=1}^{N-1} \frac{1}{2c_n} (\ep^{4nj}-1) R_n &\text{ odd } N
            \\
            -2ijR_\nu +\sum_{{n=1 \atop n \neq \nu}}^{N-1} \frac{1}{2c_n} (\ep^{4nj}-1) R_n&\text{ even } N=2 \nu
            \end{dcases}
\end{align}
\end{cor}
To derive $a_{D,j}^{(s)} $, we simply substitute the expansion (\ref{4.char}) in terms of characters into the expression for the dual period in (\ref{Rperiod}), and apply the definition of $s_n$. To derive $a_j^{(s)} $, we begin by substituting the expansion (\ref{4.char}) of $R(\zeta)$ into characters,
\begin{align}\label{aExp}
    a_j ^{(s)}&= \sum_{k=0}^{j-1}\sum_{n=1}^{N-1} \ep^{4nk} (\ep^{3n} - \ep^n)R_n
\end{align}
We then apply the following key identity
\begin{align}\label{keyId}
    \sum_{j=0}^{k-1} \ep^{4jn}= \begin{dcases}
        k&\text{ when }\ep^{4n} = 1\\
        \frac{1-\ep^{4nk}}{1-\ep^{4n}} &\text{ otherwise}
    \end{dcases} 
\end{align}
Since the range of $n$ is given by $0\leq n\leq N-1$, the instance $\ep^{4n}=1$ can occur  only if $n=0$ or $n=\frac{N}{2}$. The former does not give a contribution to the sum (\ref{aExp}) thanks to the factor $(\ep^{3n} - \ep^n)$ in the summand, while the latter can only contribute when $N$ is even. It is now clear that performing the sum over $k$ produces the different expressions for $a_j^{(s)}$ depending on whether $N$ is even or odd, thereby completing the proof of  Corollary \ref{ADper}.

\subsection{Analytical results for the K\"ahler potential}

We now turn to the analytical results for the intrinsic K\"ahler potential, and begin by evaluating  $K_\text{AD}$ of (\ref{KAD}) in terms of the character coefficients $R_n$, using the results of Corollaries \ref{simpR} and \ref{ADper}. The results are given by the following theorem. 
\begin{thm}
\label{thm:KAD}
    The K\"ahler potential $K_\text{AD}$ of $(\ma_1, \ma_{N-1})$ theories admits the following decomposition in terms of the characters $R_n$
\begin{align}
\label{4.KADform}
K_\text{AD}  = 
\begin{dcases}
\frac{N}{4\pi} \sum_{n=1}^{N-1}\frac{s_n}{c_n}\abs{R_n}^2 &\text{ odd } N
\\
\frac{N}{4\pi} \sum_{{n=1 \atop n\not= \nu}}^{N-1}\frac{s_n}{c_n}\abs{R_n}^2 
+ \frac{N}{8\pi}\left( R_\nu\sum_{{ n=1 \atop n \not= \nu }}^{N-1}\frac{\ep^{2n}-1}{c_n} \bar{R}_n + {\rm c.c.}\right) &\text{ even } N=2 \nu 
\end{dcases}
    \end{align}
\end{thm}

$\bullet$ For odd $N$, the proof proceeds by substituting the expressions for the periods obtained in Corollary \ref{ADper} into the definition of the intrinsic K\"ahler potential in (\ref{KAD}), and we obtain,
\begin{align}
K_\text{AD} = \frac{1}{4\pi} \sum_{j=1}^\mr\sum_{m,n=1}^{N-1}\frac{s_n}{c_m}R_m\bar{R}_n (\ep^{4(m-n)j}-\ep^{-4nj}) + {\rm c.c.} 
\end{align}
Applying the summation identity (\ref{keyId}) shows that the first term receives contributions only from $m=n$ since $m-n\neq N/2$ for any $1 \leq m,n \leq N-1$ when $N$ is odd. This is a key simplification for odd $N$. After several further elementary manipulations, we find,
\begin{align}
    K_\text{AD} = \frac{2r+1}{4\pi}\sum_{n=1}^{N-1}\frac{s_n}{c_n}\abs{R_n}^2 
    - \frac{i}{8\pi} \sum_{{m,n=1\atop m\neq n}}^{N-1} \frac{s_ms_n}{c_mc_nc_{m-n}} \Big ( R_m \Bar{R}_n - \Bar R_m {R}_n \Big ) 
\end{align}
The second term on the right side cancels, since the prefactor in the summand is symmetric under $m\leftrightarrow n$ while the combination in the parentheses is anti-symmetric. 

\sm

$\bullet$ For even $N=2 \nu$, with $\nu \in \NN$ and $\mr = \nu-1$, the proof proceeds by separating the term $n=\nu$ in $a_{D,i}^{(s)}$ from the other terms, and splitting the expression (\ref{KAD}) accordingly into the following four contributions, $K_{AB} = K_1 + K_2+ K_3+ K_4$,
\begin{align}
    K_1 & = -\frac{i}{\pi} R_\nu \sum_{n=1}^{N-1} s_n\bar{R}_n \sum_{j=1}^{\nu-1} j \ep^{-4nj}  + {\rm c.c.}
    \no \\
     K_2 &=\frac{1}{4\pi} \bar{R}_\nu \sum_{\nu\neq m=1}^{N-1}\frac{s_\nu}{c_m} R_m\sum_{j=1}^{\nu-1} (\ep^{4mj}-1) + {\rm c.c.} 
    \no \\
    K_3 &= -\frac{1}{4\pi} \sum_{\nu\neq m,n=1}^{N-1} \frac{s_n}{c_m}R_m\Bar{R}_n \sum_{j=1}^{\nu-1} \ep^{-4nj} + {\rm c.c.}
         \no \\
    K_4 &= \frac{1}{4\pi} \sum_{\nu\neq m,n=1}^{N-1}\frac{s_n}{c_m} R_m\Bar{R}_n\sum_{j=1}^{\nu -1} \ep^{4(m-n)j} + {\rm c.c.} 
    \end{align}
To evaluate the sum over $j$ in $K_1$, we use the following identity in $x = \ep^{-4n}$ with $x^\nu=1$,
\begin{align}
    \sum_{j=0}^{\nu-1} jx^j= - \frac{\nu x^\nu}{1-x} + \frac{x(1-x^\nu)}{(1-x)^2}  = -\frac{\nu}{1-x}
\end{align}
The contribution from the $n = \nu$ term is purely real, and hence cancels. The sums over $j$ in $K_2$ and $K_3$ are evaluated using the identity (\ref{keyId}) to give,
\begin{align}
\label{K123}
    K_1 & = \frac{\nu}{4\pi} R_\nu \sum_{\nu \neq n =1}^{N-1}\Bar{R}_n\frac{\ep^{2n}}{c_n} + {\rm c.c.}
\no \\
    K_2 &= -\frac{\nu}{4\pi} \bar{R}_\nu \sum_{\nu\neq m = 1}^{N-1} \frac{s_\nu}{c_m}R_m + {\rm c.c.}
\no \\
    K_3 &= \frac{1}{4\pi} \sum_{\nu\neq m,n =1}^{N-1} \frac{s_n}{c_m}R_m\bar{R}_n + {\rm c.c.}
\end{align}
The evaluation of $K_4$ is a bit more subtle because, 
\begin{align}
    \sum_{j=1}^{N-1} \ep^{4(m-n)j} =\begin{dcases}
    \nu-1&\text{ if } m-n\equiv 0\pmod \nu\\
    -1 &\text{ otherwise}
    \end{dcases}
\end{align}
Since the condition $m-n\equiv 0\pmod\nu$ can be satisfied only if $m-n \in\{0,\pm\nu\}$ we have,
\bea
K_4 & = & 
-\frac{1}{4\pi} \sum_{{m,n=1 \atop m-n\neq 0, \pm \nu}}^{N-1} \frac{s_n}{c_m}\left(R_m\Bar{R}_n + {\rm c.c.}\right)
+ \frac{2(\nu-1)}{4\pi} \sum_{\nu\neq m=1}^{N-1} \frac{s_m}{c_m} \abs{R_m}^2 
\no \\ &&
+ \frac{\nu-1}{4\pi} \sum_{\nu\neq m,n=1}^{N-1} \frac{s_n}{c_m}R_m\bar{R}_n\left[\delta_{m-n,\nu}+ \delta_{m-n,-\nu}\right] + {\rm c.c.}
\eea
Using the fact that $c_{n+\nu}s_{n+\nu} = -c_n s_n$ for any $n$, we see that the sum on the second line above cancels. Rearranging the  contributions from $m-n = 0,\pm \nu$ in the sum of $K_1, K_2, K_3, K_4$  gives the second line in (\ref{4.KADform}) and completes the proof of the case when $N$ is even.

\sm

Next, we prove analytical results on the positivity and convexity of the $(\ma_1,\ma_{N-1})$ K\"ahler potential using the results of Theorem \ref{thm:KAD} and Corollary \ref{simpR}, assembled in the theorem below.

\begin{thm}
    The K\"ahler potential $K_\text{AD}$ of the $(\ma_1,\ma_{N-1})$ theories is bounded from below by zero provided we only turn on moduli corresponding to operators with unitary scaling dimensions, i.e. $\Delta(\cO_k)>1$. Furthermore, in the absence of deformations with non-unitary scaling dimensions, the expansion of the K\"ahler potential in the rescaled moduli $v_k$ begins at quadratic order with positive coefficients for $k\geq 1$ when $N$ is not divisible by 4. When $N$ is divisible by 4, an additional linear term arises. \label{Kpc}
\end{thm}
To prove the theorem, we shall need the values of $R_n$ to leading orders in $v_1, \cdots, v_{N-2}$. This information  may be read off from Corollary \ref{simpR},
\bea
\label{4.R1Rn}
    R_n & = &  \frac{\Gamma(\tfrac{n}{N})\, v^{\thalf+\tfrac{1}{N}} }{\sqrt{2\pi}N \Gamma(\thalf+\tfrac{n}{N})} v_{n-1} + \cO(v_i^2)
    \hskip 0.8in n =2,\cdots, N-1
    \no \\
    R_1 & = &  - \frac{v^{\thalf+\tfrac{1}{N}} }{\sqrt{2\pi}N } \frac{\Gamma(\tfrac{1}{N})}{\Gamma(\tfrac{3}{2} + \tfrac{1}{N}) } 
    \left [ 1    + (\tfrac{1}{N^2}+\tfrac{1}{2N})\sum_{n=1}^{N-2}  v_n v_{N-n}
    \right ] + \cO(v_i^3)
\eea
where $\cO(v_i^2)$ and $\cO(v_i^3)$ stand for any bilinear or trilinear terms in $v_1, \cdots, v_{N-2}$. To confirm the absence of linear terms in $R_1$, we use the fact that its contributions arise from combinations for which $L\equiv 1 \pmod N$. A term linear in $v_n$ has $\ell_n=1$ and all other $\ell=0$, so that $L=1+n$. Since $n\leq N-2$, there are no solutions to the equation $L \equiv 1 ~ ({\rm mod} ~N)$, and hence no linear terms.

\sm

To investigate the positivity and local convexity  of $K_\text{AD}$, we consider first the case of odd~$N$, for which the K\"ahler potential is given by Theorem \ref{thm:KAD},
\bea
K_\text{AD} = { N \over 4 \pi} \sum _{n=1}^{N-1} { s_n \over c_n} |R_n|^2
\eea
Since we manifestly have the following inequalities,
\bea
{ s_n \over c_n} >0 \hbox{ for } n=1,\cdots, \tfrac{N-1}{2}
\hskip 0.8in
{ s_n \over c_n} <0 \hbox{ for } n= \tfrac{N+1}{2}, \cdots, N-1
\eea
it is clear that the contributions from $R_n$ and thus $v_{n-1}$ for $n= \tfrac{N+1}{2}, \cdots, N-1$ are negative and not convex. Setting the corresponding parameters $v_n=0$, we retain only those $v_n$ for which $n=0,\cdots, \tfrac{N-3}{2}$. In this case, the bilinear terms in $R_1$ automatically vanish, and $R_1$ has contributions in $v_1, \cdots, v_{N-2}$ to order zero, but no linear or bilinear contributions. Therefore, the K\"ahler potential, locally  near the AD point, is positive and convex. By inspection of  (\ref{4.dim}) and (\ref{4.rank}), we find, remarkably, that these values precisely correspond to operators whose dimension is larger than 1 and thus obey the unitarity bound, while the other values of $n$ correspond to operators whose dimension is below the unitarity bound, thereby proving the first part of the theorem. By contrast, turning on the deformations $v_{n-1}$ for $n= \tfrac{N+1}{2}, \cdots, N-1$  renders the K\"ahler potential non-positive and non-convex. 

\sm

The situation is more subtle for even $N=2 \nu$. The argument for the positivity of the diagonal part of the $K_\text{AD}$ is identical to the case of odd $N$. The rest of the proof has two steps. First, we show that the non-diagonal part of $K_\text{AD}$ for even $N$ gives a vanishing contribution when all non-unitary deformations are turned off. Second, we prove the absence of linear terms provided $N$ is not divisible by 4. To prove the first claim, note that all contributions to $R_\nu$ come from the $L\equiv \nu\pmod N$ sector. Parameterizing $L = \nu + Nk$ for $k\in\integers_{\geq 0}$, the coefficients in the expansion of $R$ are proportional to,
\begin{align}
    \frac{\Gamma(\tfrac{L}{N})}{\Gamma(\tfrac{3}{2}+ \tfrac{L}{N} - M)} = \frac{\Gamma(k+\thalf)}{\Gamma(2+k-M)}
\end{align}
Furthermore, we observe that $k-M<0$. Hence, the only non-vanishing contribution can come from $k-M=-1$. This is equivalent, for a single fixed $j$, to $(2\nu-j)\ell_j = \nu+1$. Such a term contributes linearly only if $2\nu-j\leq \nu+1$ or $j\geq \nu-1 = r$. All such terms are killed if we restrict to deformations with $\Delta>1$. To prove the second claim, we examine the contributions to $R_1$ from $L\equiv 1\pmod \nu$ when $\nu$ is odd. This gives rise to a linear term only if $j=\nu$, which is excluded by the constraint $\Delta > 1$. This concludes the proof of the theorem. We remark that the K\"ahler potential is positive even when $N$ is divisible by 4, provided we only turn on unitary deformation. Indeed, the non-diagonal part of the K\"ahler potential then vanishes, and we are left with a sum of absolute-squares with positive coefficients.

\sm

{\bf Remarks:} To study the global structure of the K\"ahler potential, we use numerics, which agree perfectly with our analytical results. At rank-1, convexity of $K_\text{AD}$ over intrinsic slices follows directly from the scaling of the periods: $a(v)\sim v^\frac{1}{\Delta(v)}$, which is required by scale-invariance (cf. \cite{Martone:2020hvy}). We numerically analyze the effect of turning on deformations $u_k$ with $\Delta(\cO_k)\leq 1$. Generically, such deformations introduce points where the second-derivative test on the K\"ahler potential fails to yield a unique minimum, and the determinant of the Hessian is vanishing. In the rank-$2$ case, we primarily consider the intrinsic Coulomb branch, which is now $2$-complex-dimensional. On the intrinsic Coulomb branch, the K\"ahler potential is a positive and convex function provided there are no deformations with $\Delta\leq 1$.

\subsection{Rank-1, example 1: $(\ma_1, \ma_2)$}

This is a rank-$1$ theory with $N = 3$, and has an elliptic SW curve,
\begin{align}
    \hat{y}^2 = x^3 - ux + v. 
\end{align}
The function $R(\zeta)$ for this theory is given as follows ($\zeta^3 = -1$),
\begin{align}
    R(\zeta) &= \frac{ v^\frac{5}{6}}{3\sqrt{2\pi}} \sum_{\ell = 0}^\infty \frac{(-\zeta) ^{\ell+1} \Gamma(\frac{\ell+1}{3}) \, }{\Gamma(\frac{11}{6} - \frac{2\ell}{3})\, \ell!} \, w^\ell
\hskip 1in 
w =  \frac{u}{v^\frac{2}{3}}
\end{align}
The sum may be reorganized  into hypergeometric functions of various degrees,    
 \begin{align}
 R(\zeta)   &=
 \frac{ \, v^\frac{5}{6}}{3\sqrt{2\pi}}\Bigg[
- \zeta  \frac{ \Gamma \left(\tfrac{1}{3}\right)}{ \Gamma \left(\tfrac{11}{6}\right)} 
    \,   {}_2F_1\left(-\tfrac{5}{12},\tfrac{1}{12};\tfrac{2}{3};\tfrac{4 w^3}{27}\right)
+ \zeta^2 \frac{ \Gamma \left(\frac{2}{3}\right)}{\Gamma \left(\frac{7}{6}\right)} 
   \,   {}_2F_1\left(-\tfrac{1}{12},\tfrac{5}{12};\tfrac{4}{3};\tfrac{4 w^3}{27}\right) w
   \no \\
   &\hskip 1in 
+  \frac{1}{2 {\Gamma ( \thalf)}} \, {} _3F_2 \left(\tfrac{1}{4},\tfrac{3}{4},1;\tfrac{4}{3},\tfrac{5}{3};\tfrac{4 w^3}{27}\right) w^2
   \Bigg]
\end{align}
This presentation explicitly produces exact expressions for the characters $R_0$, $R_1$ and $R_2$ from the last, first, and middle terms, respectively, and allows us to compute the periods and the K\"ahler potential. Before proceeding to the necessary numerical analysis, it is instructive to  evaluate the K\"ahler potential at low-orders in $u$, and we find,  
\begin{align}
\label{small-u}
K_\text{AD}  = \frac{27\sqrt{3}}{50 \pi^4}  \left(\Gamma(\tfrac{1}{3})^2\Gamma(\tfrac{7}{6})^2 \abs{v}^\frac{5}{3} - \Gamma(\tfrac{2}{3})^2\Gamma(\tfrac{11}{6})^2 \abs{v}^\frac{1}{3} \, \abs{u}^2 \right)+ \cO(u^3)
        \end{align}
While the K\"ahler potential for $u=0$ is convex and positive with a unique vanishing point at $v=0$ by Theorem  \ref{Kpc}, convexity is immediately lost as soon as we turn on $u$. Beyond the analytical result for the small $u$ approximation, numerics are required to explore the K\"ahler potential away from the AD points, as shown in Figure 
 \ref{K (A1, A2)}.

\begin{figure}
    \centering
    \begin{subfigure}[t]{0.3\textwidth}
        \centering
        \includegraphics[width=\linewidth]{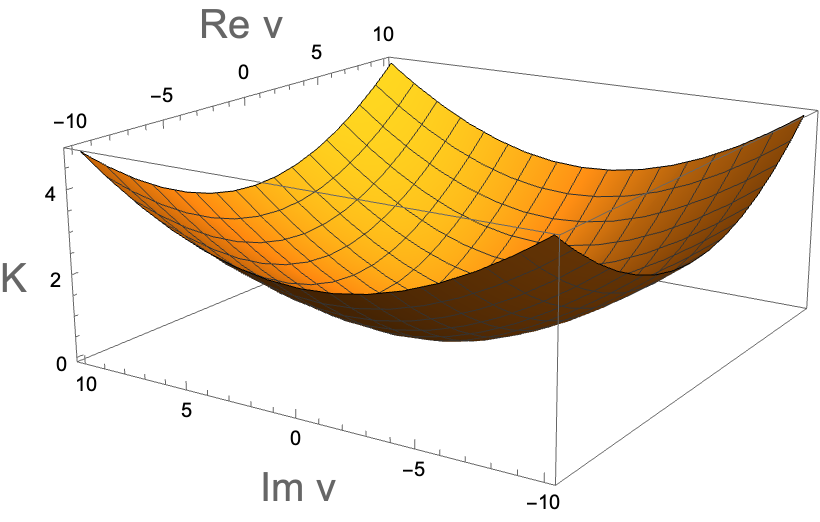} 
        \caption{$(\Re v, \Im v)$} 
    \end{subfigure}
    \hfill
    \begin{subfigure}[t]{0.3\textwidth}
        \centering
        \includegraphics[width=\linewidth]{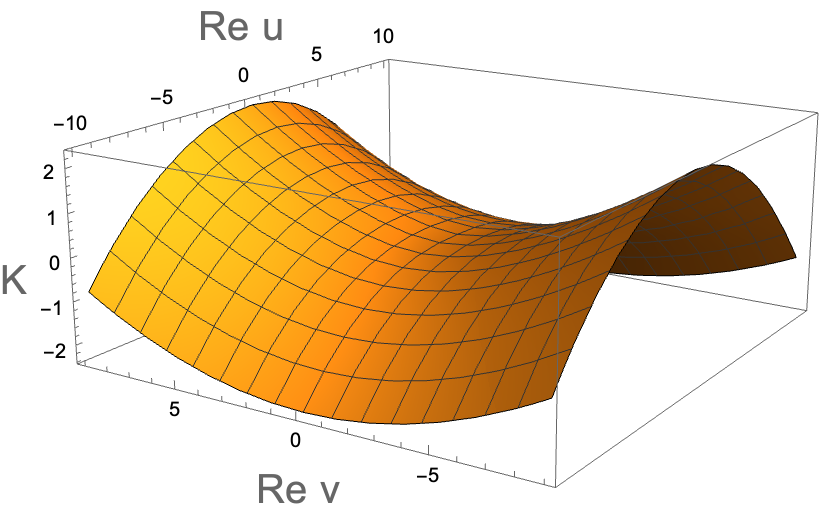} 
        \caption{$(\Re u, \Re v)$}
    \end{subfigure}
    \hfill
    \begin{subfigure}[t]{0.3\textwidth}
        \centering
        \includegraphics[width=\linewidth]{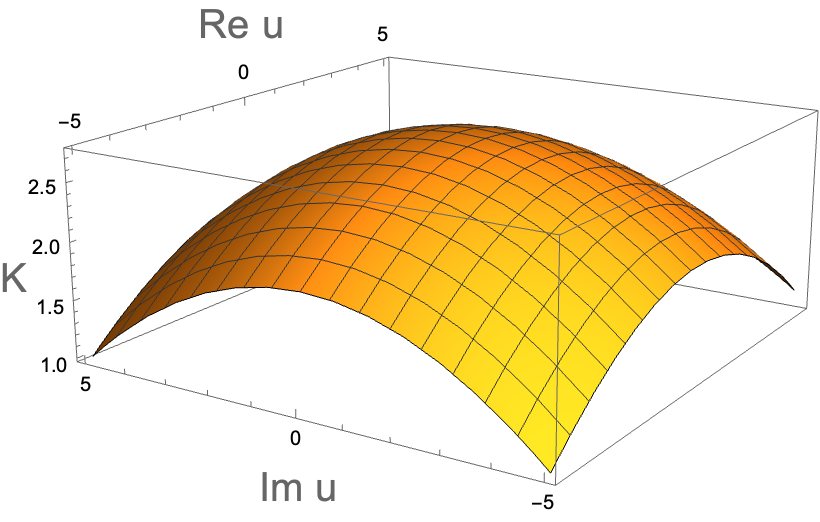} 
        \caption{$(\Re u, \Im u)$} 
    \end{subfigure}
    \caption{A plot of the $(\ma_1,\ma_2)$ K\"ahler potential for various slices of parameter space $u_1,v$.\label{K (A1, A2)}}
\end{figure}

\subsection{Rank-1, example 2: $(\ma_1,\ma_3)$}

In this subsection, we consider the AD theory that lives in the moduli-space of pure $\SU(4)$ SW theory. This theory is defined by a quartic SW curve,
\begin{align}
    \hat{y}^2 = x^4 - u_2 x^2 - u_1 x + v
\end{align}
It is clear from Theorem \ref{Kpc} that the K\"ahler potential is convex and positive-definite in the absence of any non-unitary Coulomb branch deformations with $\Delta \leq 1$. Here, we gather further evidence that turning on such deformations spoils both positivity and convexity. 

\sm

For $v>0$ the expansion in small $u_1 \in \RR$ is given as follows, 
    \begin{align}
        K_\text{AD} = \frac{\Gamma \left(\frac{1}{4}\right)^2v^\frac{3}{2}}{32 \pi ^2 \Gamma\left(\frac{7}{4}\right)^2}+\frac{u_1 \Gamma \left(\frac{1}{4}\right)v^\frac{3}{4}}{16\sqrt{2} \pi ^{3/2} \Gamma \left(\frac{7}{4}\right)}-\frac{u_1^3 \Gamma\left(\frac{3}{4}\right)}{32 \sqrt{2} \pi ^{3/2} \Gamma \left(\frac{1}{4}\right)v^\frac{3}{4}}+\cO(u_1^4)
    \end{align}
The expression shows that turning on $u_1$ spoils convexity; we find an analogous  result for $u_2\neq 0$. These results are qualitatively analogous to the results we obtained for $\SU(3)$, and we shall refrain from presenting numerical plots for this case.

\subsection{A rank-2 example: $(\ma_1, \ma_4)$}

Numerical analysis confirms, here as well, that turning on any non-intrinsic moduli, such as $u_2$ and $ u_3$, spoils both convexity and positivity.  We shall now concentrate on the numerical analysis of the dependence of the K\"ahler potential on the intrinsic moduli $u=u_1, v$, with no other deformations turned on. The SW curve is given by,  
\begin{align}
    \hat{y}^2 = x^5 - u x + v
\end{align}
and the Taylor expansion of $R(\zeta)$  in $u$ for $v \not=0$, with $\zeta ^5=-1$, is given by, 
\begin{align}
    R(\zeta) = \frac{ v^\frac{7}{10}}{5\sqrt{2\pi}} \sum_{\ell = 0}^\infty \frac{ (- \zeta)^{\ell+1} \Gamma(\frac{1+\ell}{5})}{\Gamma(\frac{17}{10} - \frac{4\ell }{5} )\ell!} \left(\frac{u}{v^\frac{4}{5}}\right)^{\ell}.
\end{align}
This series may be summed in terms of the hypergeometric functions $_5 F_4$ and $_4F_3$ with argument proportional to $u^5/v^4$. Such a closed form for the characters $R_n$ is useful  extract the small-$v$ behavior of $K_\text{AD}$ by expanding the hypergeometric functions around $u_1=\infty$. We shall not produce these lengthy explicit formulas here.  Instead we concentrate on the numerical results when only the intrinsic moduli $u=u_1$ and $v$ are turned on. 

\begin{figure}
    \centering
    \begin{subfigure}[t]{0.3\textwidth}
        \centering
        \includegraphics[width=\linewidth]{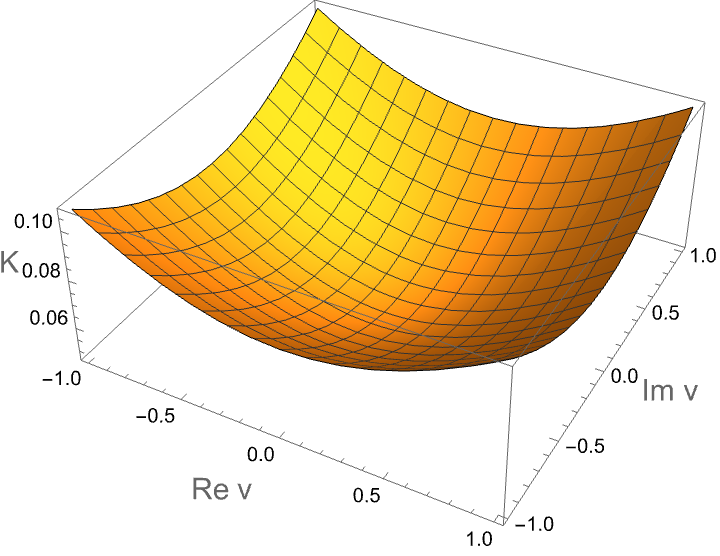} 
        \caption{$v$-plane}
    \end{subfigure}
     \hfill
    \begin{subfigure}[t]{0.3\textwidth}
        \centering
        \includegraphics[width=\linewidth]{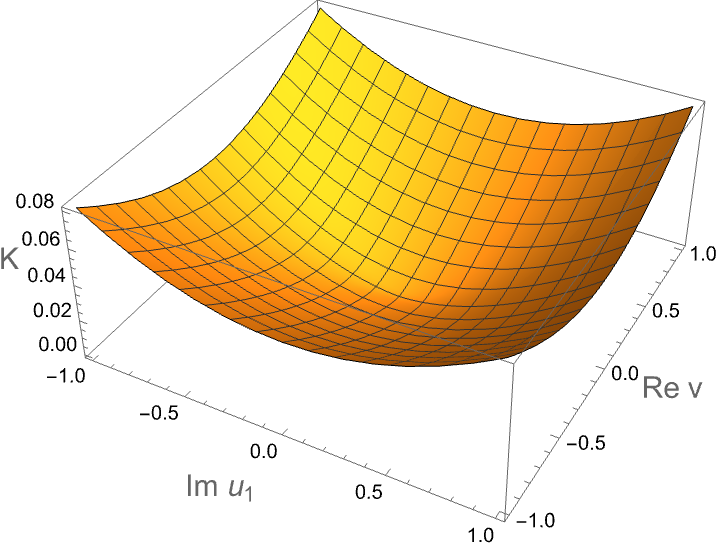} 
        \caption{$(\Im u, \Re v)$-plane}
    \end{subfigure}
     \hfill
    \begin{subfigure}[t]{0.3\textwidth}
        \centering
        \includegraphics[width=\linewidth]{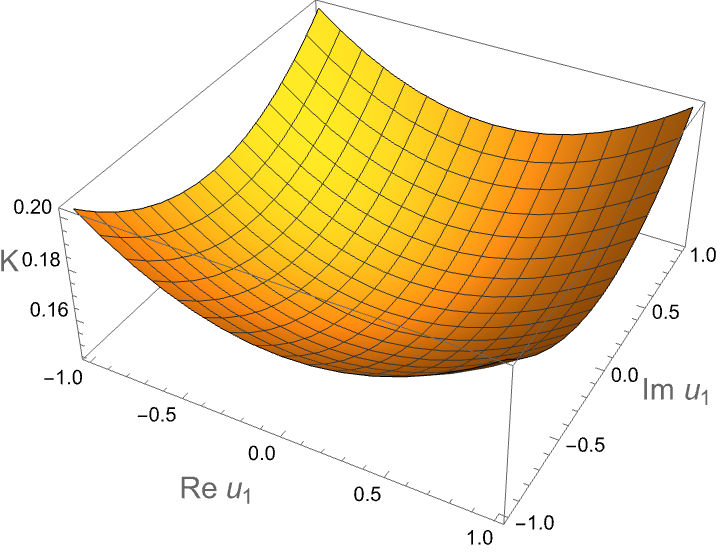} 
        \caption{$u$-plane}
    \end{subfigure}
    \caption{Plots of $K_\text{AD}$ for $(\ma_1, \ma_4)$ on various slices of the intrinsic moduli space. \label{K (A1, A4)2}}
\end{figure}

In figure \ref{K (A1, A4)2} the K\"ahler potential is plotted in various slices of the intrinsic Coulomb branch moduli $(u,v)$: versus $(\Re(v), \Im(v))$ for fixed $u$ in panel (a); versus  $(\Re(v), \Im(u))$ for fixed $\Re(u), \Im(v)$ in panel (b); and $(\Re(u), \Im(u))$ for fixed $v$ in panel (c). One observes in each case that, for the domain plotted, the K\"ahler potential is manifestly convex. More detailed numerical analysis, not manifestly visible form the plots, establishes that $K_{AB}$ is also positive definite for all values of $u,v$ studied, and vanishes only for $u=v=0$. 

\sm

In figure \ref{K (A1, A4)1} we provide a more detailed numerical analysis of the precise positivity and convexity properties, by taking different representative slicing of the intrinsic moduli space.  In panel (a) of figure \ref{K (A1, A4)1} we set $u=2$, and plot $K_\text{AD}$ as a function of $v= e \, e^{i \theta}$ as a function of $r \in [-1,2]$ for a number of discrete values of $\theta$. Convexity and positivity  is observed for every such slice. In  panel (b) of figure \ref{K (A1, A4)1} we plot $K_\text{AD}$ as a function of real $u \in [-2,2]$ for 10 evenly spaced discrete values of $v \in [1,2]$. Again, we observe positivity and convexity on each slice. Finally, in  panel (c) of figure \ref{K (A1, A4)1} we plot $K_\text{AD}$ as a function of real $v \in [-2,2]$ for 10 evenly spaced real values of $u \in [1,2]$, further confirming positivity and convexity. 

\begin{figure}
    \centering
    \begin{subfigure}[t]{0.3\textwidth}
        \centering
        \includegraphics[width=\linewidth]{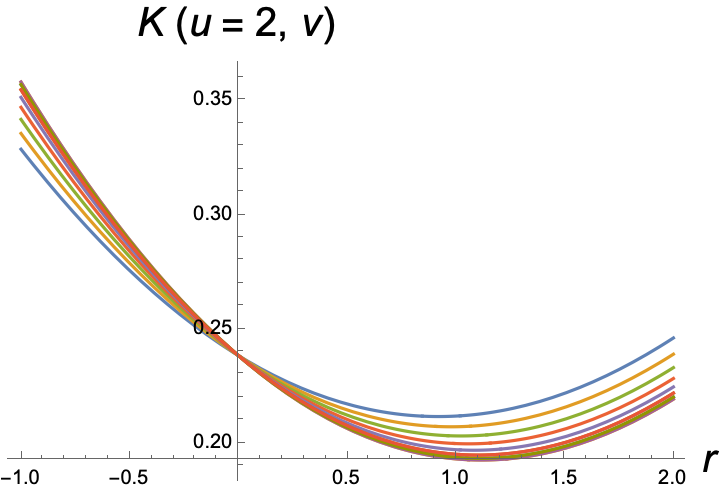} 
        \caption{$v$-plane with $v=re^{i\theta}$}
    \end{subfigure}
     \hfill
    \begin{subfigure}[t]{0.3\textwidth}
        \centering
        \includegraphics[width=\linewidth]{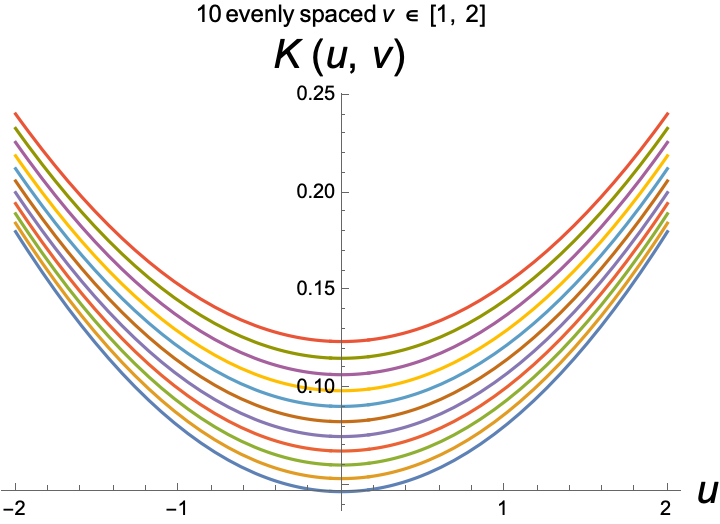} 
        \caption{$u$-plane for discrete $v$}
    \end{subfigure}
     \hfill
    \begin{subfigure}[t]{0.3\textwidth}
        \centering
        \includegraphics[width=\linewidth]{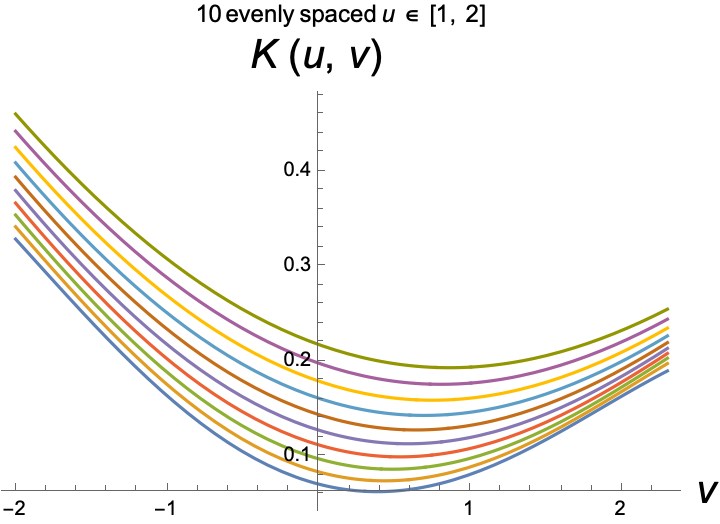} 
        \caption{$v$-plane for discrete $u$}
    \end{subfigure}
    \caption{Plots of $K_\text{AD}$ for $(\ma_1, \ma_4)$ on discrete slices of the intrinsic moduli space. \label{K (A1, A4)1}}
\end{figure}

Additional observations include the following.  One verifies  numerically, for example in panel (b) of figure \ref{K (A1, A4)1},  that the minimum of the K\"ahler potential over the $u$-plane always occurs at the origin $u=0$ for any $v \not=0$. From panel (c) of figure \ref{K (A1, A4)1}, we also see that for various real (as well as) complex values of $u$, the K\"ahler potential has a local minimum that is shifted from $v=0$ for any $u\neq 0$, though still respecting positivity and convexity. Finally, the K\"ahler potential is always positive in all these cases, and the minimum of $K$ is strictly bigger than $0$ if $u\neq 0$.
    
\subsection{A rank-3 example: $(\ma_1, \ma_6)$}

We consider the dependence of the intrinsic K\"ahler potential on the three intrinsic moduli $u_2, u_1, v$, where the SW curve is given by, 
\begin{align}
    \hat{y}^2 = x^7 - u_2x^2-u_1x+v 
\end{align}
The evaluation of the periods and K\"ahler potential proceeds very similarly to what we have already described up to rank 2, so we will not go into that here but rather only present the results. We study the regime of fixed $v$, and small $u_{1,2}$ using our expansion. This allows us to produce plots in figure \ref{K(A1, A6)} that further support our conjecture that the intrinsic K\"ahler potential is a convex and positive function with a unique minimum at $K_\text{AD}=0$ that is located at the $\integers_7$-symmetric point.

\begin{figure}[htb]
    \centering
    \begin{subfigure}[t]{0.3\textwidth}
        \centering
        \includegraphics[width=\linewidth]{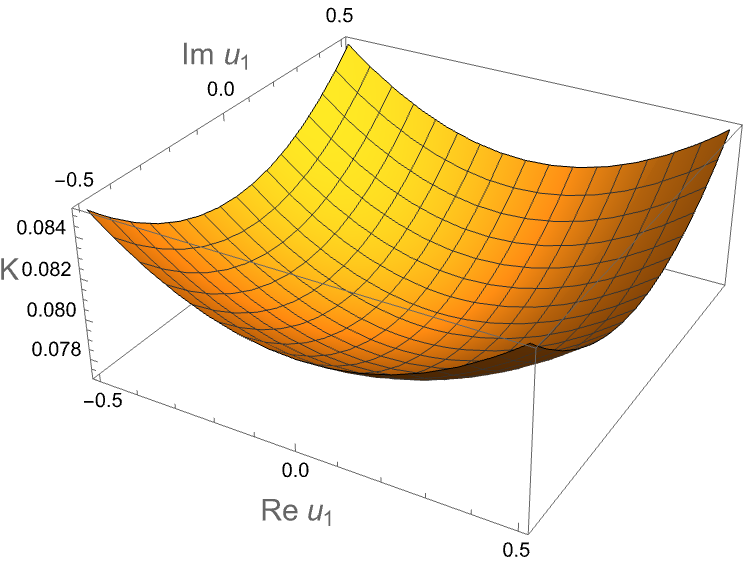} 
        \caption{$u_1$-plane}
    \end{subfigure}
     \hfill
    \begin{subfigure}[t]{0.3\textwidth}
        \centering
        \includegraphics[width=\linewidth]{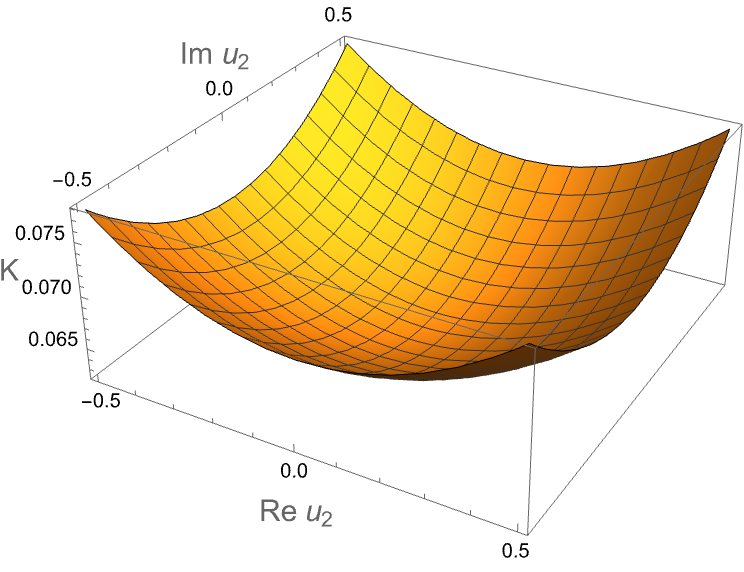} 
        \caption{$u_2$-plane}
    \end{subfigure}
     \hfill
    \begin{subfigure}[t]{0.3\textwidth}
        \centering
        \includegraphics[width=\linewidth]{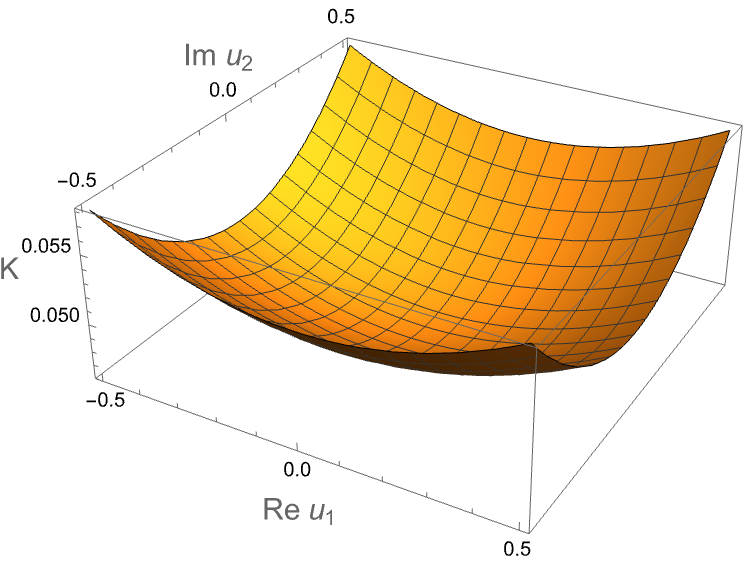} 
        \caption{$(\Re u_1, \Im u_2)$-plane}
    \end{subfigure}
    \caption{Plots of $K_\text{AD}$ for $(\ma_1, \ma_6)$ on various slices of the intrinsic moduli space for fixed value of $v$ and different splicings of the intrinsic moduli $u_1, u_2$. \label{K(A1, A6)}}
\end{figure}

\newpage
\section{Conclusions and future directions}
\label{sec5}
\setcounter{equation}{0}

In this paper, we have analyzed three different aspects of Argyres-Douglas (AD) theories, and their embeddings into the moduli-spaces of asymptotically-free gauge theories: the SW periods near the AD points; the marginal stability of mutually local BPS states in a near the AD points; and the intrinsic periods and K\"ahler potential of $(\ma_1,\ma_{N-1})$ theories.   

\subsection{Summary of results}

\begin{enumerate}
    \item 
    For gauge-group $\SU(N)$, we evaluated SW periods near the maximal AD points for $N\geq 3$ by a non-trivial analytic continuation of the expansion of the periods obtained near the $\integers_{2N}$-symmetric point studied in \cite{DDN}. Since our expansion is around one or the other maximal AD point, it allows us to access a neighborhood of the maximal AD points that includes the intrinsic Coulomb branch of the AD theory. On regions of overlap, we showed that our expansion has better convergence properties than the expansion considered in \cite{DDN}.
    \item 
    For gauge-group $\SU(3)$, we revisited the structure of the walls for marginal stability, which were analyzed in \cite{DDN} only on the restricted slice with $u=0$. Utilizing a combination of our expansion around the $\integers_3$ point and the numerical integration methods of \cite{DDN}, we mapped out the walls of marginal stability in the 3D space $(\Re u,\Im u, \text{Arc})$, where $\text{Arc}$ refers to a real-dimension 1 curve in the $v$-plane. This adds a more complete understanding of the walls, and explains the degeneracy-lifting associated with the $\integers_{3}$-symmetry of the $v$-plane. We provide partial results for $\SU(4)$ on the $u_2=u_1=0$ slice, and point out some generic features for $N\geq3$.
    \item In the last part of this paper, we explore the intrinsic Coulomb branch of $(\ma_1, \ma_{N-1})$ theories. We applied the decoupling limit $\Lambda\rightarrow\infty$ to obtain the intrinsic periods and their expansion around the AD point. We then apply this understanding of the periods to exactly compute the intrinsic K\"ahler potential and prove its positivity and convexity near $v_n=0$ on intrinsic slices, except when $4\nmid N$. We numerically test the global positivity and convexity of the K\"ahler potential over intrinsic slices in a variety of examples up to and including rank-3, i.e. $N=3,4,5,6,7$. We find broad agreement with our analytic results, and  find that convexity and positivity are  spoiled if we allow non-unitary deformations to be turned on, namely $u_{k\geq \mr}\neq 0$. This is in full agreement with Theorem \ref{Kpc}.
\end{enumerate}

\newpage

\subsection{Future directions}

Let us close this section with some concrete directions for future work.
\begin{enumerate}
    \item \textbf{Dynamics of wall-crossing.} A key question that we have not addressed in our considerations is whether genuine bound BPS states are formed when we follow a trajectory in the Coulomb branch that crosses a kinematically allowed wall. It would be interesting to explicitly determine the spectrum of BPS states after such a wall is crossed, and determine under what conditions formation of bound BPS states is possible. Work along these lines has been recently undertaken in \cite{Alim:2023doi}, and it would be interesting to apply these methods to the walls we obtain here. It is worth emphasizing, however, that more complicated phenomena can take place upon wall-crossing that we have also not addressed here -- see \cite{Gaiotto:2009hg}. 
    
    \item \textbf{Integrated correlation functions of the $(\ma_1, \ma_{N-1})$ stress-tensor.} There is a large body of work on integrated correlators in $3d$ ABJM and $4d$ $\cN=4$ super-Yang-Mills \cite{Binder:2019jwn,Chester:2019jas,Chester:2020vyz}. The key motivation behind such work is to provide non-trivial non-perturbative checks of holographic duals in eleven and ten dimensions respectively. Such works often rely on supersymmetric localization, superconformal symmetry, and $S$-duality: all of which can be accessed for AD theories. Recently, holographic duals have been proposed for AD theories in M-theory \cite{Bah:2021mzw,Bah:2021hei}, and it would interesting to compute observables on the field theory side to test these duals. On a related note, there are investigations of the correlators of chiral ring operators in AD theories using supersymmetric localization and the intrinsic SW periods \cite{Bissi:2021rei}. 
    
    In parallel to the work on $\cN=4$ super-Yang-Mills, is there a combination of supersymmetric localization and our expansion of the SW periods that can be used to compute the stress-tensor correlators of AD theories? If yes, what does this correspond to in the proposed holographic dual?
    
    \item \textbf{SUSY breaking of Argyres-Douglas theories.} As was already emphasized explicitly in \cite{Cordova:2018acb, DHoker:2020qlp, DDN, DDGN}, the convexity of the K\"ahler potential plays a key role in our understanding of the IR phases of $\SU(N)$ adjoint $\text{QCD}_4$, which is obtained by deforming pure $\cN=2$ gauge-theory by $\cT_{UV}\sim M^2\tr\bar{\phi}\phi$ near the monopole point. On representation-theoretic grounds, we expect to find a non-supersymmetric interacting CFT if we deform AD theories in an analogous manner \cite{Xie:2019aft}. 
    
    What can be said about the phases and spectrum of this IR CFT? Can this non-supersymmetric CFT be analyzed by a Lagrangian description obtained by deforming the $\cN=1$ Lagrangian that flows to Argyres-Douglas theory (cf. \cite{Maruyoshi:2016aim,Maruyoshi:2016tqk}) in an intermediate step? If such a theory does not admit a Lagrangian realization, can one analyze it using SW theory? 
\end{enumerate}

\newpage
\appendix

\section{Proof of Theorem 2.1}
\label{sec:A}
\setcounter{equation}{0}

To prove Theorem \ref{thm:2.1} we use the SW differential given in (\ref{2.lambda}) and the definition of the integrals $R(\zeta)$ in (\ref{Rdef}).  Substituting the multinomial expansion for $V(z)^M$ in powers of $z$,
\begin{align}
V(z)^{M} &= \sum _{\ell_1, \cdots, \ell_{N-2}=0} ^{M} 
\binom{M}{\ell_1, \cdots, \ell_{N-2}} \, 
 v_1^{\ell_1} \cdots v_{N-2}^{\ell_{N-2} } \, z^{L-1}
 \end{align}
where we use the combinations $L$ and $M$ familiar from (\ref{2.LMa}), gives the following expansion of the SW differential, 
\bea
\label{A.exp}
\lambda & = & { v^{\half + \frac{1}{N}} \over \sqrt{-2}} \sum _{k=0}^ \infty { \Gamma (k+\half) \over \Gamma (\half) \, k!} \, 
\left ( {v \over 2} \right ) ^{k} 
 \sum _{{\ell_n=0 \atop n=1,\cdots, N-2}} ^\infty
 { \Gamma (\half -k+M) \over \Gamma (\half -k) }
{ v_1^{\ell_1} \cdots v_{N-2}^{\ell_{N-2} } \over \ell_1! \cdots \ell_{N-2} !} 
\no \\ && \hskip 0.5in \times 
 \left ( N z^{N} -  \sum_{n=1}^{N-2} n v_n z^n \right  ) { z^{L-1}    dz \over \big ( z^N  + 1 \big ) ^{\half -k+M} }
\eea
The integral over $z$ from 0 to $\zeta$ greatly simplifies for $\zeta ^N=-1$ and we obtain, 
\bea
\int _0 ^\zeta { z^{\gamma - 1} \, dz \over (z^N + 1)^p} =
{\xi^\gamma \over \gamma} 
F( p, \tfrac{\gamma}{N}; 1+\tfrac{\gamma}{N} ; 1)
={\xi^\gamma \over N} 
{\Gamma (\tfrac{\gamma}{N}) \Gamma (1-p)
\over \Gamma ( 1+\tfrac{\gamma}{N} - p)  }
\eea
The integrals of $\lambda$ given by the expansion in (\ref{A.exp})  correspond to the values $p=\thalf -k+M$, and $\gamma = L+N$ or $\gamma = L+n$. As  a result, we obtain, 
\bea
i \pi R(\xi) & = &
{ v^{\half + \frac{1}{N}} \over \sqrt{-2}} 
\sum _{k=0}^ \infty {(-)^M}{ \Gamma (k+\half)^2 \over \Gamma (\half) \, k!} \, 
\left ({v \over 2} \right ) ^{k} 
 \sum _{{\ell_n=0 \atop n=1,\cdots, N-2}} ^\infty
{ v_1^{\ell_1} \cdots v_{N-2}^{\ell_{N-2} } \over \ell_1! \cdots \ell_{N-2} !} 
\no \\ &&  \times 
\bigg (  
{\xi^{L+N}  \, \Gamma (1+\tfrac{L}{N}) 
\over \Gamma ( \frac{3}{2}+\tfrac{L}{N} +k-M)  }
  -  \sum_{n=1}^{N-2} {n \over N}  \, v_n  \, 
{\xi^{n+L}  \, \Gamma (\tfrac{n+L}{N}) 
\over \Gamma ( \half+\tfrac{n+L}{N} +k-M)  } \bigg ),
\eea
where we have made use of the reflection formula for $\Gamma$-function $\Gamma (z) \Gamma (1-z) \sin (\pi z) = \pi$. The term labelled by $n$ under the finite sum over $n$ in the second line corresponds to shifting $\ell_n \to \ell_n-1$ and adding these contributions simplifies the sum as follows, 
\bea
i \pi R(\xi) =
{ v^{\half + \frac{1}{N}} \over \sqrt{- 2 \pi} N} 
 \sum _{{\ell_n=0 \atop n=1,\cdots, N-2}} ^\infty \!\!\!\! (-)^M
 \xi^{L}  { v_1^{\ell_1} \cdots v_{N-2}^{\ell_{N-2} } \over \ell_1! \cdots \ell_{N-2} !} 
\sum _{k=0}^ \infty { \Gamma (k+\half)^2 \,  \Gamma (\tfrac{L}{N})   \over   \Gamma ( \frac{3}{2}+\tfrac{L}{N} +k-M) \, k!} \, 
\left ( {v \over 2} \right ) ^{k} 
\quad
\eea
Relabelling $k \to \ell_0$, $v  \to v_0$ and choosing the branch $\sqrt{-2\pi}=i\sqrt{2\pi}$ produces formula (\ref{2.thm1a}) and thereby completes the proof of Theorem \ref{thm:2.1}.

\newpage

\section{Convergence, long periods, and elliptic form for $\SU(3)$}
\label{ExplicitSU(3)}
\label{sec:B}
\setcounter{equation}{0}

We explicitly display the coefficients of the $\integers_3$ expansion in this appendix, and examine the convergence criterion for the $\SU(3)$ series. Then we demonstrate that an appropriate choice of homology basis makes the long periods analytic.

\subsection{The $\integers_6$ expansion}

Explicit formulas for the periods in the case~$N = 3$ were obtained in~\cite{Klemm:1995wp} using Picard-Fuchs equations. The authors expressed their results in terms of Appell $F_4$ functions \cite{Appell,BatemanI}, which can be defined by the following series expansion,
\bea
F_4(a,b,c_1,c_2; x,y) = \sum _{m,n=0}^\infty {\Gamma (m+n+a) \Gamma (m+n+b) \Gamma (c_1) \Gamma (c_2) 
\over \Gamma (a) \Gamma (b) \Gamma (m+c_1) \Gamma (n+c_2) {\, m! \, n!}} \, x^m y^n
\eea
The $\SU(3)$ periods can be expressed as follows \cite{DDN},
\begin{align}
a_1 & =  Q(\ep^1) - Q(\ep^0)& 
a_{D,1} & =  Q(\ep^2) -  Q(\ep^1)
\no \\
a_2 & =  Q(\ep^1) - Q(\ep^0) +  Q(\ep^3) -  Q(\ep^2)&
 a_{D,2} & = Q(\ep^4) -  Q(\ep^3)
\end{align}
The function~$Q(\xi)$ can be expanded in characters of $\ZZ_{6}$
\begin{align}
    Q(\xi) &= \sum_{n = 0}^5\xi^n Q_n &Q_3&=0
\end{align}
The formula for $Q(\xi)$ given in (\ref{Qexp}) may be recast in terms of Appell functions $F_4$ expressed as follows in terms of the variables $x=4u_1^3/27$ and $y=u_0^2$
\begin{align}
    Q_{1} & =  {  2 \pi \over 2^{{1 \over 3}} \, 3^{{3 \over 2}} \, \Gamma(\tfrac{2}{3})^3  } \, 
v \, F_4(\tfrac{1}{3}, \tfrac{1}{3}, \tfrac{2}{3}, \tfrac{3}{2}; x,y)&Q_{2} & =   { 2 \pi \over 2^{{1 \over 3}} \, 3^2 \, \Gamma(\tfrac{2}{3})^3  } \, 
u \, F_4(\tfrac{1}{6}, \tfrac{1}{6}, \tfrac{4}{3}, \tfrac{1}{2}; x,y)\no\\
Q_{4} & =  { 2^{{1 \over 3}} \, 3^{{3 \over 2}} \, \Gamma(\tfrac{2}{3})^3 \over 4 \pi^2 } \,
F_4(-\tfrac{1}{6}, -\tfrac{1}{6}, \tfrac{2}{3}, \tfrac{1}{2}; x,y)&Q_{5} & =  { 2^{{1 \over 3}} \, \Gamma(\tfrac{2}{3})^3 \over 4 \pi^2 } \, 
u v \, F_4(\tfrac{2}{3}, \tfrac{2}{3}, \tfrac{4}{3}, \tfrac{3}{2}; x,y)
\end{align}
Additionally, $Q_{3}=0$, while $Q_{0}$ cancels out of all periods. Note that the double infinite series for the Appell function is absolutely convergent for $\sqrt{|x|} + \sqrt{|y|} <1$ which gives the following region of absolute convergence in terms of $u$ and $v$,
\bea
\label{2.b1}
\tfrac{ 2}{\sqrt{27}} \,  |u|^{{3 \over 2}} + |v| <1\,.
\eea
Beyond this region, partial analytic continuation formulas are known for $F_4$,\footnote{~These are obtained by expressing $F_4$ as an infinite sum of  hypergeometric functions, such as
\bea
F_4(a,b,c_1,c_2; x,y) = \sum _{n=0}^\infty {\Gamma (n+a) \Gamma (n+b) \Gamma (c_2) 
\over \Gamma (a) \Gamma (b) \Gamma (n+c_2) {\, n!}} \, y^n\, F(n+a,n+b;c_1;x) \,,
\eea
and applying inversion formulas for the hypergeometric functions.
}
 \bea
 \label{analF4}
 F_4(a,b,c_1,c_2;x,y) &  & =
{ \Gamma (c_1) \Gamma (b-a) \over \Gamma (b) \Gamma (c_1-a)} \, 
(-x)^{-a} F_4 (a,a+1-c_1,a+1-b,c_2; \tfrac{1}{x},\tfrac{y}{x})
\no \\ && \ +
{ \Gamma (c_1) \Gamma (a-b) \over \Gamma (a) \Gamma (c_1-b)} \, 
(-x)^{-b} F_4 (b,b+1-c_1, b+1-a, c_2;\tfrac{1}{x},\tfrac{y}{x})
\qquad
\eea
which  gives the following region in terms of $u_1$ and $u_0$,
\bea
\label{2.b2}
1+|u_0|< \tfrac{ 2}{\sqrt{27}} \,  |u_1|^{{3 \over 2}} 
\eea
allowing us to explore the region of large $|u_1|$ and small $|u_0|$. 
Recent progress on the analytic continuation of $F_4$ may be found in \cite{Ananthanarayan:2020xut}.

\subsection{The $\integers_6$ decomposition from the $\integers_3$ expansion}

The above-mentioned decomposition in terms of the characters of $\integers_{6}$ reduces to
    \bea
    Q(\xi) &=   (Q_0+Q_3) \xi^0 + (Q_1+Q_4) \xi^1 + (Q_2+Q_5)\xi^2 \qquad \text{ for } \xi^3 = +1
    \no \\
    Q(\xi) &=   (Q_0-Q_3) \xi^0 + (Q_1-Q_4) \xi^1 + (Q_2-Q_5)\xi^2 \qquad \text{ for }\xi^3 = -1 
    \eea
    These coefficients may be extracted from the expression for $Q(\xi)$ in (\ref{2.Q+})  by parametrizing $m = 3 \mu +\nu$ where $\nu=0,1,2$ and $\mu \geq 0$. 
\begin{itemize}
    \item For $\xi^3=-1$ we obtain, 
\bea\label{bp}
Q_0 - Q_3 & = & 
 { u^2 \, v^{-\half} \over 3 \sqrt{2} \, \pi } \sum_{\mu=0}^\infty 
 { \mu! \,  \Gamma (\half) \over \Gamma (\thalf-2 \mu ) \, (3\mu+2)!} 
 \, F(\thalf, \thalf; \thalf-2\mu ; \tfrac{v}{2})
  \left ( { u^3 \over v^2} \right )^\mu
\no \\
Q_1 - Q_4 & = &
-{ v^\frac{5}{6} \over 3 \sqrt{2} \, \pi }  \sum_{\mu=0}^\infty 
 { \Gamma(\tfrac{1}{3}+\mu ) \Gamma (\half) \over \Gamma (\tfrac{11}{6}-2 \mu ) \, (3\mu)!} 
 \, F(\thalf, \thalf; \tfrac{11}{6}-2\mu ; \tfrac{v}{2})
  \left ( { u^3 \over v^2} \right )^\mu
  \no \\
Q_2 - Q_5 & = &
{ u \, v^\frac{1}{6} \over 3 \sqrt{2} \, \pi } \sum_{\mu=0}^\infty 
 { \Gamma(\tfrac{2}{3}+\mu ) \Gamma (\half) \over \Gamma (\tfrac{7}{6}-2 \mu ) \, (3\mu+1)!} 
 \, F(\thalf, \thalf; \tfrac{7}{6}-2\mu ; \tfrac{v}{2})
  \left ( { u^3 \over v^2} \right )^\mu  
\eea
\item For $\xi^3=1$ we obtain, 
\bea\label{bm}
Q_0+Q_3 & = &
{u^2 \ v^{-\half} \over 3\sqrt{2} \, \pi } \sum_{\mu=0}^\infty 
 {\mu ! \, \Gamma (\half+2\mu) \over \Gamma(\half) (3\mu+2)!}
F(\thalf, \thalf; \thalf -2\mu; \tfrac{v}{2}) \left ( { u^3 \over v^2} \right )^\mu 
 \\
Q_1+Q_4 & = &
 \sum_{\mu=0}^\infty 
 { \Gamma(\tfrac{1}{3}+\mu) \over (3\mu)!}
\bigg [ {v^{\frac{5}{6}}    \over 3 \sqrt{2} \, \pi } \, { \Gamma (- \frac{5}{6}+2\mu) \over   \Gamma(\half)}
F(\thalf, \thalf; \tfrac{11}{6} -2\mu; \tfrac{v}{2}) \left ( { u^3 \over v^2} \right )^\mu
\no \\ && \hskip 1.1in 
{+} { 2^{2/3} \Gamma (\half) \Gamma (\frac{5}{6}-2 \mu ) \, (4 u^3)^\mu \over 6 \, \Gamma (\frac{2}{3}-\mu )^2 \Gamma (\frac{7}{6}-\mu )^2} F( 2 \mu - \tfrac{1}{3} , 2 \mu - \tfrac{1}{3} ; \tfrac{1}{6} +2 \mu ; \tfrac{v}{2})  \bigg ] 
\no \\
Q_2+Q_5 & = &
u \sum_{\mu=0}^\infty 
 { \Gamma(\tfrac{2}{3}+\mu) \over (3\mu+1)!}
\bigg [ 2^{1/3} {\Gamma (-\frac{1}{6}+2\mu) \, (4 u^3)^\mu  \over 6 \, \Gamma(\half)^3} (2v)^{\frac{1}{6} -2\mu} 
F(\thalf, \thalf; \tfrac{7}{6} -2\mu; \tfrac{v}{2})
\no \\ && \hskip 1.1in 
{+} { 2^{4/3} \Gamma (\half) \Gamma (\frac{1}{6}-2 \mu )\, (4 u^3)^\mu   \over 6 \, \Gamma (\frac{1}{3}-\mu )^2 \Gamma (\frac{5}{6}-\mu )^2} F(\tfrac{1}{3}+2 \mu , \tfrac{1}{3}+2 \mu ; \tfrac{5}{6} +2 \mu ; \tfrac{v}{2})  \bigg ] 
\no
\eea
\end{itemize}
Using the reflection formula, $Q_0+Q_3=Q_0-Q_3$ so that $Q_3=0$, consistent with \cite{DDN}.

\subsection{Convergence of the $\integers_3$ series}
\setcounter{equation}{0}
\label{sec:3}

To study the convergence properties of the series for $Q(\xi)$ when $\xi^3=-1$ we use the reflection and multiplication formulas for $\Gamma$-functions to obtain,
\bea
Q(\xi) &  = &  {v^\frac{5}{6} \over 3 \pi}    \sum_{\nu=0,1,2} (-)^{\nu+1} \xi ^{\nu +1} 
\sin \pi (\tfrac{4\nu-5}{6} )  \cQ_\nu
\no \\
\label{calQ}
\cQ_\nu & = & { 1 \over 2^\frac{4}{3} 3^\half} \sum_{\mu=0}^\infty 
 { \Gamma (\mu + \frac{4\nu-5}{12}) \Gamma (\mu + \frac{4\nu+1}{12}) 
 \over   \Gamma(\mu+\frac{\nu+2}{3}) \Gamma(\mu+\frac{\nu+3}{3}) } 
 \, F(\thalf, \thalf; \tfrac{11-4\nu}{6}-2\mu ; \tfrac{v}{2})
  \left ( { 4u^3 \over 27v^2} \right )^{\mu+ \frac{\nu}{3}}
\eea
We begin by considering the $\lambda \to +\infty$ behavior of the hypergeometric function which is given by the following Taylor series for fixed value of $|z|<1$, 
\bea
F(\thalf, \thalf; \lambda; z) = \sum _{k=0}^\infty 
{ \Gamma (k+\half)^2 \, \Gamma ( \lambda)  \over \Gamma (\half)^2 \, \Gamma (k+\lambda) \, k! } \, z^k
\eea
The first few terms are given by,
\bea
F(\thalf, \thalf; \lambda; z) = 1 + { z \over 4 \lambda} + { 9 \, z^2 \over 32 \, \lambda (\lambda +1)} 
+{ 75 \, z^3 \over 128 \, \lambda (\lambda+1) (\lambda+2)} + \cO(z^4)
\eea
The series is absolutely convergent for any compact subset of the open disc $|z|<1$ uniformly in $\lambda$ greater than any fixed number strictly greater than one; for our purposes it suffices to choose $\lambda \geq 1$. For fixed $|z|<1$, the absolute value of each term in the series strictly decreasing as $\lambda \to + \infty$ and therefore the limit of the series  as $\lambda \to +\infty$ is simply given by,
\bea
\lim _{\lambda \to + \infty} F(\thalf, \thalf; \lambda; z)  =1 \qquad \hbox{ for all } \quad |z|<1
\eea
Next, we consider the case where $ \lambda \to - \lambda$ in the hypergeometric function, and use the general  formula below to relate this case to the previous one,\footnote{The procedure of relating the cases for positive and negative $\lambda$  was followed in \cite{Temme}, but the coefficient of the first term is incorrect there. To establish the correct  relation, one easily verifies that all three functions satisfy the hypergeometric differential equation $z(1-z) f'' +(-\lambda -(a+b+1)z) f' -abf=0$. The coefficient of the second term on the right side may be determined by setting $z=0$ and using Gauss's formula for the hypergeometric function at unit argument, while the coefficient of the first term may be determined using the asymptotics of the left side $z \to 1$, using the formulas in section 2.7.1 of \cite{BatemanI}.}
\bea
F(a,b;-\lambda;z) & = &  {z^{\lambda+1} \Gamma (-\lambda) \Gamma (a+1+\lambda) \Gamma (b+1+\lambda) 
\over (1-z)^{\lambda +a+b} \Gamma (a) \Gamma (b) \Gamma (\lambda+2)}
 F(1-a,1-b;\lambda+2; z)
\no \\ &&
+ {\Gamma (a+1 +\lambda) \Gamma (b+1+\lambda) \over \Gamma (a+b+1+\lambda) \Gamma (\lambda+1)}
F(a,b;a+b+1+\lambda;1-z)
\eea
For the special case  $a=b=\thalf$ of interest  here, we have the following simplification,
\bea
F(\thalf, \thalf ;-\lambda;z) & = & - {z^{\lambda+1} \over (1-z)^{\lambda +1}}
 {\Gamma (-1-\lambda) \Gamma (\frac{3}{2} +\lambda)^2 
\over \Gamma (\half)^2 \Gamma (\lambda+1)}   F(\thalf,\thalf;\lambda+2; z)
\no \\ &&
+ {\Gamma (\tfrac{3}{2} +\lambda)^2  \over \Gamma (\lambda+2) \Gamma (\lambda+1)}
F(\thalf, \thalf ;\lambda+2;1-z)
\eea
For the first term to admit a finite limit as $\lambda \to + \infty$, we must require $|z|<|1-z|$, in which case the contribution from the first term tends to zero. That this sufficient condition is also necessary may be established numerically by taking the limit $z\to \half$ and verifying that the limit to be established below does not hold. In the absence of the first term, the limit of the second term is then given by,
\bea
F(\thalf, \thalf ;-\lambda;z) = {\Gamma (\tfrac{3}{2} +\lambda)^2  \over \Gamma (\lambda+2) \Gamma (\lambda+1)} 
\left ( 1 +{1-z \over 4 \lambda} + \cO(\lambda^{-2}) \right )
\eea
and thus, 
\bea
\lim _{\lambda \to + \infty} F(\thalf, \thalf ;-\lambda;z) = 1 \qquad \hbox{ for all } \quad |z|<\hbox{min}(1,|1-z|)
\eea

\subsubsection*{Convergence of the series for $\cQ_\nu$}

We shall now use the results of the preceding subsection to investigate the convergence properties of the series given in (\ref{calQ}) for $\cQ_\nu$. The large $\mu$ behavior of the prefactor of $\Gamma$-functions is as follows,
\bea
 { \Gamma (\mu + \frac{4\nu-5}{12}) \Gamma (\mu + \frac{4\nu+1}{12}) 
 \over   \Gamma(\mu+\frac{\nu+2}{3}) \Gamma(\mu+\frac{\nu+3}{3}) }  
= { 1 \over \mu^2} \left ( 1 + \cO(\lambda^{-1}) \right )
 \eea
In view of the asymptotics  of the hypergeometric function for $\mu \to \infty$ in the domain,
\bea
|\tfrac{v}{2}| < \thalf \hskip 1in |\tfrac{v}{2}| < |1-\tfrac{v}{2}|
\eea
the large $\mu$ behavior of the summand in $\cQ_\nu$ is as follows,
\bea
{ 1 \over \mu^2}  \left ( { 4u^3 \over 27v^2} \right )^{\mu+ \frac{\nu}{3}}
\eea
and the series is convergent provided,
\bea
\left |{ 4 u^3 \over 27} \right | < |v|^2=|1-u_0|^2  <1
\eea
since the condition $|\tfrac{v}{2}| < \thalf $ implies the condition $ |\tfrac{v}{2}| < |1-\tfrac{v}{2}|$.

\subsection{Elliptic expression for the $(\ma_1,\ma_2)$ K\"ahler potential}

{For completeness, we add here the exact expression for the intrinsic periods in terms of the elliptic formulation developed in subsection 2.4.3. The SW differential is given by, 
\begin{align}
    \lambda = -i \frac{dz}{2 \sqrt{2} (2\omega)^{5}}\left( 12 \wp(z)^3 -g_2\wp(z)\right),
\end{align}
Using the homology basis of subsection 2.4.3, $\mA = [0, 2\pi i],\mB = [0,2 \pi i \tau]$,  the SW periods may be read off from  Theorem \ref{2.thm:4} by setting $k=\ell=m=0$, and we have, 
\begin{align}
    a &= -i \, \frac{\HE_2 \HE_4 - \HE_6}{720 \sqrt{2 \pi }(2\omega)^5} &
    a_D - \tau a &= \frac{-\HE_4}{60 (2 \pi)^\frac{3}{2} (2\omega)^5} 
\end{align}
The right formula confirms that $a_D=\rho a$ for the AD theory since $\HE_4(\rho)=0$. As one approaches the AD point, $u, v \to 0$, which forces $\om \to \infty$ since $\HE_6(\rho)$ is non-vanishing. Thus, both periods tend to zero at the AD point, as expected.  The K\"ahler potential is given by,
\begin{align}
    K_\text{AD} = \frac{\Im (\tau) }{2 \pi} |a|^2 - {\Re \big (\bar \HE_4 ( \HE_2 \HE_4 - \HE_6) \big ) \over
    675 \, \pi^3 |4 \om|^{10}}
\end{align}
One verifies that upon setting  $\tau = \rho$ and then letting $\omega\rightarrow\infty$, the intrinsic K\"ahler potential $K_\text{AD}$ tends to zero as expected.}

\end{document}